\documentclass[10pt,twoside]{article}
\usepackage{amsmath}
\usepackage{textcomp}
\usepackage{fancyhdr}
\usepackage{multicol}
\usepackage{epsfig,subfigure}
\usepackage[authoryear,sort]{natbib}
\usepackage{SPE}
\usepackage{color}
\usepackage{bm}
\usepackage{hyphenat}
\usepackage[scaled=1]{helvet}
\usepackage[normalem]{ulem}
\usepackage{caption}

\usepackage{lineno}
\nolinenumbers
\usepackage{IEEEtrantools}
\usepackage{epstopdf}
\usepackage{url}
\usepackage{multirow}

\definecolor{myBlue}{rgb}{0,0,0.56}
\definecolor{myRed}{rgb}{0.5,0,0}
\makeatletter
\newcommand*{\rom}[1]{\uppercase\expandafter{\romannumeral #1\relax}}
\makeatother
\begin{document}

\setlength{\headheight}{15.2pt}
\pagestyle{fancy}
\lhead[\small \sffamily \thepage]{\small \sffamily}
\rhead[\small \sffamily ]{\small \sffamily \thepage}
\rfoot[]{}
\cfoot[]{}
\lfoot[]{}

\thispagestyle{empty}

\newcommand{\mean}[1]{\langle{#1}\rangle}

\newcommand\blueuline{\bgroup\markoverwith
{\textcolor{blue}{\rule[-0.5ex]{2pt}{0.4pt}}}\ULon}

\begin{figure}
 \includegraphics{SPEInt.eps}
\end{figure}

\vspace*{0cm}
{\noindent \Large \sffamily \textbf{}}

\vspace*{0.8cm} {\noindent \sffamily \Large
  \textbf{\nohyphens{Efficient big data assimilation through sparse representation: A 3D benchmark case
  		study in seismic history matching}}}

\vspace*{-.3cm} \bigskip {\noindent \sffamily \nohyphens{Xiaodong Luo,
    SPE, IRIS; Tuhin Bhakta, IRIS; Morten Jakobsen, UoB \& IRIS; and Geir N\ae{}vdal, IRIS. }

\vspace*{.7cm}
\renewcommand{\scriptsize}{\fontsize{6.3}{7.2}\selectfont}
\noindent \parbox[t]{19cm}{
\scriptsize \sffamily

\medskip

\noindent 

\medskip

\noindent
}

\smallskip

\vspace{-0.3cm}
\begin{figure}[hp]
  \includegraphics[width=19cm,height=0.2cm]{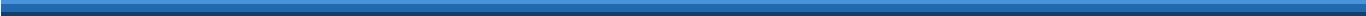}
\end{figure}

{\fontfamily{ptm}\selectfont
\section{Abstract}\label{sec:abstract}
\noindent In a previous work \citep{luo2016sparse2d_spej}, the authors proposed an ensemble-based 4D seismic history matching (SHM) framework, which has some relatively new ingredients, in terms of the type of seismic data in choice, the way to handle big seismic data and related data noise estimation, and the use of a recently developed iterative ensemble history matching algorithm. 

In seismic history matching, it is customary to use inverted seismic attributes, such as acoustic impedance, as the observed data. In doing so, extra uncertainties may arise during the inversion processes. The proposed SHM framework avoids such intermediate inversion processes by adopting amplitude versus angle (AVA) data. In addition, SHM typically involves assimilating a large amount of observed seismic attributes into reservoir models. To handle the big-data problem in SHM, the proposed framework adopts the following wavelet-based sparse representation procedure: First, a discrete wavelet transform is applied to observed seismic attributes. Then, uncertainty analysis is conducted in the wavelet domain to estimate noise in the resulting wavelet coefficients, and to calculate a corresponding threshold value. Wavelet coefficients above the threshold value, called leading wavelet coefficients hereafter, are used as the data for history matching. The retained leading wavelet coefficients preserve the most salient features of the observed seismic attributes, whereas rendering a substantially smaller data size. Finally, an iterative ensemble smoother is adopted to update reservoir models, in such a way that the leading wavelet coefficients of simulated seismic attributes better match those of observed seismic attributes. 

As a proof-of-concept study, \cite{luo2016sparse2d_spej} applied the proposed SHM framework to a 2D case study, and numerical results therein indicated that the proposed framework worked well. However, the seismic attributes used in \cite{luo2016sparse2d_spej} are 2D datasets with relatively small data sizes, in comparison to those in practice. In the current study, we extend our investigation to a 3D benchmark case, the Brugge field. The seismic attributes used are 3D datasets, with the total number of seismic data in the order of $\mathcal{O}(10^6)$. Our study indicates that, in this 3D case study, the wavelet-based sparse representation is still highly efficient in substantially reducing the size of seismic data, whereas preserving the information content therein as much as possible. Meanwhile, the proposed SHM framework also achieves reasonably good history matching performance. This investigation thus serves as an important preliminary step towards applying the proposed SHM framework to real field case studies.

\section*{Introduction}
\label{sec:introduction}
\noindent
Seismic is one of the most important tools used for reservoir exploration, monitoring, characterization and management in the petroleum industry. Compared to conventional production data used in history matching, seismic data is less frequent in time, but much denser in space. Therefore, complementary to production data, seismic data provide valuable additional information for reservoir characterization.

There are different types of seismic data that one can use in history matching. Figure \ref{fig:seis_type} provides an outline of the relation of some types of seismic data to reservoir petro-physical parameters (e.g., permeability and porosity) in forward simulations. As indicated there, using petro-physical parameters as the inputs to reservoir simulators, one generates fluid saturation and pressure fields. Through a petro-elastic model (PEM), one obtains acoustic and/or shear impedance (or equivalently, compressional and/or shear velocities (velocity) and formation density) based on fluid saturation and/or pressure, and porosity. Finally, amplitude versus angle (AVA) data are computed by plugging impedance (or velocities and density) into an AVA equation (e.g., Zoeppritz equation, see, for example, \citealp{avseth2010quantitative}).  

To reduce the computational cost in forward simulations, many seismic history matching (SHM) studies use inverted seismic attributes that are obtained through seismic inversions. Such inverted properties can be, for instance, acoustic impedance (see, for example, \citealp{emerick2012history,emerick2013history,fahimuddin2010ensemble,skjervheim2007incorporating}) or fluid saturation fronts (see, for example, \citealp{abadpour20134d,leeuwenburgh2014distance,trani2012seismic}). One issue in using inverted seismic attributes as the observational data is that, they may not provide uncertainty quantification for the observation errors, since inverted seismic attributes are often obtained using certain deterministic inversion algorithms.

\begin{figure*}[!htb]
	\centering
	\includegraphics[scale=0.45]{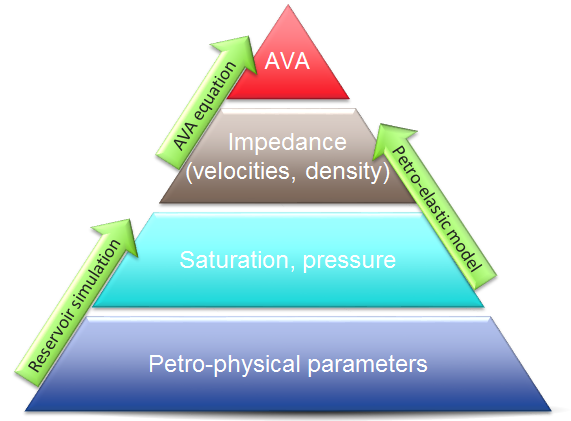}
	\caption{\label{fig:seis_type} Some types of seismic data and their relation to reservoir petro-physical parameters.}
\end{figure*}

Typically the volume of seismic data is huge, therefore SHM often constitutes a ``big data assimilation'' problem. For ensemble-based history matching algorithms, a big data size may lead to certain numerical problems, e.g., ensemble collapse and high costs in computing and storing Kalman gain matrices \citep{Aanonsen-ensemble-2009,emerick2012history}. In addition, many history matching algorithms are developed for under-determined inverse problems, whereas a big data size could make the inverse problem become over-determined instead. This may thus affect the performance of history matching algorithms, as demonstrated in our previous study \citep{luo2016sparse2d_spej}. 

\begin{figure*}[!htb]
	\centering
	\includegraphics[scale=0.4]{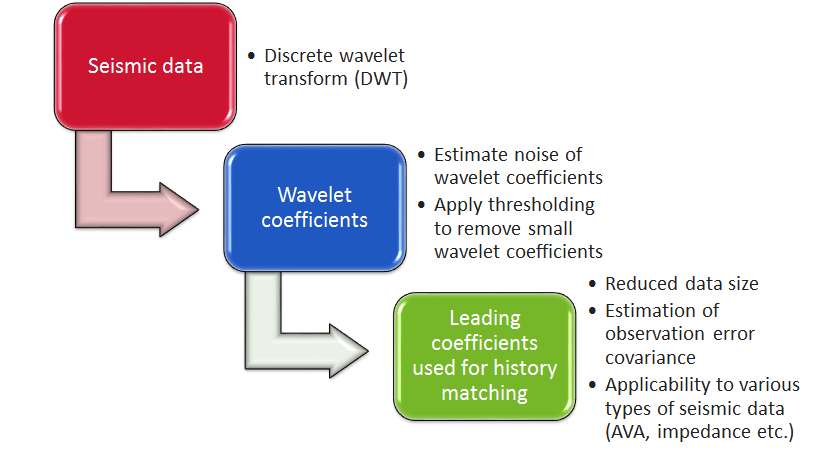}
	\caption{\label{fig:sparse_representation} Workflow of wavelet-based sparse representation.}
\end{figure*}

In \cite{luo2016sparse2d_spej}, we proposed an ensemble-based 4D SHM framework in conjunction with a wavelet-based sparse representation procedure. We take AVA attributes as the seismic data to avoid the extra uncertainties arising from a seismic inversion process. To address of the issue of big data, we adopt a wavelet-based sparse representation procedure. Figure \ref{fig:sparse_representation} explains the idea behind wavelet-based sparse representation. Given a set of seismic data (which can be 2D or 3D), one first applies a multilevel discrete wavelet transform (DWT) to the data. DWT is adopted for the following two purposes: one is to reduce the size of seismic data by exploiting the sparse representation nature of wavelet basis functions, and the other is to exploit its capacity of noise estimation in the wavelet domain \citep{donoho1995adapting}. Based on the estimated standard deviation (STD) of the noise, one can construct the corresponding observation error covariance matrix that is needed in many (including ensemble-based) history matching algorithms. 

For a chosen family of wavelet basis functions, seismic data are represented by the corresponding wavelet coefficients. When dealing with 2D data, DWT is similar to singular value decomposition (SVD) applied to a matrix. In the latter case, the matrix is represented by the corresponding singular values in the space spanned by the products of associated left and right singular vectors. Likewise, in the 2D case, one can also draw similarities between wavelet-based sparse representation and truncated singular value decomposition (TSVD). Indeed, in DWT, small wavelet coefficients are typically dominated by noise, whereas large coefficients mainly carry signal information \citep{jansen2012noise}. Therefore, as will be demonstrated later, it is possible for one to use only a small subset of leading wavelet coefficients to capture the main features of the signal, while significantly reducing the number of wavelet coefficients. We remark that TSVD-based sparse representation is not a suitable choice in the context of history matching, since the associated basis functions (i.e., products of left and right singular vectors) are data-dependent, meaning that in general it is not meaningful to compare and match the singular values of observed and simulated data. 

Wavelet-based sparse representation involves suppressing noise components in the wavelet domain. To this end, we first estimate the STD of the noise in wavelet coefficients, and then compute a threshold value that depends on both the noise STD and data size. One can substantially reduce the data size by only keeping leading wavelet coefficients above the threshold, while setting those below the threshold value to zero. The leading wavelet coefficients are then taken as the (transformed) seismic data, and are history-matched using a certain algorithm. In the experiments later, we will examine the impact of the threshold value on the history matching performance.   

A number of ensemble-based SHM frameworks have been proposed in the literature. For instance, \cite{abadpour20134d,fahimuddin2010ensemble,katterbauer2015history,leeuwenburgh2014distance,skjervheim2007incorporating,trani2012seismic} adopt the ensemble Kalman filter (EnKF) or a combination of the EnKF and ensemble Kalman smoother (EnKS), whereas \cite{emerick2012history,emerick2013history,luo2016sparse2d_spej} employ the ensemble smoother with multiple data assimilation (ES-MDA), and regularized Levenburg-Marquardt (RLM) based iterative ensemble smoother (RLM-MAC, see \citealp{luo2015Iterative}), respectively. We note that the history matching algorithm itself is independent of the wavelet-based sparse representation procedure. Therefore, one may combine the sparse representation procedure with a generic history matching algorithm, whether it is ensemble-based or not.  

This work is organized as follows. First, we will introduce three key components of the proposed workflow, which includes: (1) forward AVA simulation, (2) 3D DWT based sparse representation procedure, and (3) regularized Levenburg-Marquardt based iterative ensemble smoother. Then, we will apply the proposed framework to the 3D Brugge benchmark case, and investigate the performance of the proposed framework in various situations. Finally, we draw conclusions based on the results of our investigation and discuss possible future works.

\section{The proposed framework}\label{sec:framework}
\begin{figure*}[!htb]
	\centering
	\includegraphics[scale=0.4]{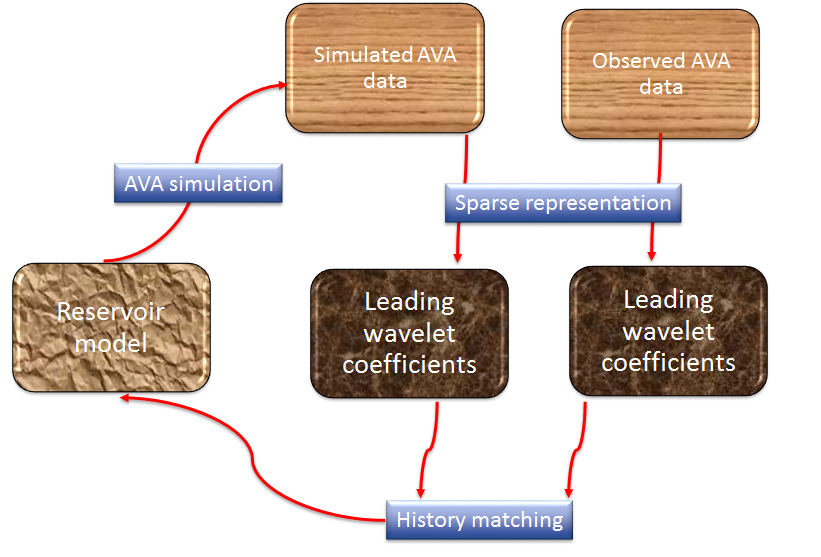}
	\caption{\label{fig:framework} The proposed 4D seismic history matching framework.}
\end{figure*}

The proposed framework consists of three key components (see Figure \ref{fig:framework}), namely, forward AVA simulation, sparse representation (in terms of leading wavelet coefficients) of both observed and simulated AVA data, and the history matching algorithm. It is expected that the proposed framework can be also extended to other types of seismic data, and more generally, geophysical data with spatial correlations. 

\subsection{Forward AVA simulation}

As can be seen in Figure \ref{fig:seis_type}, the forward AVA simulation involves several steps. First, pore pressure and fluid saturations are generated through reservoir flow simulation that takes petro-physical (e.g., permeability and porosity) and other parameters as the inputs. The generated pressure and saturation values are then used to calculate seismic parameters, such as P- and S-wave velocities and densities of reservoir and overburden formations, through a petro-elastic model (PEM). Finally, a certain AVA equation is adopted to compute the AVA attributes at different angles or offsets.  

Building a proper PEM is crucial to the success of SHM, whereas it is a challenging task at the same time. To interpret the changes in seismic response over time, an in-depth understanding of rock and fluid properties is required \citep{jack2001coming}. In this study, since our focus is on validating the performance of the proposed framework in a 3D problem, we assume that the PEM is perfect. Here, we use a soft sand model as the PEM \citep{mavko2009rock}. The model assumes that, the cement is deposited away from the grain contacts. It further considers that, the initial framework of the uncemented sand rock is densely random pack of spherical grains with the porosity (denote by $\phi$ hereafter) around 36\%, which is the maximum porosity value that the rock could have before suspension. For convenience of discussion later, we denote this value as the critical porosity $(\phi_c)$ \citep{nur1991wave, nur1998critical}. The dry bulk modulus $(K_{HM})$ and shear modulus $(\mu_{HM})$ at critical porosity can then be computed using the Hertz-Mindlin model \citep{Mindlin:1949} below
\begin{equation}
	K_{HM}= \sqrt[n]{\frac{C_{p}^2(1- \phi_c)^2 \mu_{s}^2}{18\pi^2(1-\nu_{s})^2}P_{eff}} \, ,
\end{equation}
and  
\begin{equation}
	\mu_{HM}= \frac{5-4\nu_{s}}{5(2-\nu_{s})}\sqrt[n]{\frac{3C_{p}^2(1- \phi_c)^2 \mu_{s}^2}{2\pi^2(1-\nu_{s})^2}P_{eff}} \, ,
\end{equation}
where $\mu_{s}$, $\nu_{s}$, $P_{eff}$ are grain shear modulus, Poisson's ratio, and effective stress, respectively. The coordination number $C_{p}$ denotes the average number of contacts per sphere, and $n$ is the degree of root. Here, $C_{p}$ and $n$ are set to 9 and 3, respectively.

To find the effective dry moduli for a porosity value less than $\phi_c$, the modified Lower Hashin-Shtrikman (MLHS) bound can be used \citep{mavko2009rock}. The MLHS connects two end points in the elastic modulus-porosity plane. One end point, ($K_{HM}$, $\mu_{HM}$), corresponds to critical porosity $\phi_c$. The other end point corresponds to zero porosity, taking the moduli of the solid phase, i.e. quartz mineral ($K_s$, $\mu_s$). For a porosity value $\phi$ between zero and $\phi_{c}$, the lower bound for dry rock effective bulk $K_{d}$ and shear moduli $G_{d}$ can be expressed as 
\begin{equation}\label{eqn:MUHS_K}
	K_{d} = \left[\frac{\frac{\phi}{\phi_{c}}}{K_{HM} + \frac{4}{3}\mu_{HM}} +  \frac{\frac{1 - \phi}{\phi_{c}}}{K_{s} + \frac{4}{3}\mu_{HM}} \right]^{-1} - \frac{4}{3}\mu_{HM}
\end{equation}
and 
\begin{equation}\label{eqn:MUHS_G}
	G_{d} = \left[\frac{\frac{\phi}{\phi_{c}}}{\mu_{HM} + \frac{\mu_{HM}}{6} Z} +  \frac{\frac{1 - \phi}{\phi_{c}}}{\mu_{s} + \frac{\mu_{HM}}{6} Z} \right]^{-1} - \frac{\mu_{HM}}{6} Z \, ,
\end{equation}  
respectively, where $K_{s}$ is solid/mineral bulk modulus and $ Z= (9K_{HM} + 8\mu_{HM}) / (K_{HM} + 2\mu_{HM})$. 

Further, the saturation effect is incorporated using the Gassmann model \citep{Gassmann:1951}. The saturated bulk modulus $K_{sat}$ and shear modulus $\mu_{sat}$ can be expressed as 
\begin{equation}
	K_{sat}= K_{d} + \frac{(1-\frac{K_{d}}{K_{s}})^2}{\frac{\phi}{K_{f}} + \frac{1 - \phi}{K_{s}} - \frac{K_{d}}{K_{s}^2}  }  \, , 
\end{equation}
and 
\begin{equation}
	\mu_{sat}= \mu_{d} \, ,
\end{equation}
respectively, where $K_{f}$ is the effective fluid bulk modulus, and is estimated using the Reuss average \citep{reuss1929berechnung}. For an oil-water mixture (as is the case in the Brugge field), $K_{f}$ is given by 
\begin{equation}
	K_{f}= (\frac{S_{w}}{K_{w}} + \frac{S_{o}}{K_{o}} )^{-1} \, , 
\end{equation}
where $K_{w}$, $K_{o}$, $S_{w}$ and $S_{o}$ are bulk modulus of water/brine, bulk modulus of oil, saturation of water/brine and saturation of oil, respectively.

Further, the saturated density \citep{mavko2009rock} can be written as (for the water-oil mixture)
\begin{equation}\label{eqn:density}
	\rho_{sat} = (1-\phi)\rho_{m} + \phi S_{w}\rho_{w} + \phi S_{o}\rho_{o}  \, ,
\end{equation}
where $\rho_{sat}$, $\rho_{m}$, $\rho_{w}$ and $\rho_{o}$ are saturated density of rock, mineral density, water/brine density and oil density, respectively. 

Using the above equations, we can obtain P- and S-wave velocities given by \citep{mavko2009rock} 
\begin{equation}
	V_{P} = \sqrt{\frac{K_{sat} + \frac{4}{3}\mu_{sat}}{\rho_{sat}}} \, ,
\end{equation}
and
\begin{equation}
	V_{S} =  \sqrt{\frac{\mu_{sat}}{\rho_{sat}}} \, ,
\end{equation}
where $V_{P}$ and $V_{S}$ represent P- and S-wave velocities, respectively.

After seismic parameters are generated by plugging reservoir parameters into the PEM, we can then simulate seismogram based on these seismic parameters. First, the Zoeppritz equation is used to calculate the reflection coefficient at an interface between two layers. For multi-layer cases, we need to calculate reflectivity series as a function of two-way travel time (see, for example, \citealp{buland2003bayesian,mavko2009rock}). Here, travel time is computed from the P-wave velocity and vertical thickness of each grid block. We then convolve the reflectivity series with a Ricker wavelet of 45 Hz dominant frequency to obtain the desired seismic AVA data. In the experiments later, we generate AVA data at two different angles (i.e. 10$^{\circ}$ and 20$^{\circ}$). We use these AVA attributes directly without converting them further to other attributes like intercept and gradient. In doing so, we avoid introducing extra uncertainties in the course of attribute conversion.  

\subsection{Sparse representation and noise estimation in wavelet domain}

Let $\mathbf{m}^{ref} \in \mathcal{R}^{m}$ denote the reference reservoir model. In the current study, we consider 3D AVA attributes (near- and far-offset traces) in the form of $p_1 \times p_2 \times p_3$ arrays (tensors), where $p_1$, $p_2$ and $p_3$ represent the numbers of inline, cross-line and time slices in seismic surveys, respectively. Accordingly, let $\mathbf{g}: \mathcal{R}^{m} \rightarrow \mathcal{R}^{p_1 \times p_2 \times p_3}$ be the forward simulator of AVA attributes. The observed AVA attributes $\mathbf{d}^o$ are supposed to be the forward simulation $\mathbf{g}(\mathbf{m}^{ref})$ with respect to the reference model, plus certain additive observation errors $\boldsymbol{\epsilon}$, that is, 
\begin{linenomath*} 
	\begin{equation} \label{eq:obs_system}
		\mathbf{d}^o = \mathbf{g}(\mathbf{m}^{ref}) + \boldsymbol{\epsilon} \, .
	\end{equation}     
\end{linenomath*}  
For ease of discussion below, suppose at the moment that all the tensors in Eq. (\ref{eq:obs_system}), i.e., $\mathbf{d}^o$, $\mathbf{g}(\mathbf{m}^{ref})$ and $\boldsymbol{\epsilon}$, are reshaped into vectors with $p_1 \times p_2 \times p_3$ elements. Throughout this study, we assume that, for a given AVA attribute, the elements of $\boldsymbol{\epsilon}$ are independently and identically distributed (i.i.d) Gaussian white noise, with zero mean but unknown variance $\sigma^2$, where $\sigma$ will be estimated through wavelet multiresolution analysis below. More generally, one may also assume that $\boldsymbol{\epsilon}$ follows a Gaussian distribution with zero mean and covariance $\sigma^2 \, \mathbf{R}$, where $\mathbf{R}$ is a known covariance matrix and $\sigma^2$ is a scalar to be estimated. In this case, one can whiten the additive noise by multiplying both sides of Eq. (\ref{eq:obs_system}) by $\mathbf{R}^{-1/2}$.       

As shown in Figure \ref{fig:sparse_representation}, wavelet-based sparse representation involves the following steps: (\rom{1}) Apply DWT to seismic data; (\rom{2}) Estimate noise STD of wavelet coefficients; and (\rom{3}) Compute a threshold value that depends on both noise STD and data size, and do thresholding accordingly.

\begin{figure*}[!htb]
	\centering
	\includegraphics[scale=0.4]{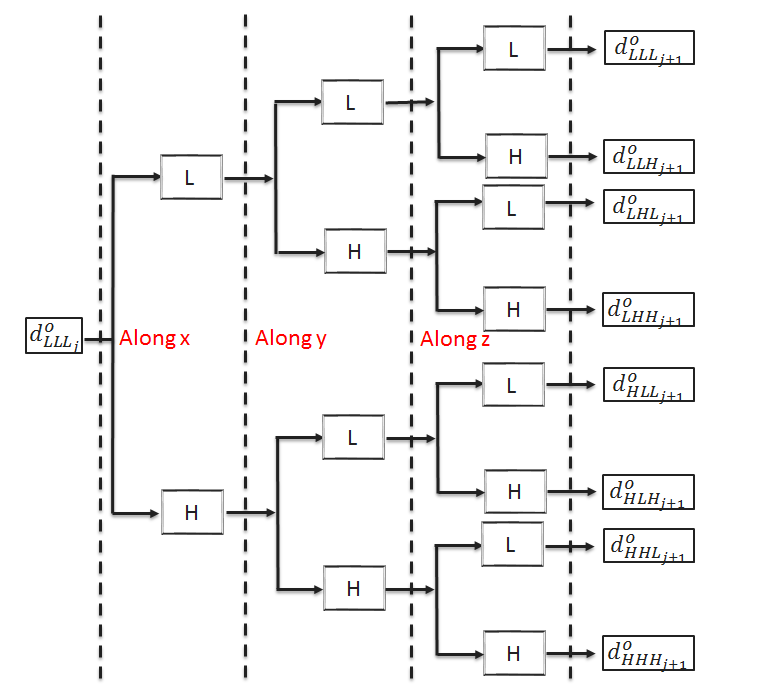}
	\caption{\label{fig:3D_DWT} A single level decomposition in 3D decimated discrete wavelet transform.}
\end{figure*}

We adopt 3D DWT to handle AVA attributes in this study. An introduction to wavelet theory is omitted here, and interested readers are referred to, for example, \cite{mallat1999wavelet}. 
Figure \ref{fig:3D_DWT} illustrates a single level decomposition in 3D DWT. For convenience, we call these three dimensions of AVA attributes $x$, $y$ and $z$, respectively. The input data is $\mathbf{d}^o_{LLL_j}$, which can be either the original observation $\mathbf{d}^o$ (when $j=0$) or the partial data recovered using the wavelet coefficients of the sub-band $LLL_j$ (when $j>0$). Without loss of generality, let us assume $j=0$, such that Figure \ref{fig:3D_DWT} corresponds to the first level 3D wavelet transform. In this case, the 3D transform is achieved by sequentially applying 1D DWT along $x$, $y$ and $z$ directions. In the 1D DWT along each direction, there are both low- (L) and high-pass (H) filters, and the transform results in one ``L'' and one ``H'' sub-band of wavelet coefficients, respectively. The ``H'' sub-band corresponds to high frequency components in wavelet domain, while the ``L'' sub-band to low frequency ones. As a result, after the first level of 3D DWT, there are 8 sub-bands in the wavelet domain, which are labelled as $LLL_1$, $LLH_1$, $LHL_1$, $LHH_1$, $HLL_1$, $HLH_1$, $HHL_1$ and $HHH_1$, respectively. The sub-band $LLL_1$ ($HHH_1$) results only from low-pass (high-pass) filters, while the others from mixtures of low- and high-pass ones. One can continue the 3D DWT to the next level by applying the transform to the data $\mathbf{d}^o_{LLL_1}$ that corresponds to the sub-band $LLL_1$. This leads to a set of new sub-bands of wavelet coefficients (labelled as $LLL_2$, $LLH_2$, $LHL_2$, $LHH_2$, $HLL_2$, $HLH_2$, $HHL_2$ and $HHH_2$, respectively), and so on. 

Since $HHH_1$ corresponds to the high frequency (typically noise) components of the original data $\mathbf{d}^o$, it can be used to infer noise STD in the wavelet domain. Specifically, let $\mathcal{W}$ and $\mathcal{T}$ denote orthogonal wavelet transform and thresholding operators, respectively; $\tilde{\mathbf{d}}^o = \mathcal{W} (\mathbf{d}^o)$ stands for the whole set of wavelet coefficients corresponding to the original data $\mathbf{d}^o$, and $\tilde{\mathbf{d}}^o_{HHH_1}$ for the wavelet coefficients in the sub-band $HHH_1$. After DWT and thresholding, the effective observation system becomes          
\begin{linenomath*} 
	\begin{equation} \label{eq:tw_obs_system}
		\mathcal{T} \circ \mathcal{W} (\mathbf{d}^o) = \mathcal{T} \circ \left( \mathcal{W} \circ \mathbf{g}(\mathbf{m}^{ref}) + \mathcal{W} (\boldsymbol{\epsilon}) \right) \, .
	\end{equation}     
\end{linenomath*} 
As will be discussed below, for leading wavelet coefficients (those above the threshold value), $\mathcal{T}$ is an identity operator such that it does not modify the values of leading wavelet coefficients. The reason for us to require orthogonal $\mathcal{W}$ is as follows: If $\mathcal{W}$ is orthogonal, then the wavelet transform preserves the energy of Gaussian white noise (e.g., the Euclidean norm of the noise term $\boldsymbol{\epsilon}$). In addition, like the power spectral distribution of white noise in frequency domain, the noise energy in the wavelet domain is uniformly distributed among all wavelet coefficients \citep{jansen2012noise}. This implies that, if one can estimate the noise STD $\sigma$ of small wavelet coefficients (e.g., those in $HHH_1$), then this estimation can also be used to infer the noise STD of leading wavelet coefficients used in history matching. Similar to our previous study \cite{luo2016sparse2d_spej}, the noise STD $\sigma$ is estimated using the median absolute deviation (MAD) estimator \citep{donoho1995adapting}:              
\begin{linenomath*} 
	\begin{equation} \label{eq:noise_std_mad}
		\sigma = \dfrac{\operatorname{median}(\operatorname{abs}(\tilde{\mathbf{d}}^o_{HHH_1}))}{0.6745} \, ,
	\end{equation}     
\end{linenomath*} 
where $\operatorname{abs}(\bullet)$ is an element-wise operator, and takes the absolute value of an input quantity. 

After estimating $\sigma$ in an $n$-level wavelet decomposition, we apply hard thresholding and select leading wavelet coefficients on the element-by-element basis, in a way such that 
\begin{linenomath*} 
	\begin{equation} \label{eq:hard_thresholding}
		\mathcal{T}(\tilde{{d}}^o_i) = 
		\begin{cases}
			0       & \quad \text{if } \tilde{{d}}^o_i < \lambda \, ,\\
			\tilde{{d}}^o_i  & \quad \text{otherwise} \, ,\\
		\end{cases}
	\end{equation}     
\end{linenomath*} 
where, without loss of generality, the scalar $\tilde{{d}}^o_i \in \tilde{\mathbf{d}}^o$ represents an individual wavelet coefficient, and $\lambda$ is a certain threshold value to be computed later. Eq. (\ref{eq:hard_thresholding}) means that, for leading wavelet coefficients above (or equal to) the threshold $\lambda$, their values are not changed, whereas for those below $\lambda$, they are set to zero. Note that in \cite{luo2016sparse2d_spej}, hard thresholding is not applied to the coarsest sub-band (i.e., the $LL_n$/$LLL_n$ sub-band for an $n$-level 2D/3D DWT) in light of the fact that the wavelet coefficients in this sub-band correspond to low-frequency components, which are typically dominated by the signal. As a result, applying thresholding to this sub-band may lead to certain loss of signal information in history matching. However, for an AVA attribute in this study, we have observed that its corresponding $LLL_n$ sub-band may contain a large amount of wavelet coefficients (e.g., in the order of $10^4$). To have the flexibility of efficiently reducing the data size, we lift the restriction such that thresholding can also be applied to the sub-band $LLL_n$.     

In \cite{luo2016sparse2d_spej}, the threshold value $\lambda$ is computed using the universal rule \citep{donoho1994ideal}
\begin{linenomath*} 
	\begin{equation} \label{eq:universal_rule}
		\lambda = \sqrt{2 \, \operatorname{ln}(\# \mathbf{d}^o)} \, \sigma \, ,
	\end{equation}     
\end{linenomath*}  
with $\# \mathbf{d}^o$ being the number of elements in $\mathbf{d}^o$. In the current work, when using Eq. (\ref{eq:universal_rule}) to select the threshold value, it is found that the resulting number of leading wavelet coefficients may still be very large. As a result, in the experiments later, we select the threshold value according to    
\begin{linenomath*} 
	\begin{equation} \label{eq:multiple_universal_rule}
		\lambda = c \, \sqrt{2 \, \operatorname{ln}(\# \mathbf{d}^o)} \, \sigma \, ,
	\end{equation}     
\end{linenomath*}
where $c>0$ is a positive scalar, and the larger the value of $c$, the less the number of leading wavelet coefficients. Therefore, the scalar $c$ can be used to control the total number of leading wavelet coefficients.  

Combining Eqs. (\ref{eq:tw_obs_system}) -- (\ref{eq:multiple_universal_rule}), the effective observation system in history matching becomes
\begin{linenomath*} 
	\begin{equation} \label{eq:reduced_obs_system}
		\tilde{\mathbf{d}}^o = \mathcal{W} \circ \mathbf{g}(\mathbf{m}^{ref}) + \mathcal{W} (\boldsymbol{\epsilon}) \, , \text{ for } \tilde{\mathbf{d}}^o \geq \lambda \, ,
	\end{equation}     
\end{linenomath*}  
where now $\tilde{\mathbf{d}}^o$ is a vector containing all selected leading wavelet coefficients, and $\mathcal{W} (\boldsymbol{\epsilon})$ the corresponding noise component in the wavelet domain, with zero mean and covariance $\mathbf{C}_{\tilde{\mathbf{d}}^o} = \sigma^2 \mathbf{I}$ (here $\mathbf{I}$ is the identity matrix with a suitable dimension).

\newcommand{\nScale}{0.33}
\begin{figure*}
	{
		\centering
		\subfigure[Reference AVA far-offset trace]{ \label{subfig:ref_data}
			\includegraphics[scale=\nScale]{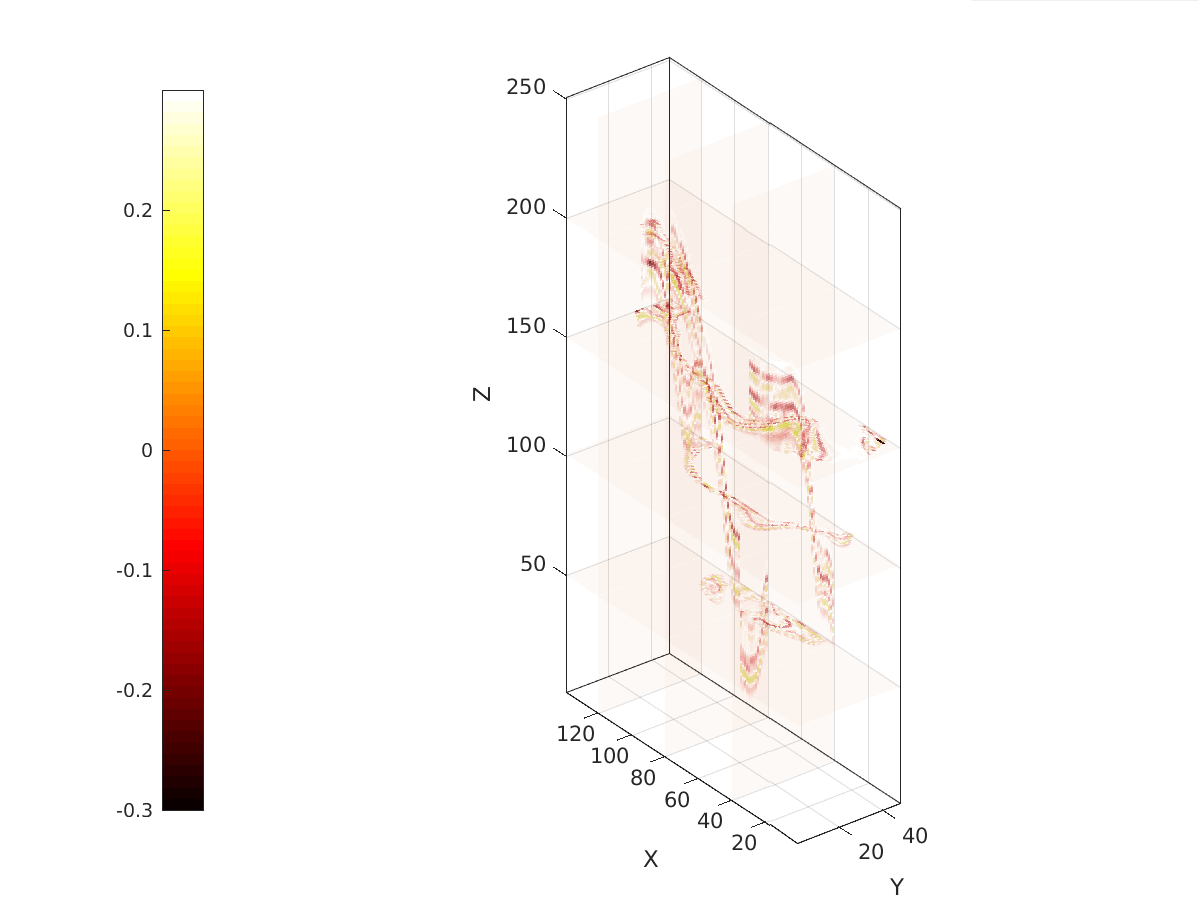}
		}
		\subfigure[Noisy AVA far-offset trace]{ \label{subfig:noisy_data}
			\includegraphics[scale=\nScale]{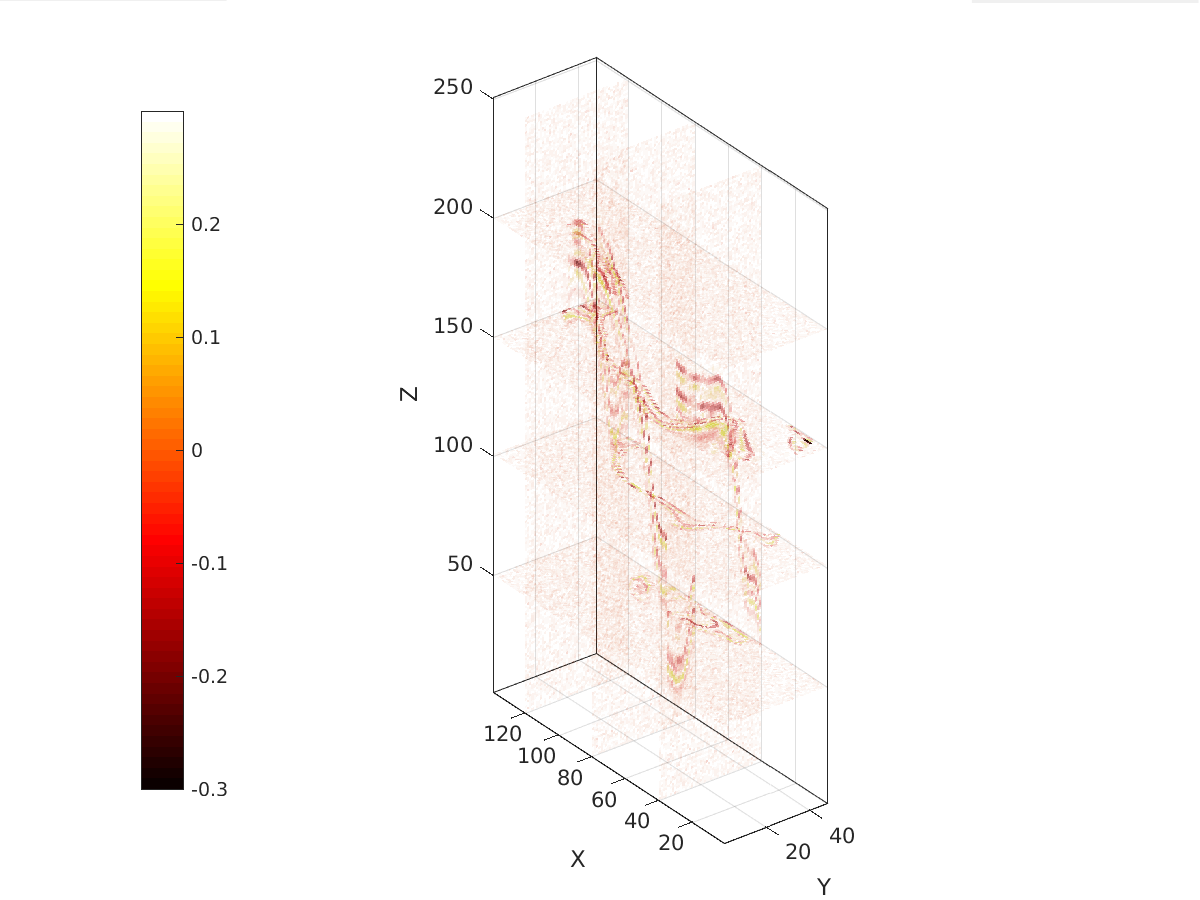}
		}
		\subfigure[Reconstructed  AVA far-offset trace]{ \label{subfig:denoised_data}
			\includegraphics[scale=\nScale]{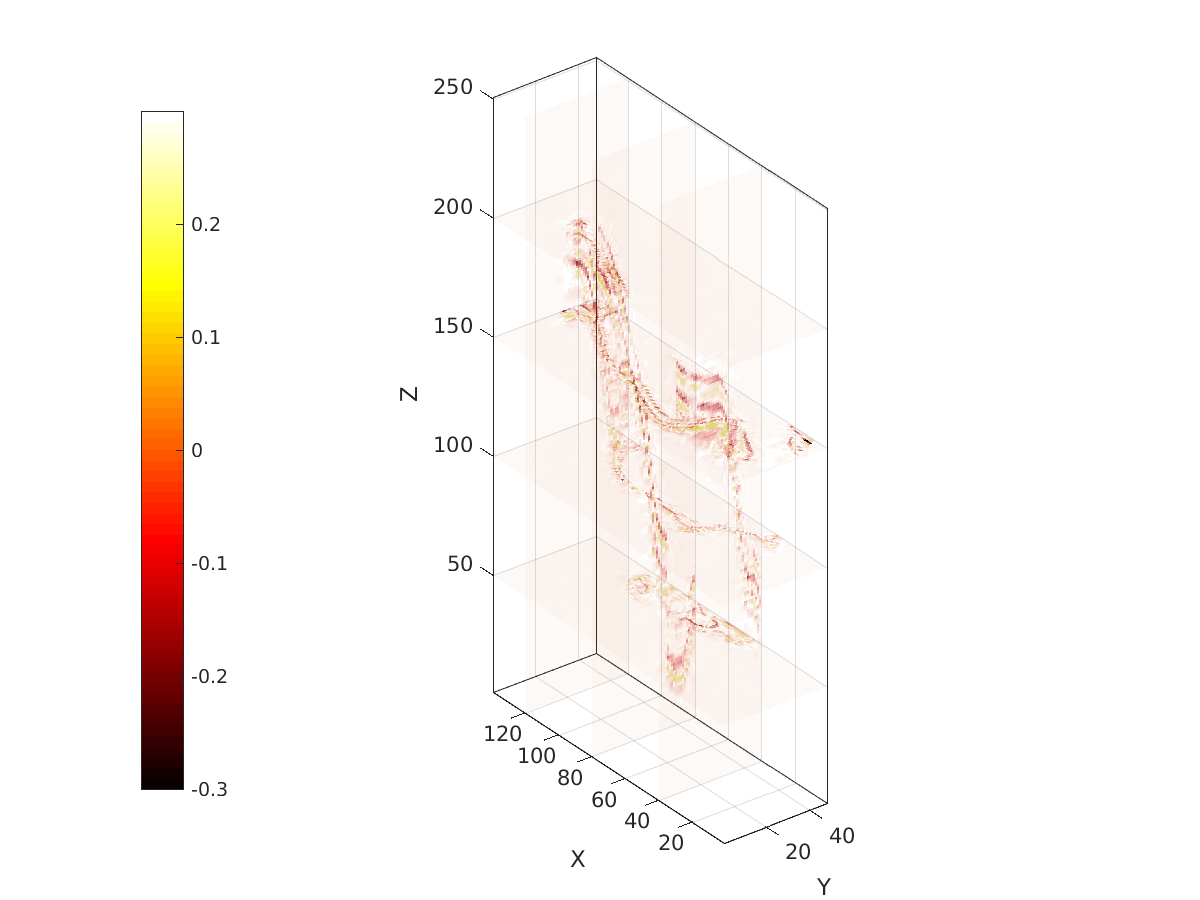}
		}
		
		\subfigure[$HHL_1$ of reference trace]{ \label{subfig:ref_HLL1}
			\includegraphics[scale=\nScale]{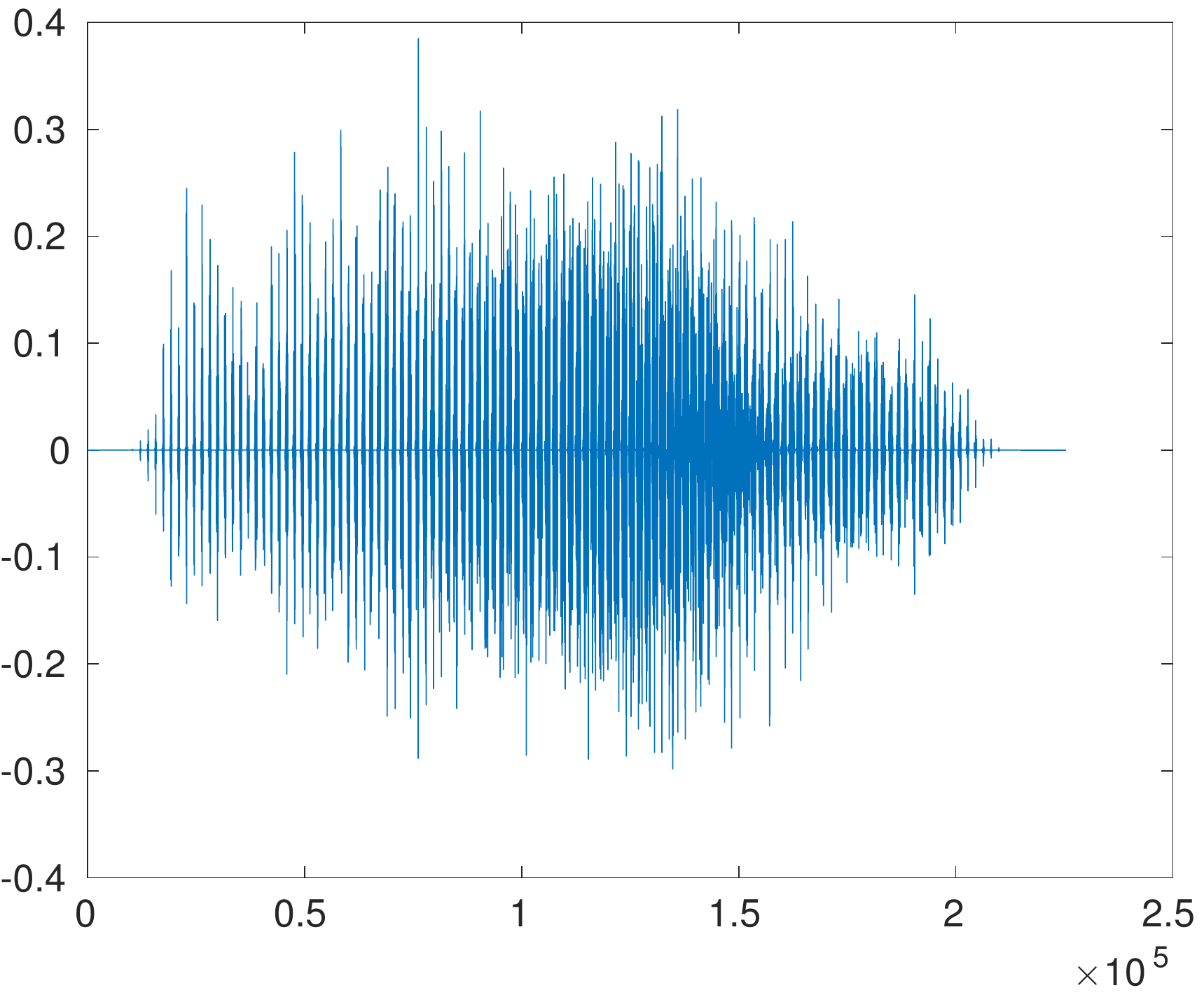}
		}\hfill
		\subfigure[$HHL_1$ of noisy trace]{ \label{subfig:noisy_HLL1}
			\includegraphics[scale=\nScale]{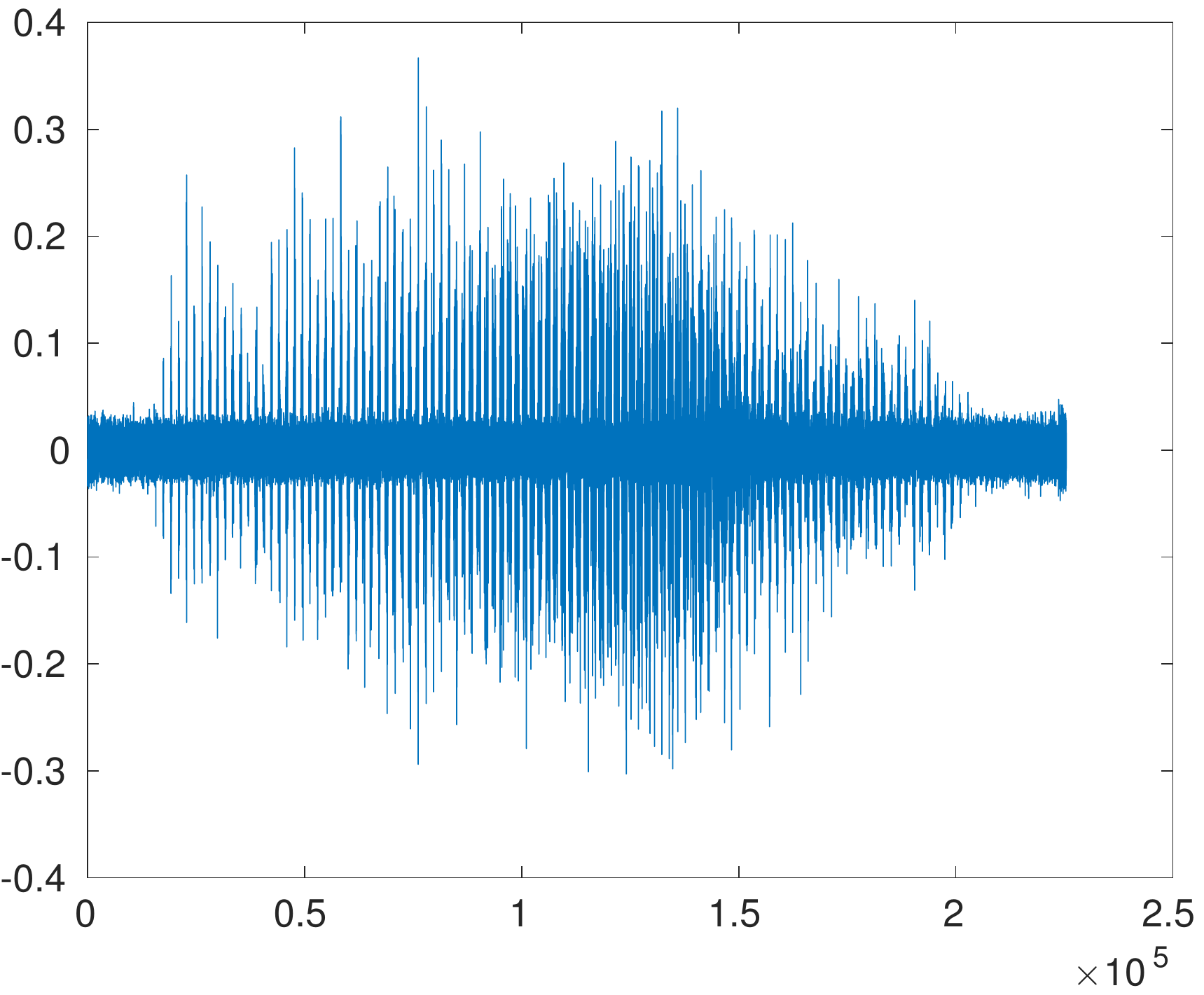}
		}\hfill	
		\subfigure[$HHL_1$ of reconstructed trace]{ \label{subfig:rec_HLL1}
			\includegraphics[scale=\nScale]{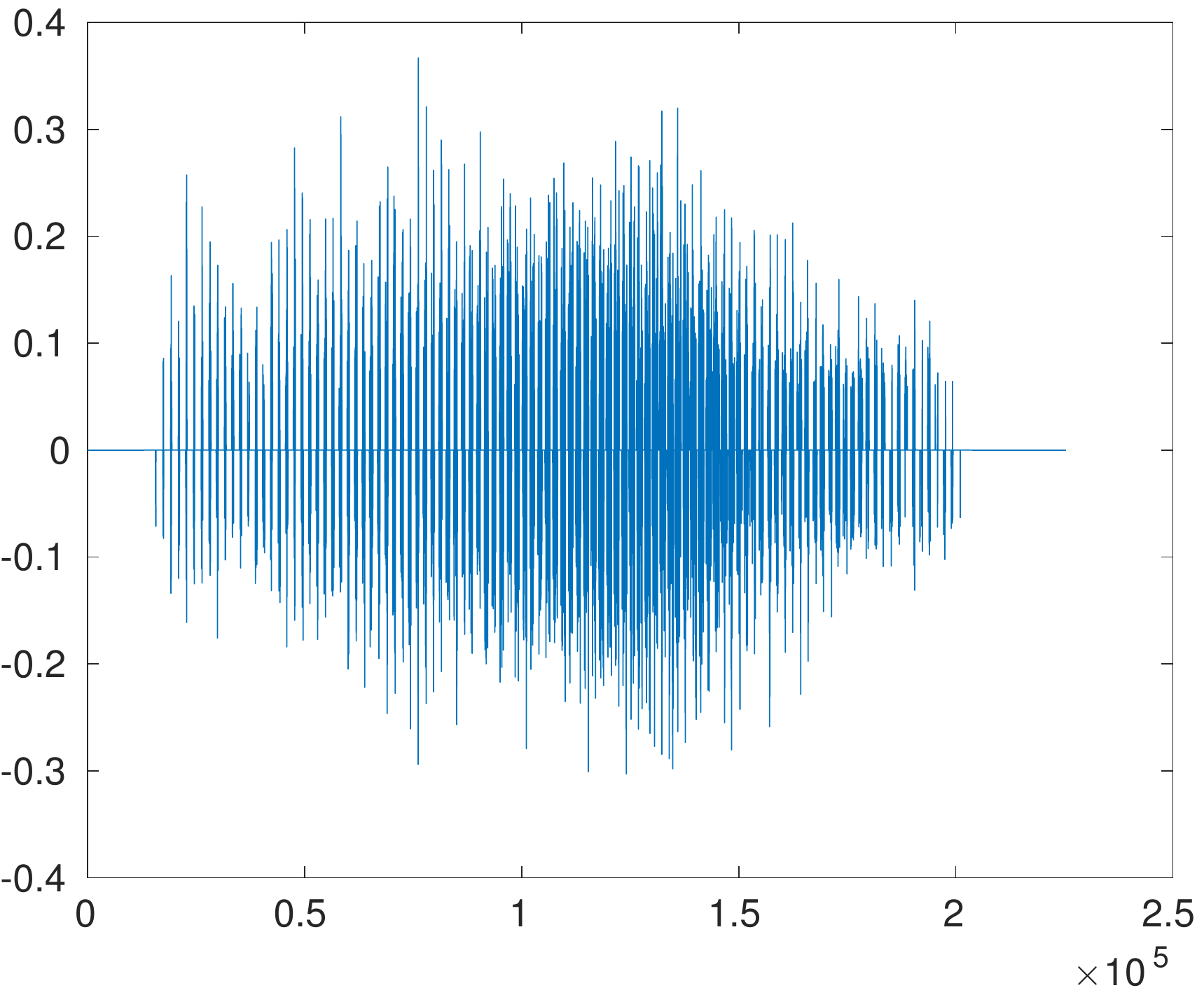}
		}\hfill
		
		\subfigure[Reference noise]{ \label{subfig:ref_noise}
			\includegraphics[scale=\nScale]{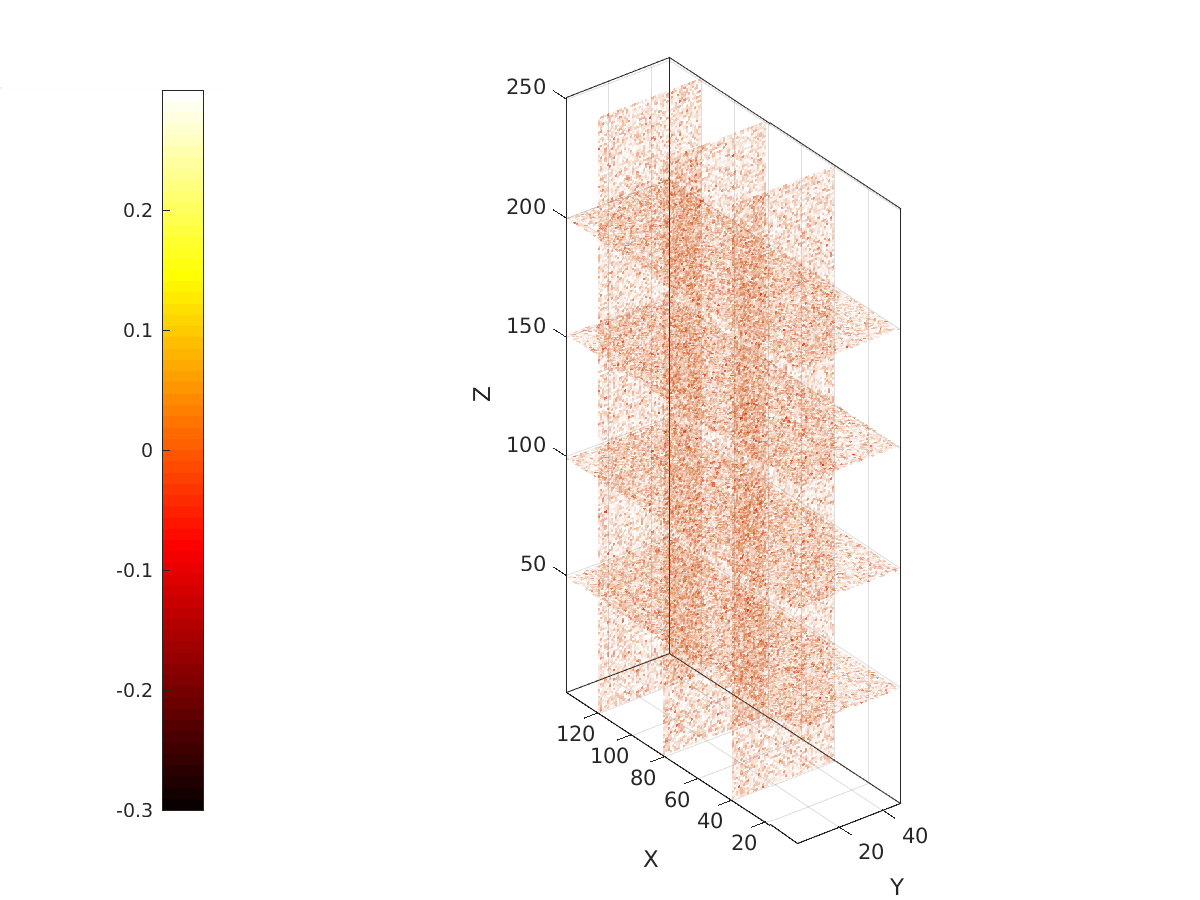}
		}
		\subfigure[Estimated noise]{ \label{subfig:rec_noise}
			\includegraphics[scale=\nScale]{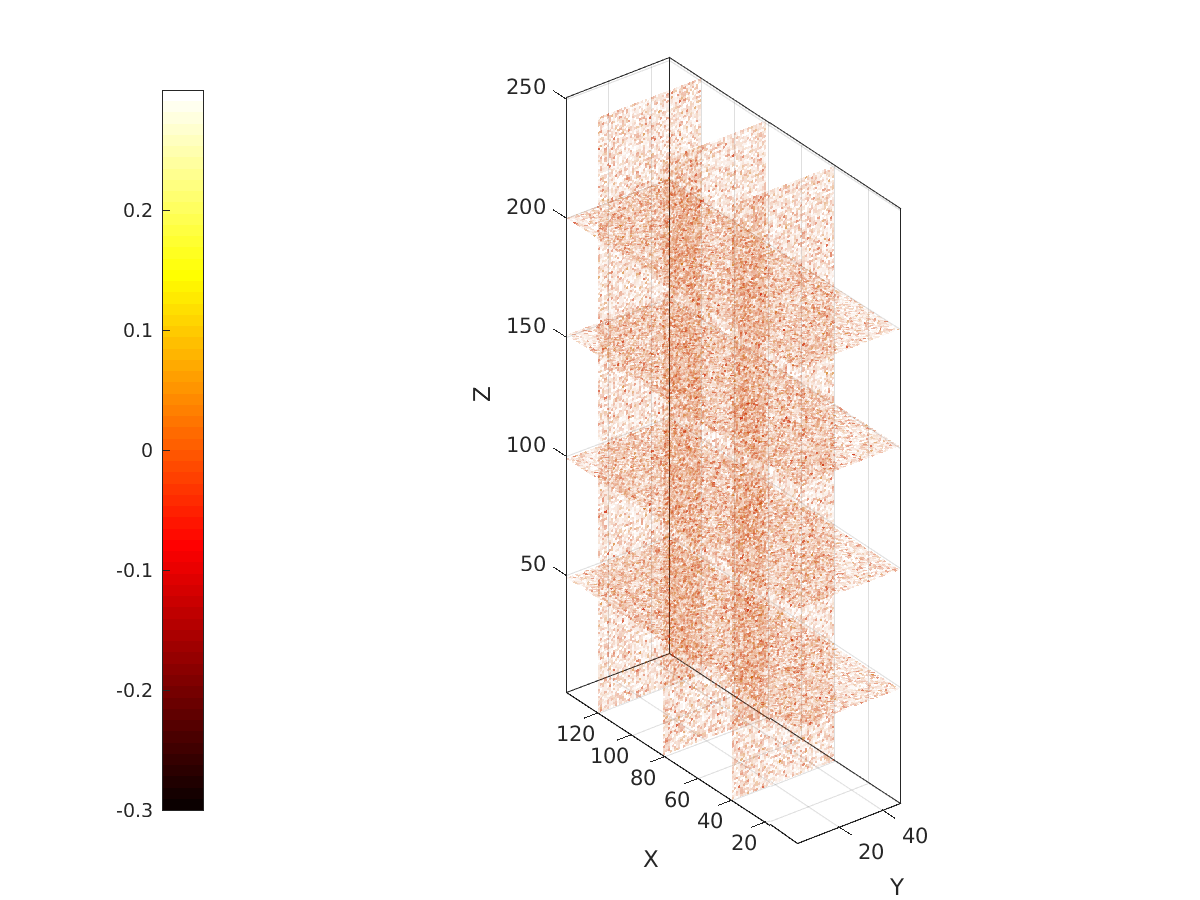}
		}
		\subfigure[Noise difference]{ \label{subfig:diff_noise}
			\includegraphics[scale=\nScale]{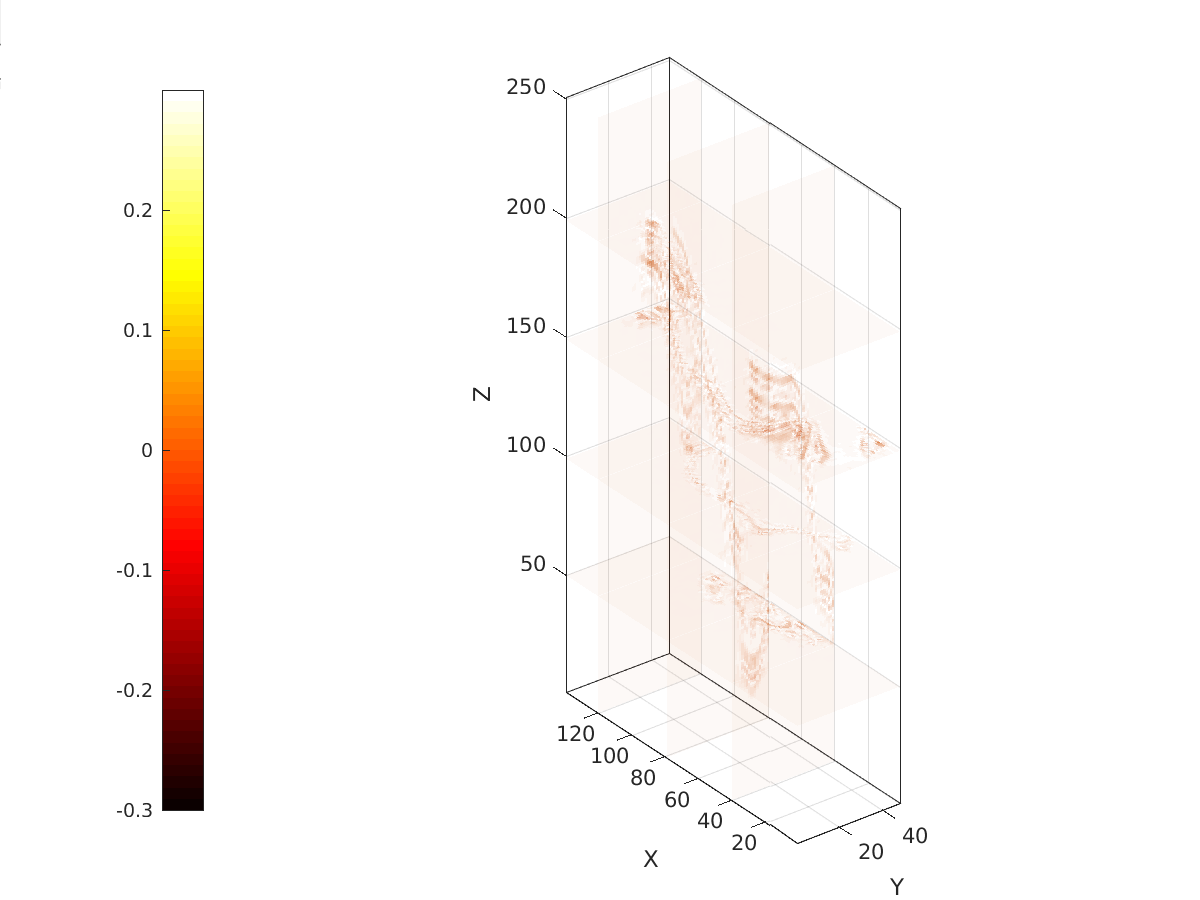}
		}
	}
	\caption{\label{fig:illustration_data} Illustration of sparse representation of a 3D AVA far-offset trace using slices at $X=40, 80, 120$ and at $Z= 50, 100, 150, 200$, respectively. (a) Reference AVA trace; (b) Noisy AVA trace obtained by adding Gaussian white noise (noise level = 30\%) to the reference data; (c) Reconstructed AVA trace obtained by first conducting a 3D DWT on the noisy data, then applying hard thresholding (using the universal threshold value) to wavelet coefficients, and finally reconstructing the data using an inverse 3D DWT based on the modified wavelet coefficients; (d) Wavelet sub-band $HHL_1$ corresponding to the reference AVA data; (e) Wavelet sub-band $HHL_1$ corresponding to the noisy AVA data; (f) Wavelet sub-band $HHL_1$ corresponding to the reconstructed AVA data; (g) Reference noise, defined as noisy AVA data minus reference AVA data; (g) Estimated noise, defined as noisy AVA data minus reconstructed AVA data; (i) Noise difference, defined as estimated noise minus reference noise. All 3D plots are created using the package \it{Sliceomatic} (version 1.1) from MATLAB Central (File ID: \#764).} 
\end{figure*} 

We use an example to illustrate the performance of sparse representation and noise estimation in 3D DWT. In this example, we first generate a reference AVA far-offset trace using the forward AVA simulator. The dimension of this trace is $139 \times 48 \times 251$, therefore the data size is $1,674,672$. Figure \ref{subfig:ref_data} plots slices of the AVA trace at $X=40$, $80$, $120$, and $Z=50$, $100$, $150$ and $200$, respectively. We then add Gaussian white noise to obtain the noisy AVA trace, with the noise level being $30\%$. Here, noise level is defined as: 
\begin{linenomath*} 
	\begin{equation} \label{eq:def_noise_level}
		\text{Noise level } = \dfrac{\text{variance of noise}}{\text{variance of pure signal}} \, .  
	\end{equation}     
\end{linenomath*} 
Figure \ref{subfig:noisy_data} shows slides of the noisy AVA trace at the same locations as in Figure \ref{subfig:ref_data}. 

We apply a three-level 3D DWT to the noisy data using Daubechies wavelets with two vanishing moments \citep{mallat1999wavelet}, and use hard thresholding combined with the universal rule (Eqs. (\ref{eq:noise_std_mad}) -- (\ref{eq:universal_rule})) to select leading wavelet coefficients. After thresholding, the number of leading wavelet coefficients reduces to $33,123$, only $2\%$ of the original data size. On the other hand, by applying Eq. (\ref{eq:noise_std_mad}), the estimated noise STD is $0.0105$, and it is very close to the true noise STD $0.0104$. By applying an inverse 3D DWT to leading wavelet coefficients, we obtain the reconstructed AVA trace (Figure \ref{subfig:denoised_data} plots slices of this trace at the same places as Figures \ref{subfig:ref_data} and \ref{subfig:noisy_data}). Comparing Figures \ref{subfig:ref_data} -- \ref{subfig:denoised_data}, one can see that, using leading wavelet coefficients that amounts to only $2\%$ of the original data size, the slices of reconstructed AVA trace well capture the main features in the corresponding slices of the reference AVA data.         

Figures \ref{subfig:ref_HLL1} -- \ref{subfig:rec_HLL1} show wavelet coefficients in the sub-bands $HHL_1$ of the reference, noisy and reconstructed AVA traces, respectively. From these figures, one can see that, after applying thresholding to wavelet coefficients of noisy data (Figure \ref{subfig:noisy_HLL1}), the modified coefficients (Figure \ref{subfig:rec_HLL1}) preserve those with large amplitudes in the reference case (Figure \ref{subfig:ref_HLL1}). In general, the modified coefficients appear similar to those of the reference case, whereas certain small coefficients of the reference case are suppressed due to thresholding. 

Finally, Figures \ref{subfig:ref_noise} -- \ref{subfig:diff_noise} depict slices of reference and estimated noise, and their difference, respectively, at the same places as in Figure \ref{subfig:ref_data}. Here, reference noise is defined as noisy AVA data (Figure \ref{subfig:noisy_data}) minus reference AVA data (Figure \ref{subfig:ref_data}), estimated noise as noisy AVA data minus reconstructed AVA data (Figure \ref{subfig:denoised_data}), and noise difference as estimated noise minus reference noise. The estimated noise appears very similar to the reference noise, although there are also certain differences according to Figure \ref{subfig:diff_noise}. This might be largely due to the fact that some small wavelet coefficients of the reference data are smeared out after thresholding, as aforementioned.

\subsection{The ensemble history matching algorithm}

We adopt the RLM-MAC algorithm \citep{luo2015Iterative} in history matching. 
Without loss of generality, let $\mathbf{d}^o$ denote $p$-dimensional observations in history matching, which stands for values in the ordinary data space (e.g., 3D AVA attributes by reshaping 3D arrays into vectors), or their sparse representations in the transform domain (e.g., leading wavelet coefficients in wavelet domain). The observations $\mathbf{d}^o$ are contaminated by Gaussian noise with zero mean and covariance $\mathbf{C}_d$ (denoted by $\mathbf{d}^o \sim N(\mathbf{0},\mathbf{C}_d)$). Also denote by $\mathbf{g}$ the forward simulator that generates simulated observations $\mathbf{d} \equiv \mathbf{g}(\mathbf{m})$ given an $m$-dimensional reservoir model $\mathbf{m}$. 

In the RLM-MAC algorithm, let $\mathbf{M}^i \equiv \{ \mathbf{m}_j^i \}_{j=1}^{{N_e}}$ be an ensemble of ${N_e}$ reservoir models obtained at the $i$th iteration step, based on which we can construct two square root matrices used in the RLM-MAC algorithm. One of the matrices is in the form of
\begin{linenomath*} 
	\begin{IEEEeqnarray}{clc} \label{eq:model_sqrt}
		& \mathbf{S}_m^i = \frac{1}{\sqrt{{N_e}-1}}\left[\mathbf{m}_1^i - \bar{\mathbf{m}}^i,\dotsb, \mathbf{m}_{N_e}^i - \bar{\mathbf{m}}^i \right] \, ; & \quad \bar{\mathbf{m}}^i = \frac{1}{{N_e}} \sum_{j=1}^{{N_e}} \mathbf{m}_j^i \, ,
	\end{IEEEeqnarray}
\end{linenomath*}   
and is called \textit{model square root matrix}, in the sense that $\mathbf{C}_{m}^{i} \equiv \mathbf{S}_m^i \left( \mathbf{S}_m^i \right)^T$ equals the sample covariance matrix of the ensemble $\mathbf{M}^i$. The other, defined as  
\begin{linenomath*} 
	\begin{IEEEeqnarray}{clc} \label{eq:data_sqrt}
		& \mathbf{S}_d^i = \frac{1}{\sqrt{{N_e}-1}}\left[\mathbf{g}(\mathbf{m}_1^i) - \mathbf{g}(\bar{\mathbf{m}}^i),\dotsb, \mathbf{g}(\mathbf{m}_{N_e}^i) - \mathbf{g}(\bar{\mathbf{m}}^i) \right] \, ,
	\end{IEEEeqnarray}
\end{linenomath*}   
is called \textit{data square root matrix} for a similar reason. 

The RLM-MAC algorithm updates $\mathbf{M}^i$ to a new ensemble $\mathbf{M}^{i+1} \equiv \{ \mathbf{m}_j^{i+1} \}_{j=1}^{{N_e}}$ by solving the following minimum-average-cost problem
\begin{linenomath*}    
	\begin{IEEEeqnarray}{lll} \label{eq:wls_rlm_mac}
		\underset{\{\mathbf{m}^{i+1}_j\}_{j=1}^{N_e}}{\operatorname{argmin}} & \dfrac{1}{N_e} \sum\limits_{j=1}^{N_e} & \, \left[ \left( \mathbf{d}^o_j - \mathbf{g} \left( \mathbf{m}^{i+1}_j \right) \right)^T \mathbf{C}_{d}^{-1} \left( \mathbf{d}^o_j - \mathbf{g} \left( \mathbf{m}^{i+1}_j \right) \right)  +  \gamma^{i} \left( \mathbf{m}^{i+1}_j - \mathbf{m}^{i}_j \right)^T \left(  \mathbf{C}_{m}^{i} \right)^{-1} \left( \mathbf{m}^{i+1}_j - \mathbf{m}^{i}_j \right) \right] \, , \nonumber
	\end{IEEEeqnarray}
\end{linenomath*}   
where $\mathbf{d}^o_j$ ($j = 1,2,\dotsb, N_e$) are perturbed observations generated by drawing $N_e$ samples from the Gaussian distribution $N(\mathbf{d}^o,\mathbf{C}_{d})$, and $\gamma^i$ a positive scalar that can be used to control the step size of an iteration step, and is automatically chosen using a procedure similar to back-tracking line search \citep{luo2015Iterative}. Through linearization, the MAC problem is approximately solved through the following iteration: 
\begin{linenomath*}  
	\begin{IEEEeqnarray}{crlc} \label{eq:rlm_mac}
		& \mathbf{m}^{i+1}_j = \mathbf{m}^{i}_j + \mathbf{S}_m^i (\mathbf{S}_d^i)^T \left( \mathbf{S}_d^i (\mathbf{S}_d^i)^T + \gamma^i \, \mathbf{C}_{d}  \right)^{-1} \left( \mathbf{d}^o_j - \mathbf{g} \left( \mathbf{m}^{i}_j \right) \right) \, , \text{ for } j = 1, 2, \dotsb, N_e \, .& &
	\end{IEEEeqnarray}
\end{linenomath*}  

The stopping criteria have substantial impact on the performance of an iterative inversion algorithm \citep{Engl2000-regularization}. \cite{luo2015Iterative} mainly used the following two stopping conditions for the purpose of run-time control:
\begin{itemize}
	\item[(C1)] RLM-MAC stops if it reaches a maximum number of iteration steps;
	\item[(C2)] RLM-MAC stops if the relative change of average data mismatch over two consecutive iteration steps is less than a certain value.
\end{itemize}
For all the experiments later, we set the maximum number of iterations to $20$, and the limit of the relative change to  $0.01\%$. 

Let 
\begin{linenomath*}    
	\begin{IEEEeqnarray}{lll} \label{eq:avg_data_mismatch}
		\boldsymbol{\Xi}^i \equiv \dfrac{1}{N_e} \sum\limits_{j=1}^{N_e} & \, \left[ \left( \mathbf{d}^o_j - \mathbf{g} \left( \mathbf{m}^{i}_j \right) \right)^T \mathbf{C}_{d}^{-1} \left( \mathbf{d}^o_j - \mathbf{g} \left( \mathbf{m}^{i}_j \right) \right) \right] & 
	\end{IEEEeqnarray}
\end{linenomath*}  
be the average (normalized) data mismatch with respect to the ensemble $\mathbf{M}^i$.
Following Proposition 6.3 of \cite{Engl2000-regularization}, a third stopping condition is introduced and implemented in \cite{luo2016sparse2d_spej}. Concretely, we also stop the iteration in Eq. (\ref{eq:rlm_mac}) when 
\begin{linenomath*}    
	\begin{IEEEeqnarray}{lll} \label{eq:stopping_criterion_ndm}
		\boldsymbol{\Xi}^i < 4 p
	\end{IEEEeqnarray}
\end{linenomath*} 
for the first time, where the factor $4$ is a critical value below which the iteration process starts to transit from convergence to divergence \citep[page 158]{Engl2000-regularization}. Numerical results in \cite{luo2016sparse2d_spej} indicate that, in certain circumstances, equipping RLM-MAC with the extra stopping condition (\ref{eq:stopping_criterion_ndm}) may substantially improve its performance in history matching. Readers are referred to \cite{luo2016sparse2d_spej} for more details. In the current study, however, the impact of the stopping criterion (\ref{eq:stopping_criterion_ndm}) is not as substantial as that in \cite{luo2016sparse2d_spej}. Nevertheless, we prefer to keep this stopping criterion as an extra safeguard procedure.    

\section{Numerical results in the Brugge benchmark case}\label{sec:results}
We demonstrate the performance of the proposed workflow through a 3D Brugge benchmark case study. Table \ref{tab:brugge_3d} summarizes the key information of the experimental settings. Readers are referred to \cite{peters2010results} for more information of the benchmark case study.

\begin{table*}
	\centering
	\caption{\label{tab:brugge_3d} Summary of experimental settings in the Brugge benchmark case study}
	\begin{tabular}{||l||l||}
		\hline  
		\multirow{2}{*}{Model dimension}  & $139 \times 48 \times 9$ (60048 gridblocks), with 44550 out of 60048 being \\
		& active cells  \\ 
		\hline   
		Parameters to estimate & PORO, PERMX, PERMY, PERMZ. Total number is $4 \times 44550 = 178200$  \\ 
		\hline   
		Gridblock size & Irregular. Average $\Delta X \approx 93 m$, $\Delta Y \approx 91 m$, and average $\Delta Z \approx 5 m$  \\ 
		\hline 
		Reservoir simulator  & ECLIPSE 100 (control mode LRAT)   \\ 
		\hline 
		Number of wells  & 10 injectors and 20 producers  \\ 
		\hline 
		Production period & 3647.5 days (with 20 report times) \\ 
		\hline 
		\multirow{2}{*}{Production data}  & Production wells : BHP, OPR and WCT; Injection wells : BHP. \\ & Total number: $20 \times 70 = 1400$  \\ 
		\hline 
		\multirow{2}{*}{Seismic survey time}    & Base: day 1; Monitor (1st): day 991;  Monitor (2nd): \\
		& day 2999   \\ 
		\hline 
		\multirow{2}{*}{4D seismic data} & AVA data from near- and far- offsets at three survey times. \\ 
		& Total number: $\sim$ 7.02 M   \\ 
		\hline 
		\multirow{2}{*}{DWT (seismic)}   & Three-level decomposition using 3D Daubechies wavelets with  \\
		& two vanishing moments   \\ 
		\hline 
		Thresholding  & Hard thresholding based on Eqs.(\ref{eq:hard_thresholding}) and (\ref{eq:multiple_universal_rule})  \\ 
		\hline 
		History matching method  & iES (RLM-MAC) with an ensemble of 103 reservoir models  \\ 
		\hline 
	\end{tabular}
\end{table*}

The Brugge field model has 9 layers, and each layer consists of $139 \times 48$ gridblocks. The total number of gridblocks is 60048, whereas among them $44550$ are active. The data set of the original benchmark case study does not contain AVA attributes, therefore we generate synthetic seismic data in the following way: The benchmark case contains an initial ensemble of 104 members. We randomly pick one of them as the reference model (which turned out to be the member ``FN-SS-KP-1-92''), and use the rest 103 members as the initial ensemble in this study. The model variables to be estimated include porosity (PORO) and permeability (PERMX, PERMY, PERMZ) at all active gridblocks. Consequently, the total number of parameters in estimation is $178200$. 

There are 20 producers and 10 water injectors in the reference model, and they are controlled by the liquid rate (LRAT) target. The production period is 10 years, and in history matching we use production data at 20 report times. The production data consist of oil production rates (WOPR) and water cuts (WWCT) at 20 producers, and bottom hole pressures (WBHP) at all 30 wells. Therefore the total number of production data is $1400$. Gaussian white noise is added to production data of the reference model. For WOPR and WWCT data, their noise STD are taken as the maximum values between $10\%$ of their magnitudes and $10^{-6}$ (the latter is used to prevent the numerical issue of division by zero), whereas for WBHP data, the noise STD is 1 bar. 

In the experiments, there are three seismic surveys taking place on day 1 (base), day 991 (1st monitor), and day 2999 (2nd monitor), respectively. At each survey, we apply forward AVA simulation described in the previous section to the static (porosity) and dynamic (pressure and saturation) variables of the reference model, and generate AVA attributes at two different angles: $10^\circ$ (near-offset) and $20^\circ$ (far-offset). Each AVA attribute is a 3D ($139 \times 48 \times 176$) cube, and consists of around $1.17 \times 10^6$ elements. Therefore the total number of 4D seismic data is around $3 \times 2 \times 1.17 \times 10^6 = 7.02 \times \times 10^6$. For convenience of discussion later, we label the dimensions of the 3D cubes by $X$, $Y$ and $Z$, respectively, such that $X = 1, 2, \dotsb, 139$, $Y = 1, 2, \dotsb, 48$ and $Z = 1, 2, \dotsb, 176$. In history matching, we add Gaussian white noise to each reference AVA attribute, with the noise level being $30\%$. Here we do not assume the noise STD is known. Instead, we first apply three-level 3D DWT to each AVA cube using Daubechies wavelets with two vanishing moments, and then use Eq. (\ref{eq:noise_std_mad}) to estimate noise STD in the wavelet domain.    

In what follows, we consider three history matching scenarios that involve: (S1) production data only; (S2) 4D seismic data only; and (S3) both production and 4D seismic data. Because of the huge volumes of AVA attributes, in scenarios (S2) and (S3), it is not convenient to directly use the 4D seismic data in the original data space. Therefore, to examine the impact of data size on the performance of SHM, in each scenario (S2 or S3), we consider two cases that have different numbers of leading wavelet coefficients. This is achieved by letting the scalar $c$ of Eq. (\ref{eq:multiple_universal_rule}) be $1$ and $5$, respectively.

\subsection{Results of scenario S1 (using production data only)}

\begin{figure*}
	\centering
	\includegraphics[scale=0.4]{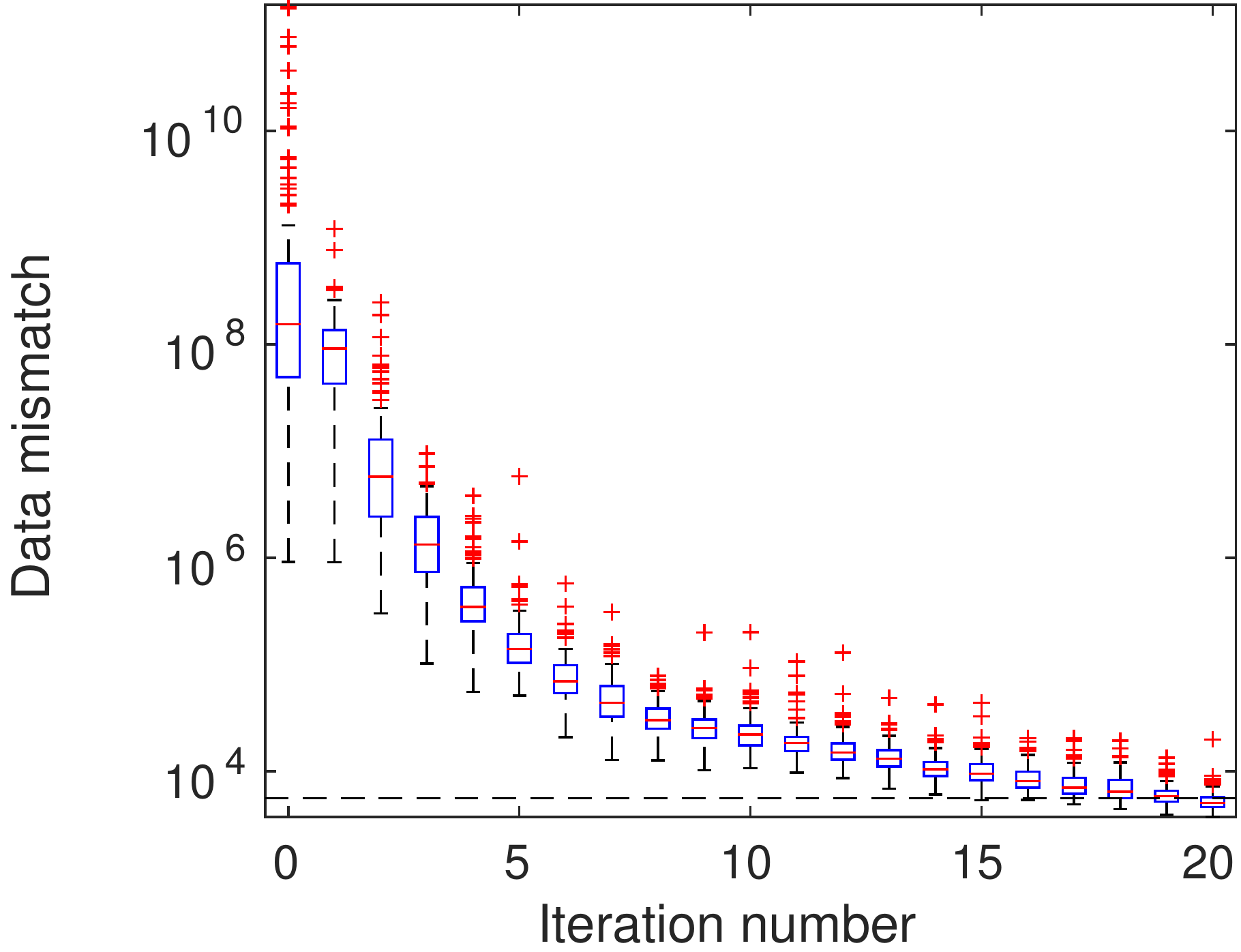}
	\caption{\label{fig:Brugge_boxplot_objRealIter_S1} Boxplots of production data mismatch as a function of iteration step (scenario S1). The horizontal dashed line indicates the threshold value ($4 \times 1400 = 5600$) for the stopping criterion (\ref{eq:stopping_criterion_ndm}). For visualization, the vertical axis is in the logarithmic scale. In each box plot, the horizontal line (in red) inside the box denotes the median; the top and bottom of the box represent the 75th and 25th percentiles, respectively; the whiskers indicate the ranges beyond which the data are considered outliers, and the whiskers positions are determined using the default setting of MATLAB$^\copyright$ R2015b, while the outliers themselves are plotted individually as plus signs (in red).}
\end{figure*}  

Figure \ref{fig:Brugge_boxplot_objRealIter_S1} shows the boxplots of data mismatch as a function of iteration step. The average data mismatch of the initial ensemble (iteration 0) is around $5.65 \times 10^9$. After 20 iteration steps, the average data mismatch is reduced to $5431.97$, lower than the threshold value $4 \times 1400 = 5600$ in (\ref{eq:stopping_criterion_ndm}) for the first time. In this particular case, the stopping step selected according to the criterion (\ref{eq:stopping_criterion_ndm}) coincides with the maximum number of iteration steps. Therefore, we take the ensemble at the 20th iteration step as the final estimation. 

\renewcommand{\nScale}{0.4} 
\begin{figure*}
	\centering
	\subfigure[RMSEs of log PERMX]{ \label{subfig:rmse_PERMX_boxplot_ensemble_S1}
		\includegraphics[scale=\nScale]{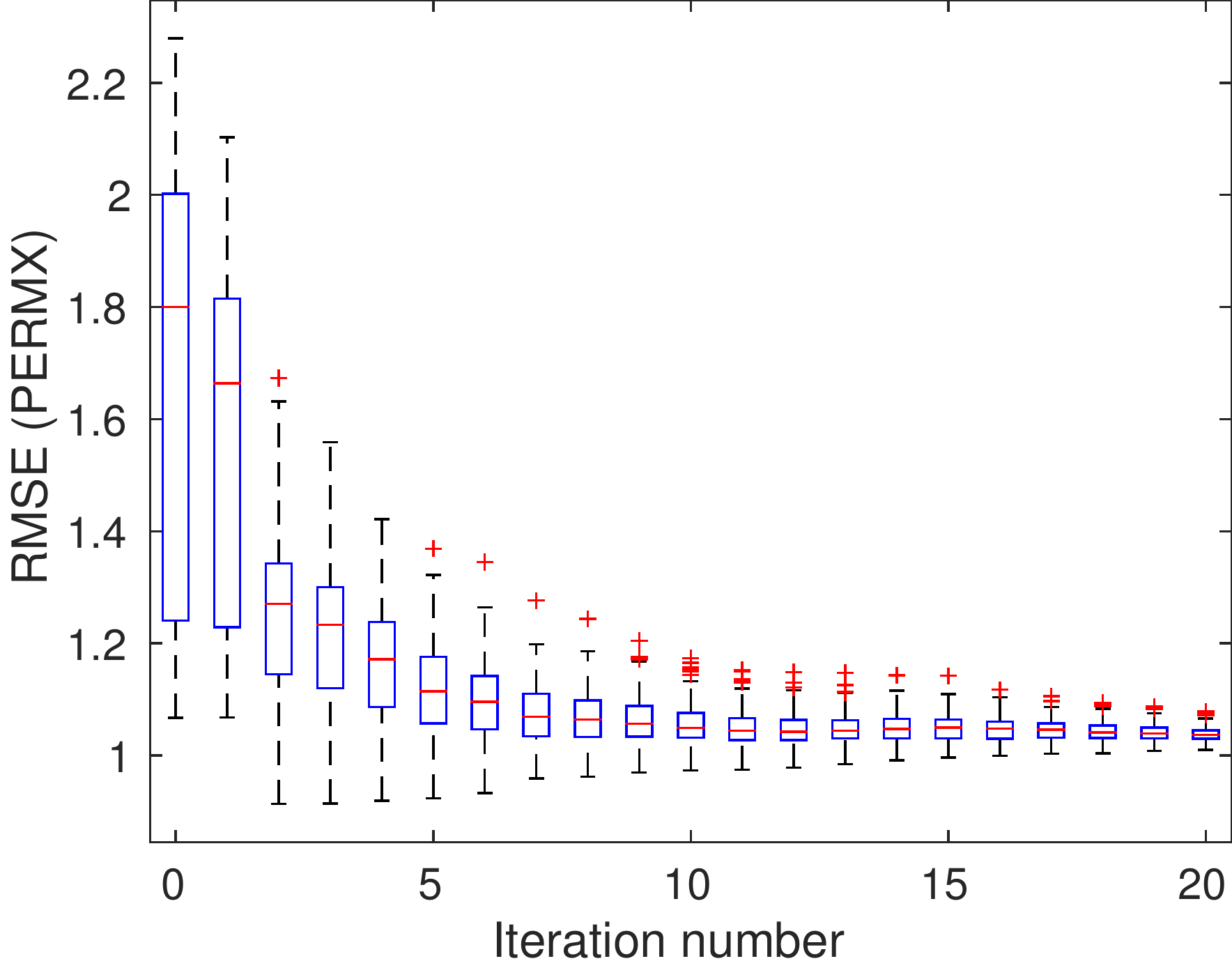}
	}
	\subfigure[RMSEs of PORO]{ \label{subfig:rmse_PORO_boxplot_ensemble_S1}
		\includegraphics[scale=\nScale]{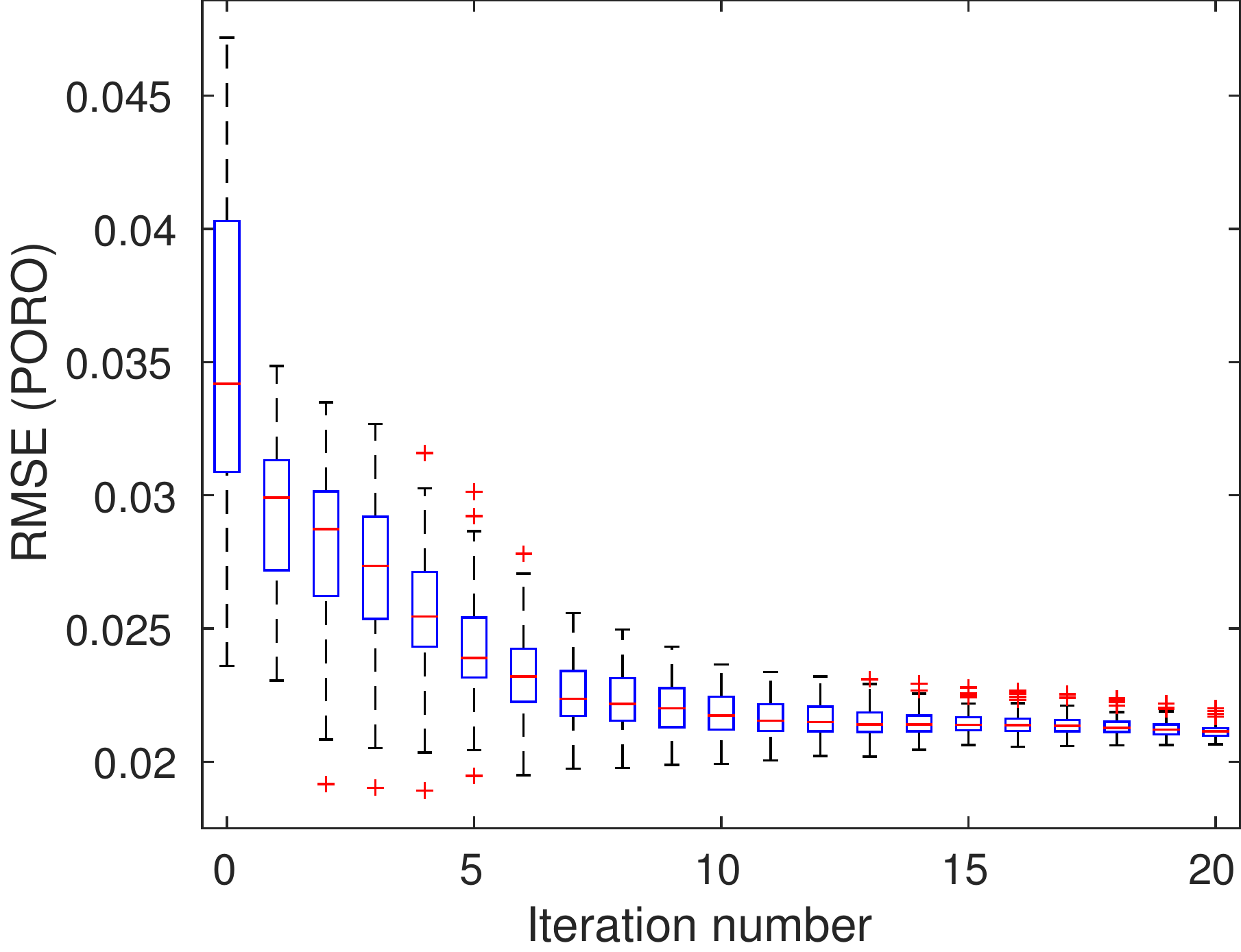}
	}
	\caption{\label{fig:Brugge_RLM-MAC_RMSE_S1} Boxplots of RMSEs of (a) log PERMX and (b) PORO as functions of iteration step (scenario S1).}
\end{figure*} 

In this synthetic study, the reference reservoir model is known. As a result, we use root mean squared error (RMSE) in the sequel to measure the $\ell_2$-distance (up to a factor) between an estimated model and the reference one. More specifically, let $\mathbf{v}^{tr}$ be the $\ell$-dimensional reference property, and $\hat{\mathbf{v}}$ an estimation, then the RMSE $e_{\mathbf{v}}$ of $\hat{\mathbf{v}}$ with respect to the reference $\mathbf{v}^{tr}$ is defined by  
\begin{linenomath*} 
	\begin{IEEEeqnarray}{lll} \label{eq:RMSE_def}
		e_{\mathbf{v}} = \dfrac{\Vert \hat{\mathbf{v}} - \mathbf{v}^{tr} \Vert_2}{\sqrt{\ell}} \, ,
	\end{IEEEeqnarray}
\end{linenomath*} 
where $\Vert \bullet \Vert_2$ denotes the $\ell_2$ norm. 

For brevity, Figure \ref{fig:Brugge_RLM-MAC_RMSE_S1} reports the boxplots of RMSEs in estimating PERMX (in the natural log scale) and PORO, at different iteration steps, whereas the results for PERMY and PERMZ are similar to that for PERMX. As can be seen in Figure \ref{fig:Brugge_RLM-MAC_RMSE_S1}, the average RMSEs of both log PERMX and PORO tend to reduce as the number of iteration steps increases.       

\renewcommand{\nScale}{0.33} 
\begin{figure*}
	\centering
	\subfigure[WBHP of the initial ensemble]{ \label{subfig:WBHP_BR-P-5_initial} 
		\includegraphics[scale=\nScale]{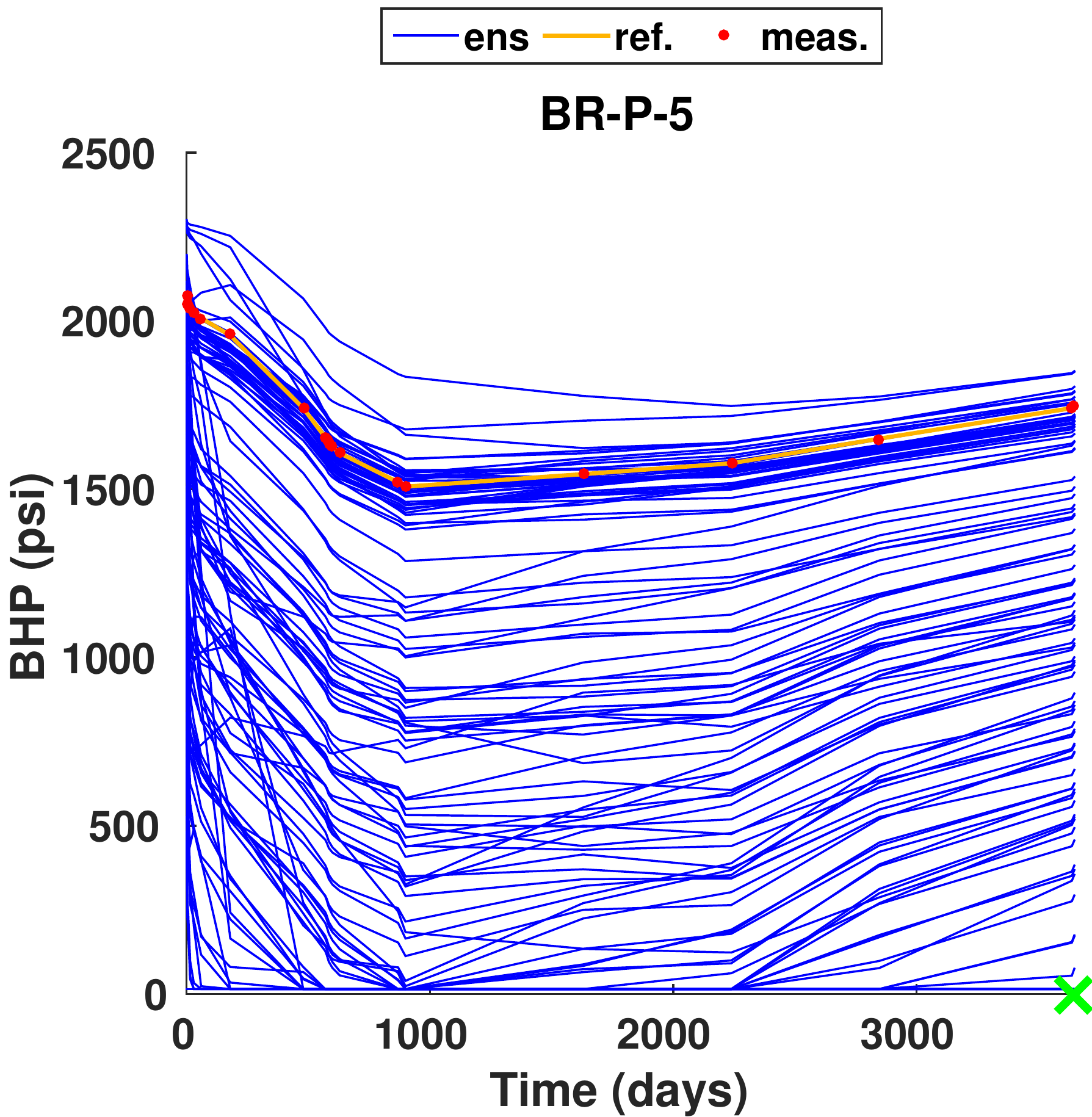}
	}%
	\subfigure[WOPR of the initial ensemble]{ \label{subfig:WOPR_BR-P-5__initial}
		\includegraphics[scale=\nScale]{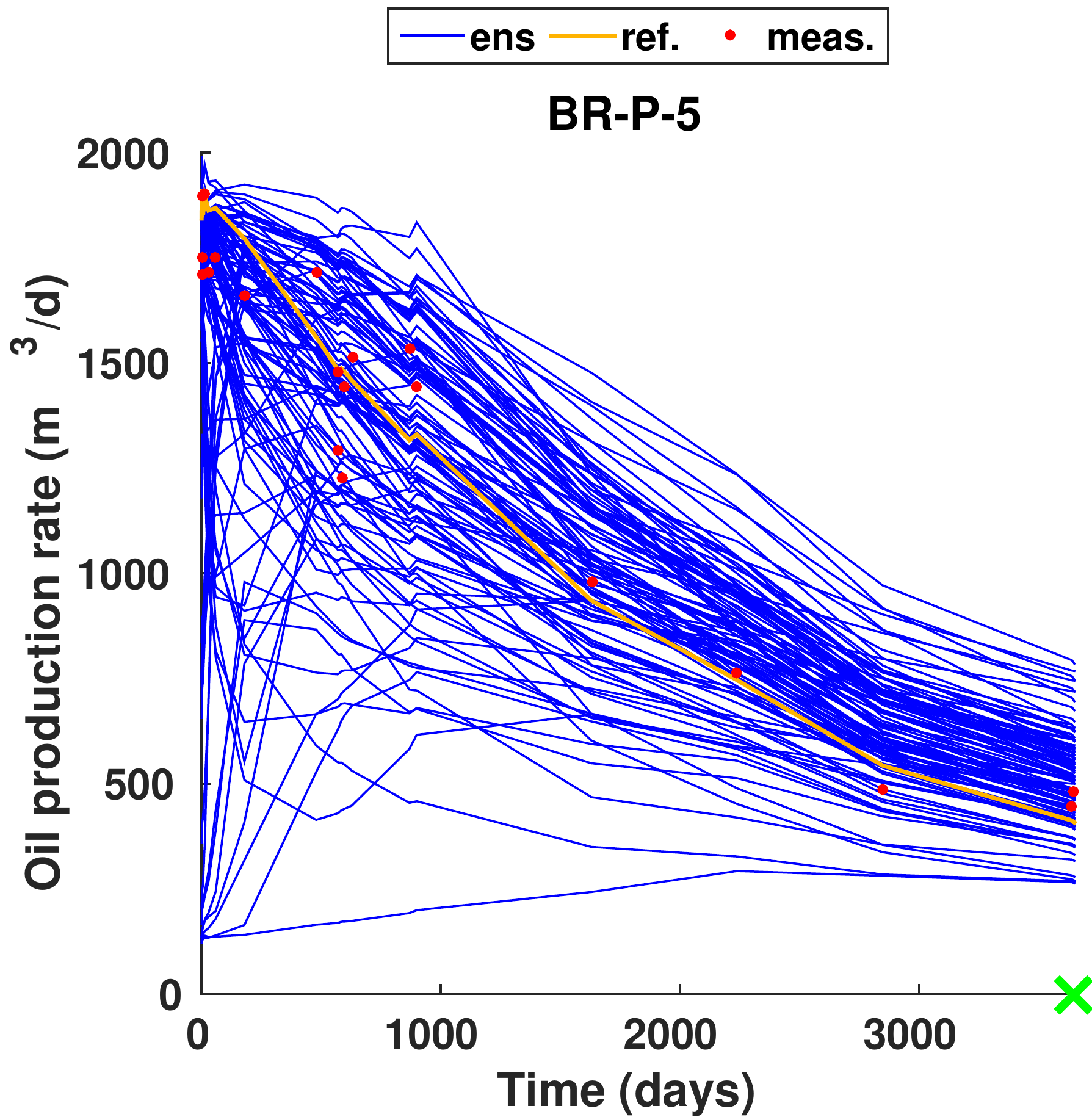}
	}%
	\subfigure[WWCT of the initial ensemble]{ \label{subfig:WWCT_BR-P-5__initial}
		\includegraphics[scale=\nScale]{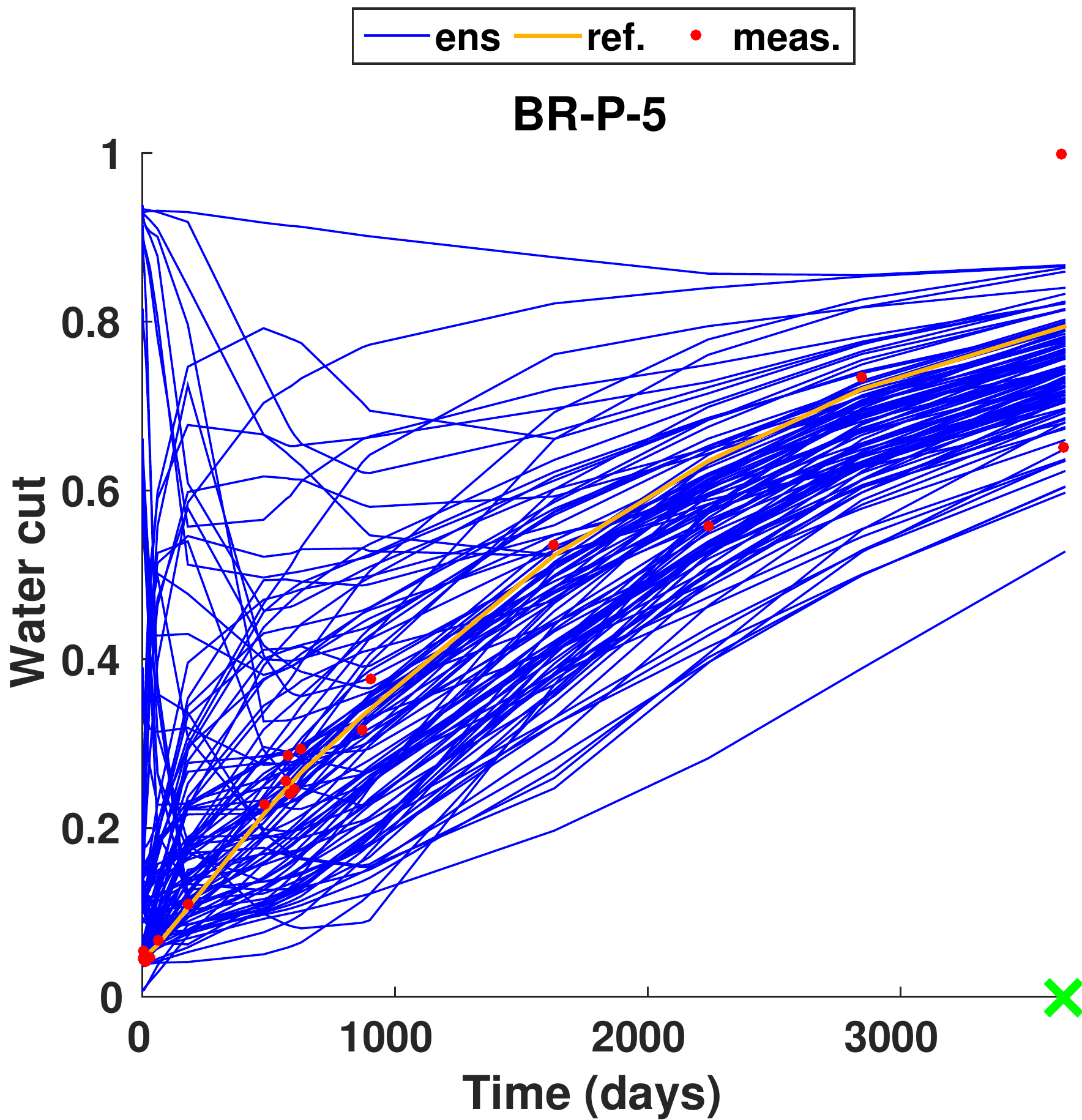}
	}%
	
	\subfigure[WBHP of the final ensemble]{ \label{subfig:WBHP_BR-P-5_S1} 
		\includegraphics[scale=\nScale]{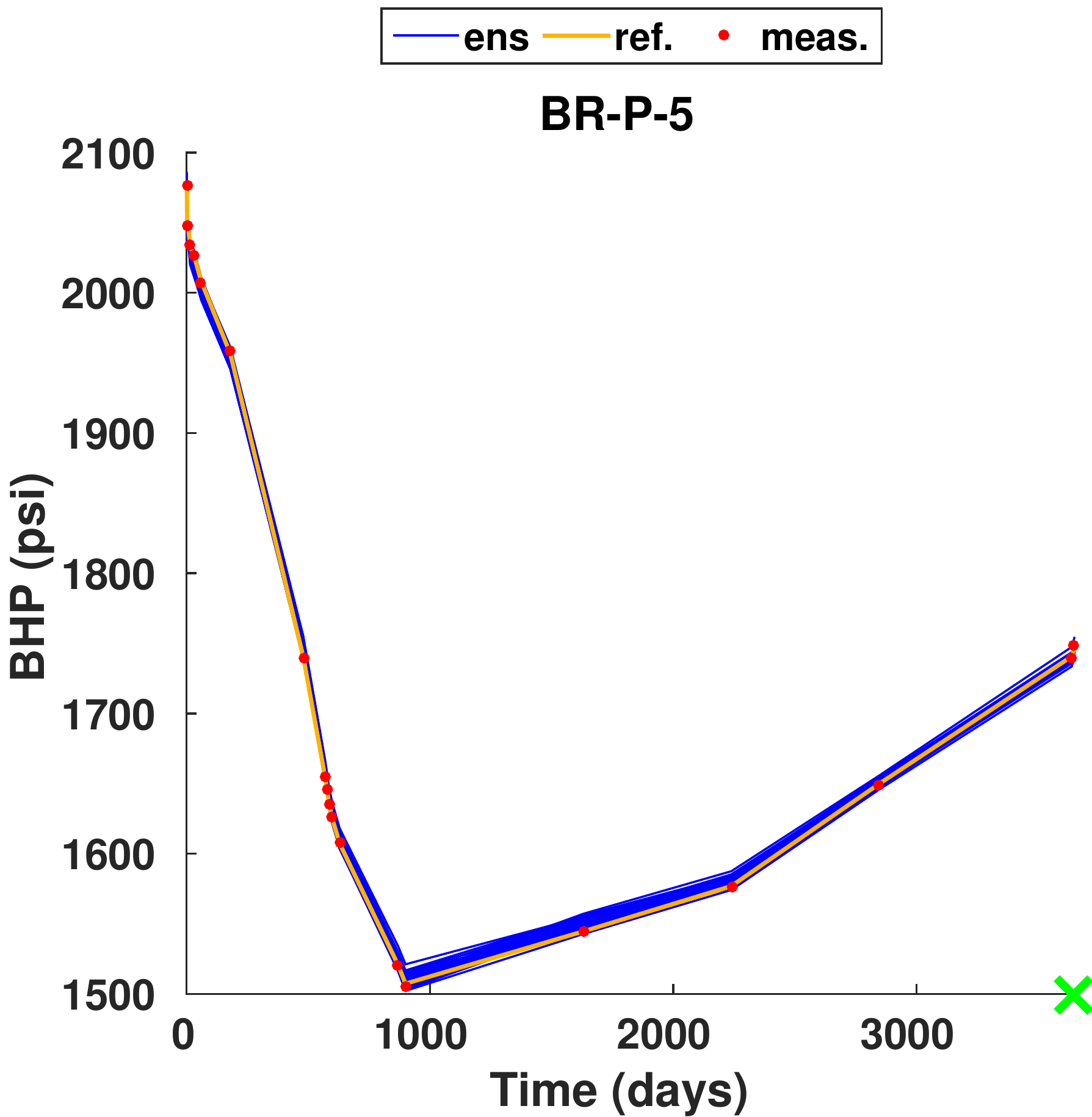}
	}%
	\subfigure[WOPR of the final ensemble]{ \label{subfig:WOPR_BR-P-5__S1}
		\includegraphics[scale=\nScale]{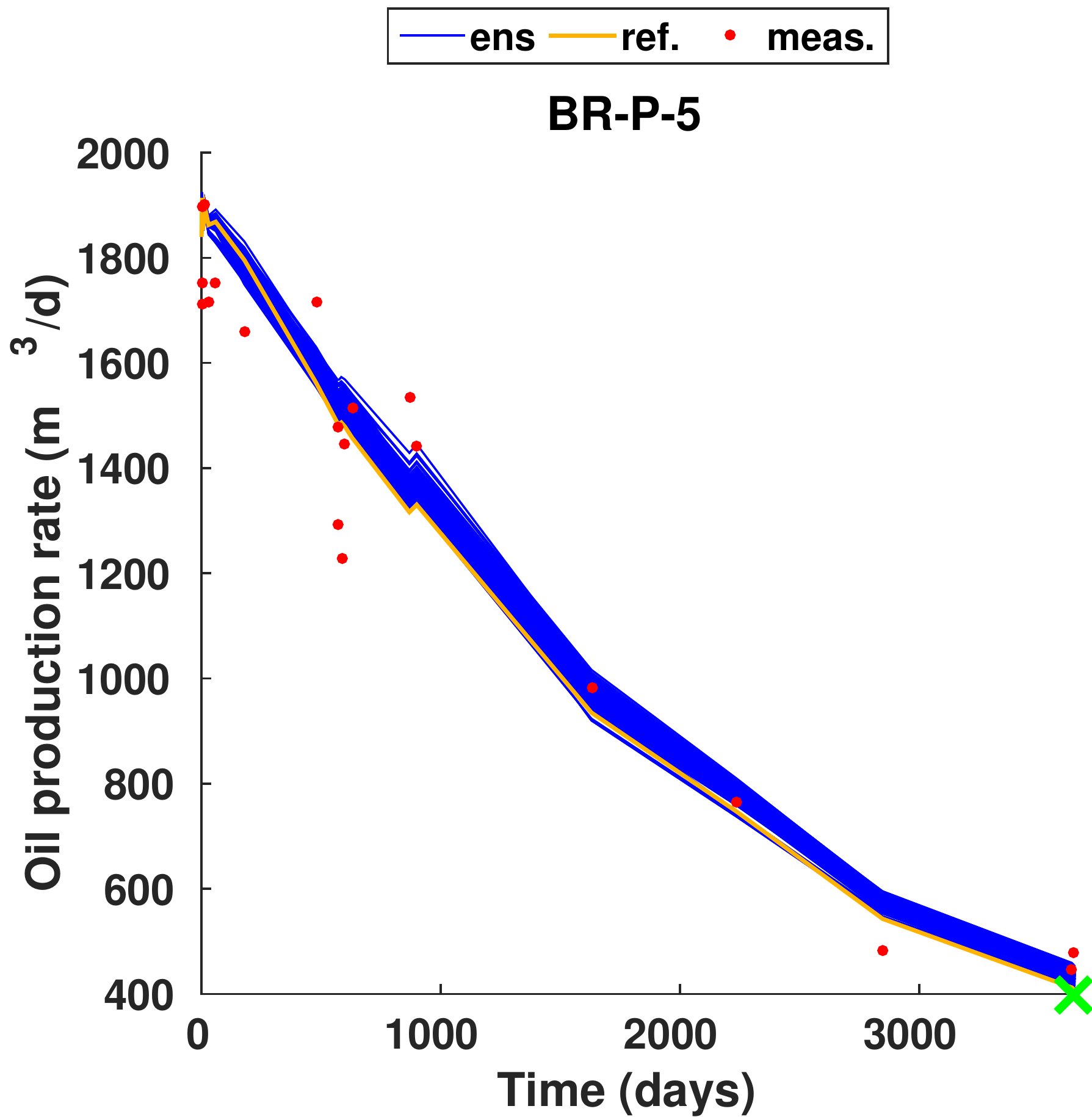}
	}%
	\subfigure[WWCT of the final ensemble]{ \label{subfig:WWCT_BR-P-5__S1}
		\includegraphics[scale=\nScale]{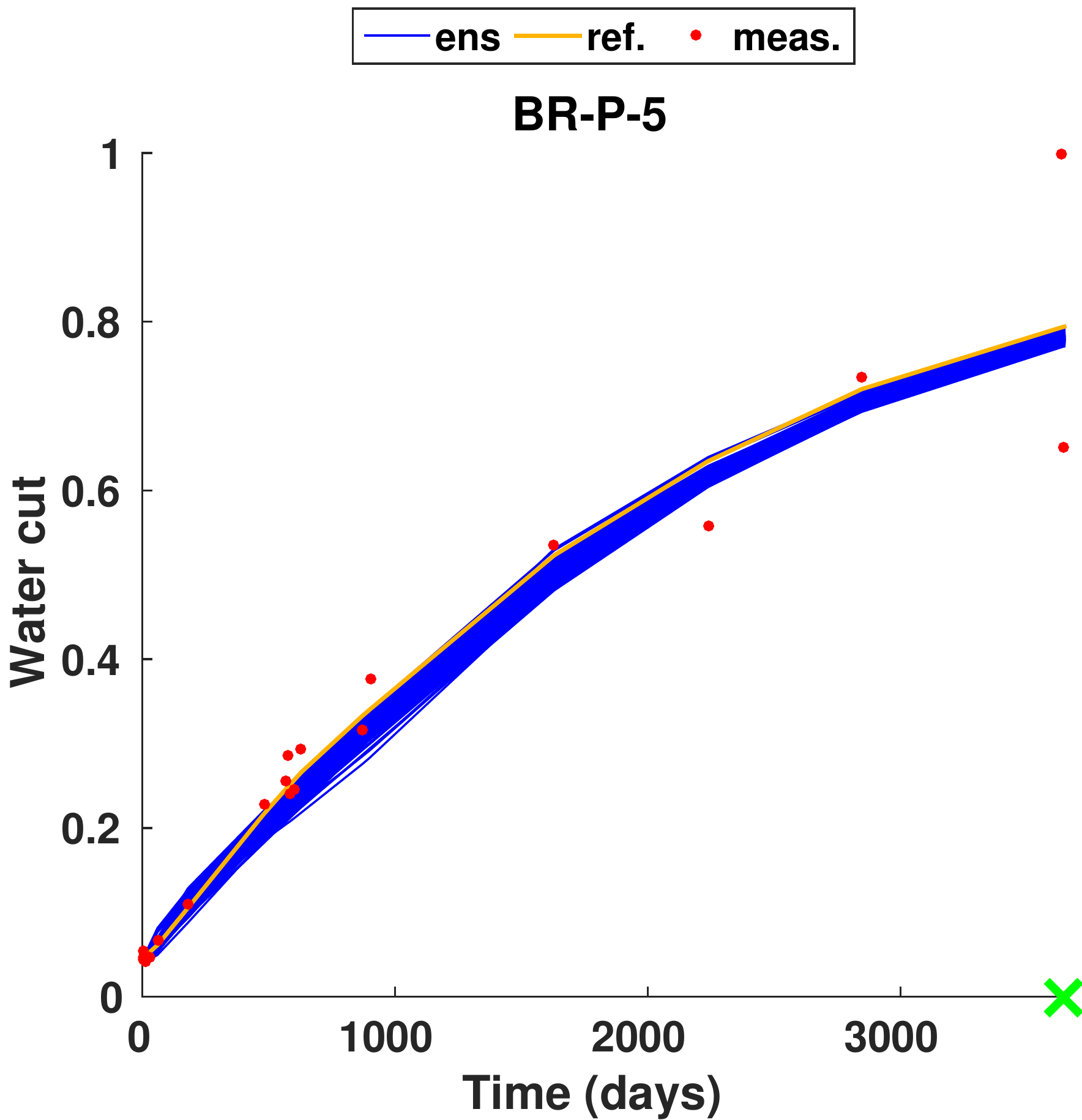}
	}%
	
	\caption{\label{fig:production_P5_S1} Profiles of WBHP, WOPR and WWCT of the initial (1st row) and final (2nd row) ensembles at the producer BR-P-5 (scenario S1). The production data of the reference model are plotted as orange curves, the observed production data at 20 report times as red dots, and the simulated production data of initial and final ensembles as blue curves.}
\end{figure*} 

\renewcommand{\nScale}{0.21} 
\begin{figure*}
	\centering
	\subfigure[Rerefrence log PERMX]{ \label{subfig:PERMX_L2_true} %
		\includegraphics[scale=\nScale]{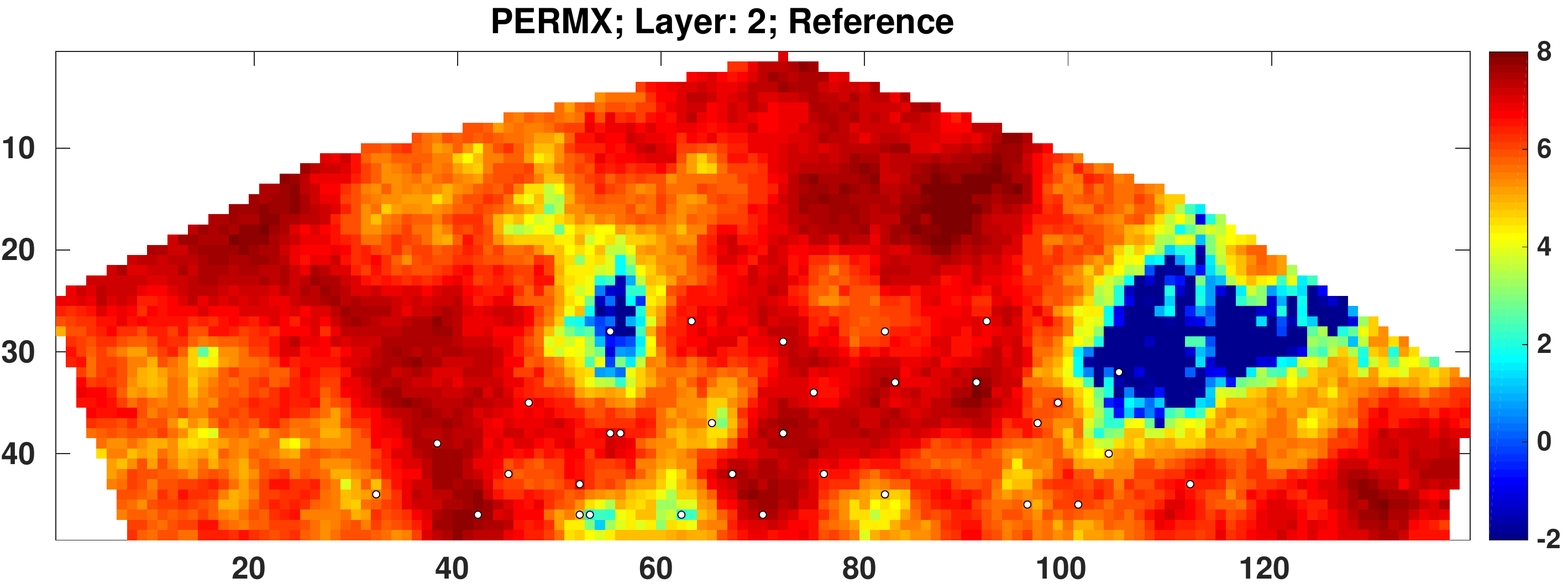}
	}
	\subfigure[Mean of initial log PERMX]{ \label{subfig:PERMX_L2_Mean_initEns}
		\includegraphics[scale=\nScale]{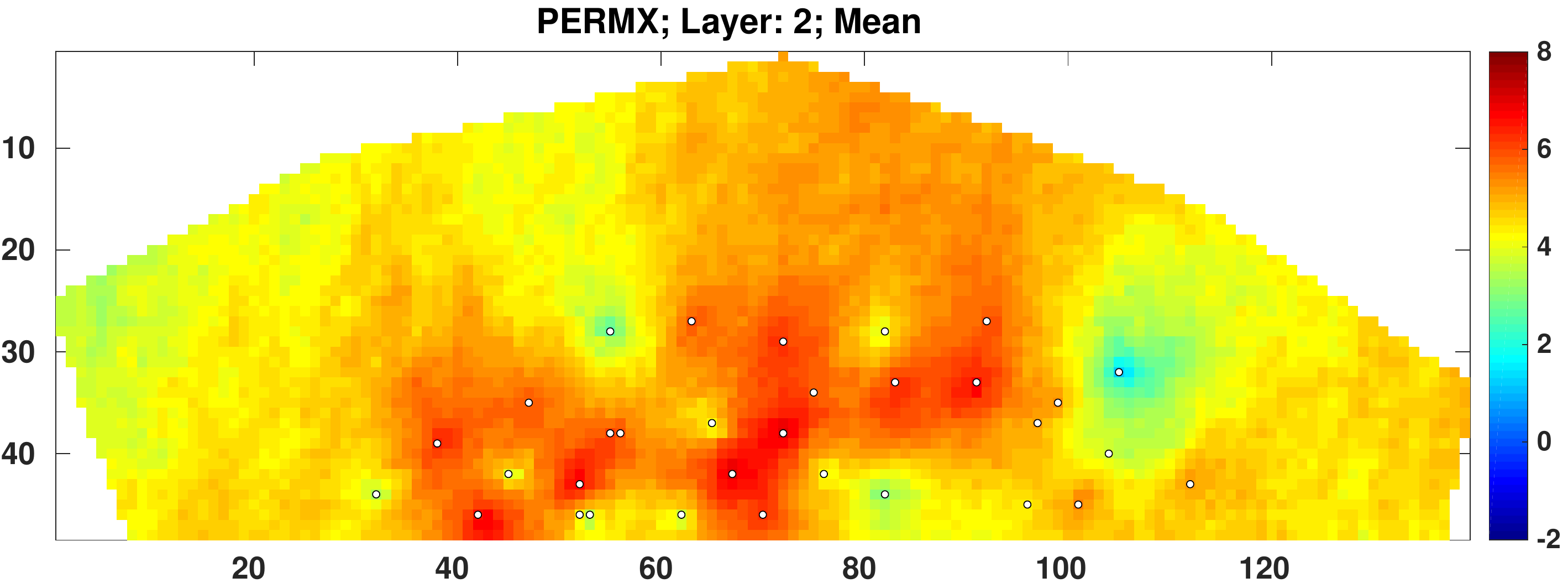}
	}
	\subfigure[Mean of final log PERMX]{ \label{subfig:PERMX_L2_Mean_ensemble20_S1}
		\includegraphics[scale=\nScale]{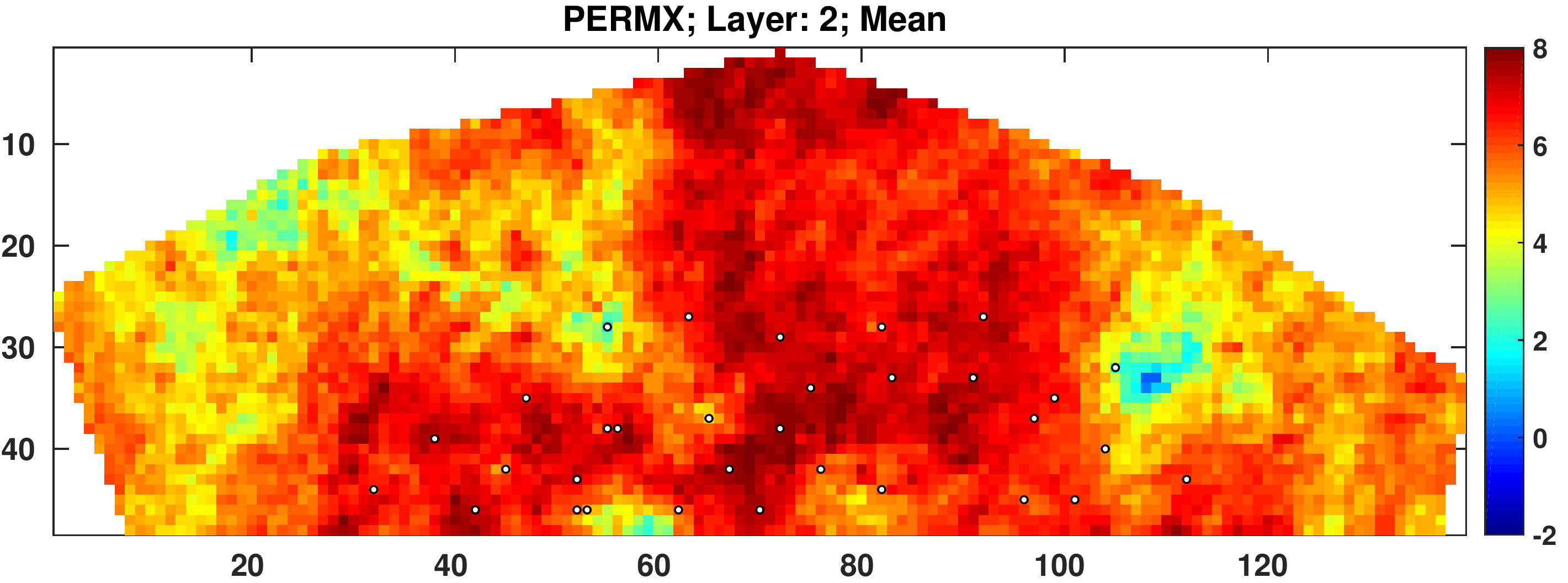}
	}%
	
	\subfigure[Rerefrence PORO]{ \label{subfig:PORO_L2_true} %
		\includegraphics[scale=\nScale]{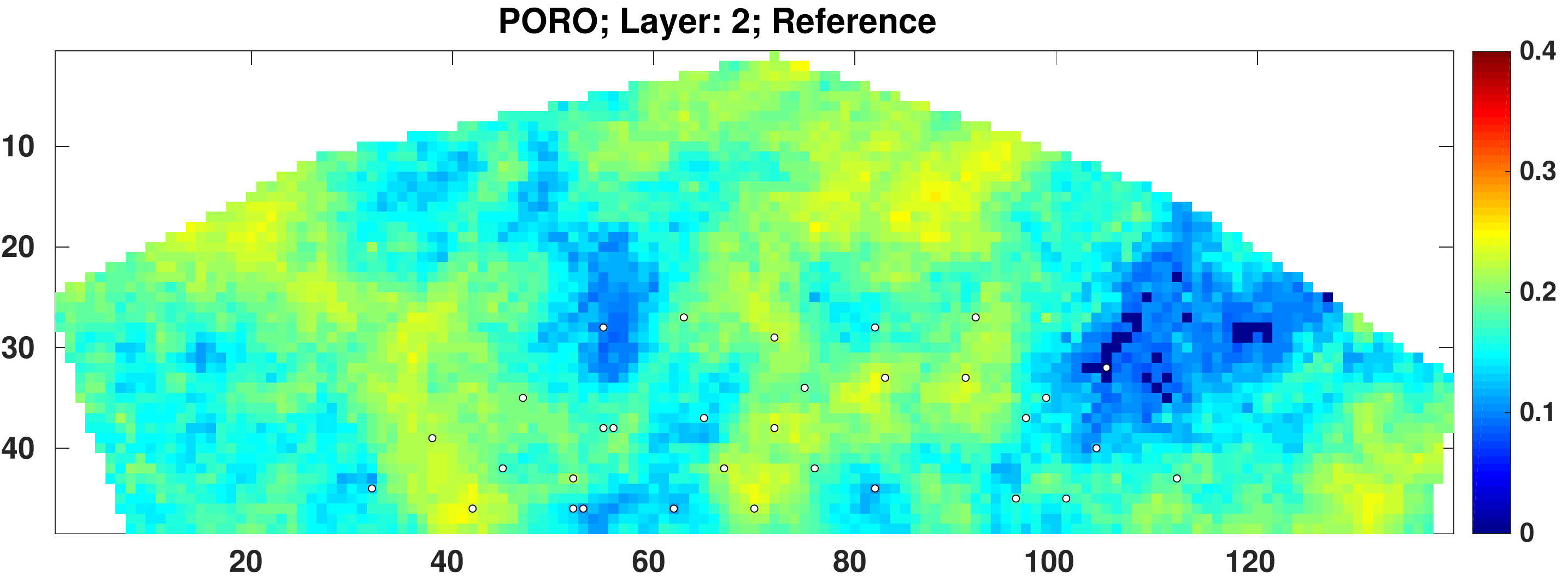}
	}
	\subfigure[Mean of initial PORO]{ \label{subfig:PORO_L2_Mean_initEns}
		\includegraphics[scale=\nScale]{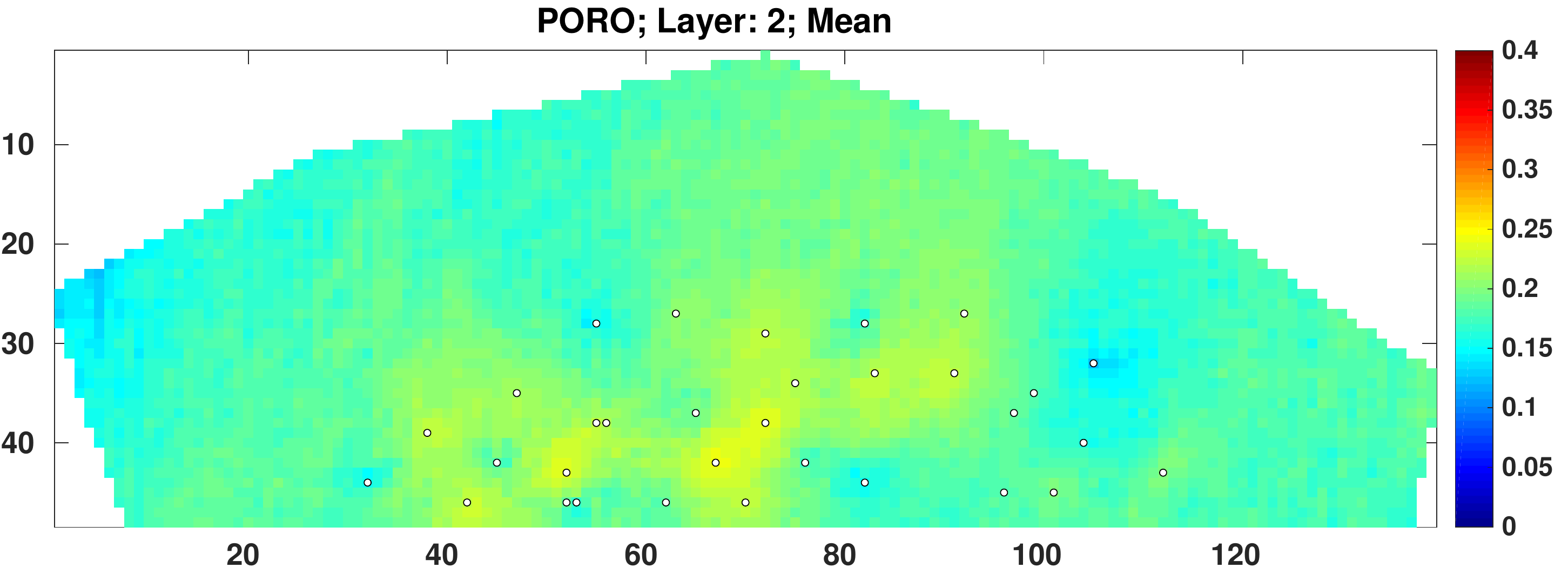}
	}
	\subfigure[Mean of final PORO]{ \label{subfig:PORO_L2_Mean_ensemble20_S1}
		\includegraphics[scale=\nScale]{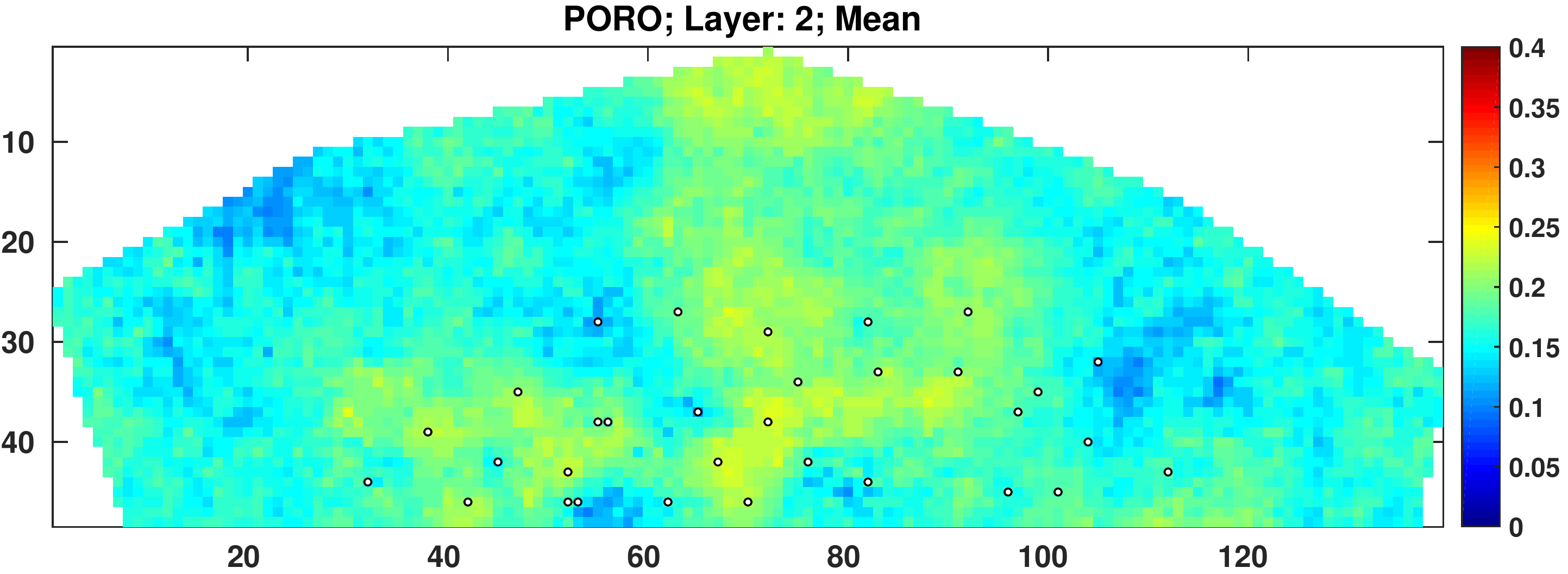}
	}%
	
	\caption{\label{fig:estimation_S1} Log PERMX (top row) and PORO (bottom row) of the reference reservoir model (1st column) and the means of the initial (2nd column) and final (3rd column) ensembles at Layer 2 (scenario S1). The black dots in the figures represent the locations of injection and production wells (top view).}
\end{figure*} 

Figure \ref{fig:production_P5_S1} shows the profiles of WBHP, WOPR and WWCT of the initial (1st row) and final (2nd row) ensemble at the producer BR-P-5. It is evident that, through history matching, the final ensemble matches the production data better than the initial one, and this is consistent with the results in Figure \ref{fig:Brugge_boxplot_objRealIter_S1}.  

For illustration, Figure \ref{fig:estimation_S1} presents the reference log PERMX and PORO at layer 2 (1st column), the mean of log PERMX and PORO at layer 2 from the initial ensemble (2nd column), and the mean of log PERMX and PORO at layer 2 from the final ensemble (3rd column). A comparison between the initial and final estimates of log PERMX and PORO indicates that the final estimates appear more similar to the reference fields, in consistence with results in Figure \ref{fig:Brugge_RLM-MAC_RMSE_S1}.  

\subsection{Results of scenario S2 (using seismic data only)}
\renewcommand{\nScale}{0.45} 
\begin{figure*}
	\centering
	\centering
	\subfigure[Seismic data mismatch ($c=1$)]{ \label{subfig:Brugge_boxplot_objRealIter_c1_S2} %
		\includegraphics[scale=\nScale]{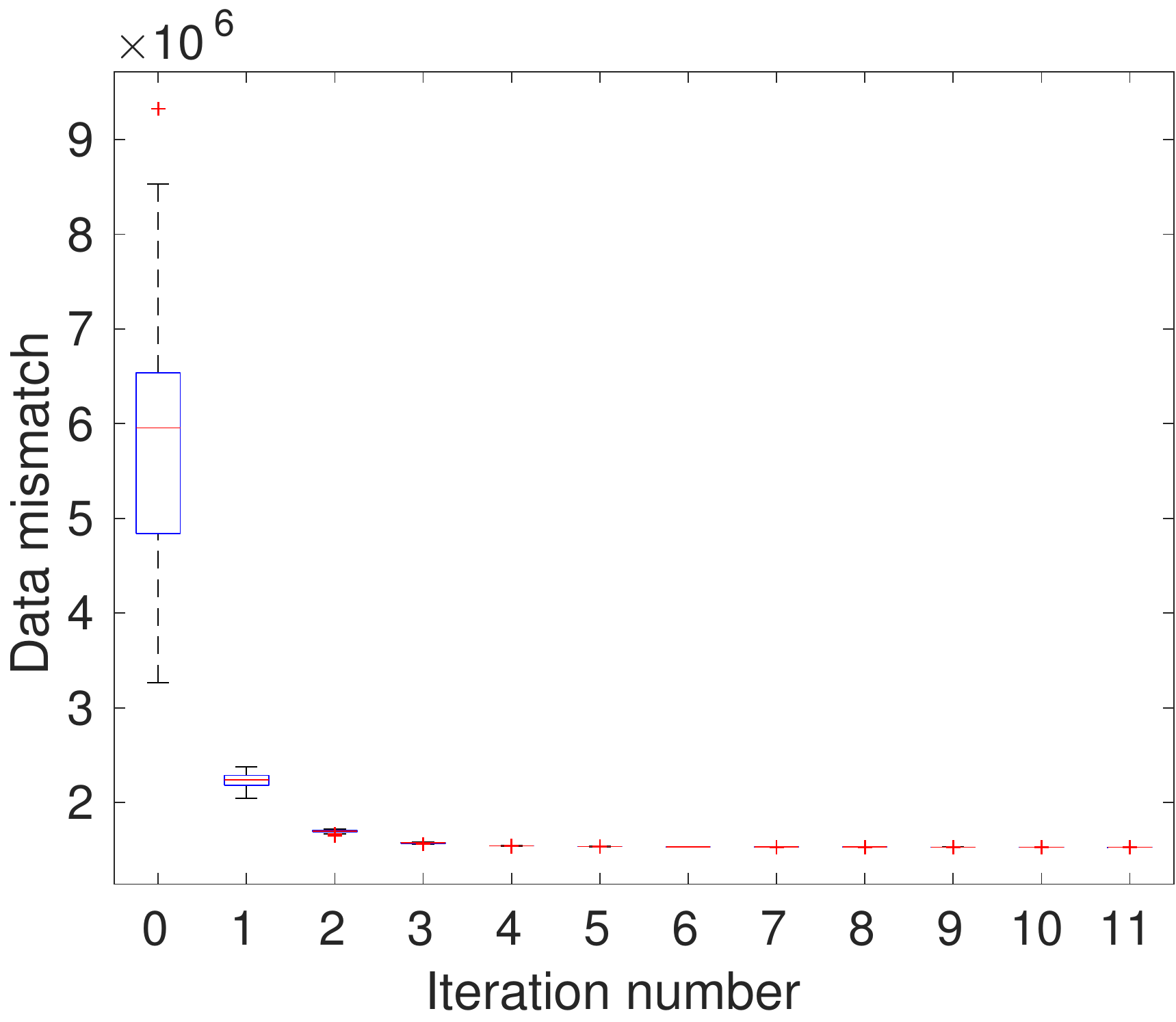}
	}%
	\subfigure[Seismic data mismatch ($c=5$)]{ \label{subfig:Brugge_boxplot_objRealIter_c5_S2}
		\includegraphics[scale=\nScale]{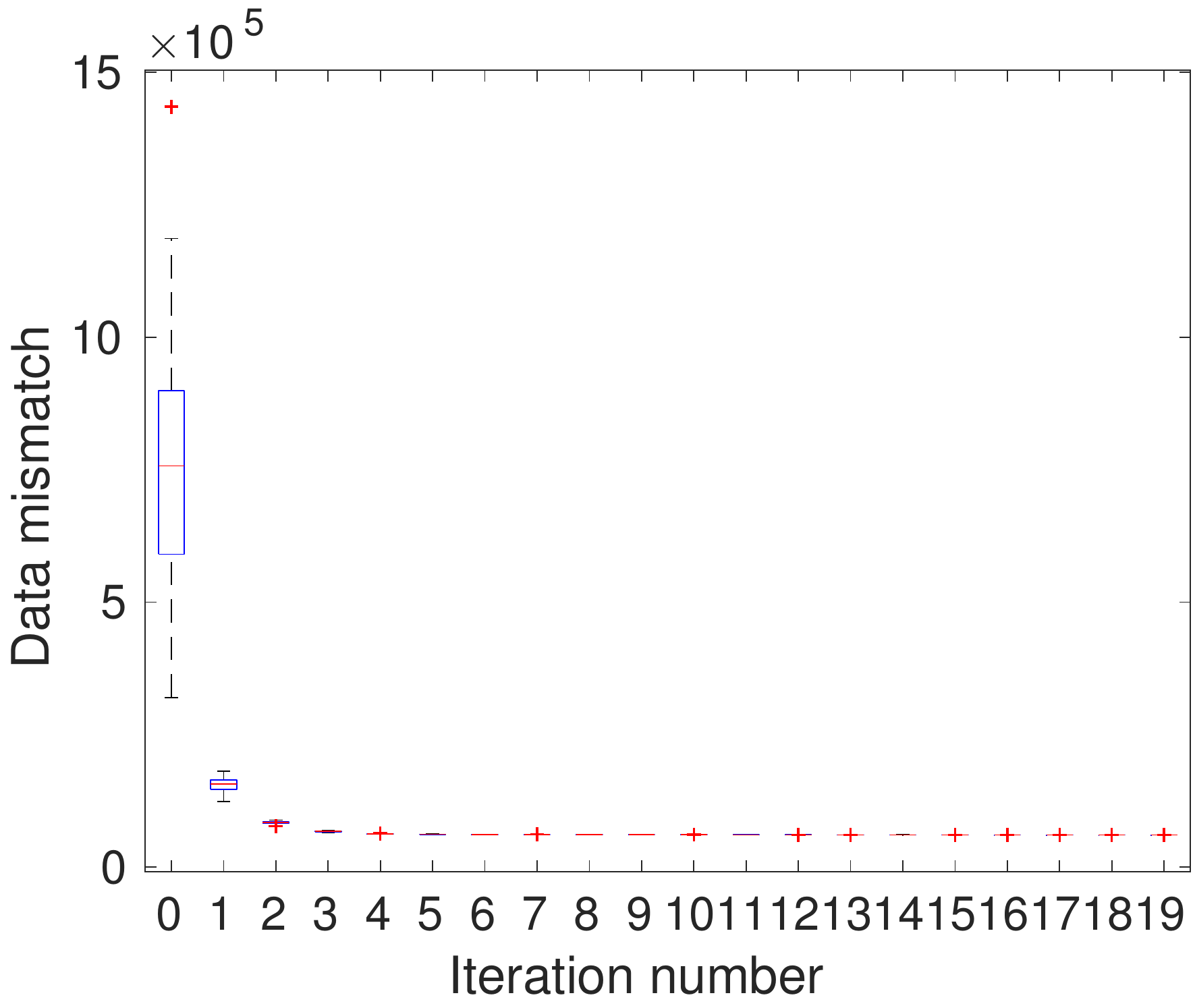}
	}%
	\caption{\label{fig:Brugge_boxplot_objRealIter_S2} Boxplots of seismic data mismatch as functions of iteration step (scenario S2). Case (a) corresponds to the results with $c=1$, for which choice the number of leading wavelet coefficients is $178332$, roughly $2.5\%$ of the original data
		size; Case (b) to the results with $c=5$, for which choice the number of leading wavelet coefficients is $1665$, more than $4000$ times reduction in data size.}
\end{figure*}  

To examine the impact of data size on the performance of SHM, we consider two cases with different threshold values chosen through Eq. (\ref{eq:multiple_universal_rule}). In the first case,  we let $c=1$, such that Eq. (\ref{eq:multiple_universal_rule}) reduces to the universal rule in choosing the threshold value \citep{donoho1994ideal}. Under this choice, the number of leading wavelet coefficients is $178332$, around $2.5\%$ of the original AVA data size ($7.04 \times 10^6$). In the second case, we increase the value of $c$ to $5$, such that the number of leading wavelet coefficients further reduces to $1665$, which is more than $4000$ times reduction in comparison to the original data size.

Figure \ref{fig:Brugge_boxplot_objRealIter_S2} indicates the boxplots of seismic data mismatch as functions of iteration step. In either case, seismic data mismatch reduces fast at the first few iteration steps, and then changes slowly afterwards. The stopping criterion (C2), monitoring the relative change of average data mismatch, becomes effective in both cases, such that the iteration stops at the $11$th step when $c=1$, and at the $19$th when $c=5$. Accordingly, the ensembles at iteration step $11$ and $19$, respectively, are taken as the final estimates in these two cases. In addition, it appears that ensemble collapse takes place in both cases, although this phenomenon is somewhat mitigated in the case $c=5$, in comparison to the case $c=1$. The mitigation of ensemble collapse is even more evident when we further increase $c$ to $8$, and accordingly, reduce the data size to $534$. By doing so, however, the history matching performance is deteriorated (results not included here), largely due to the fact that such a significant reduction of data size leads to substantial loss of information content in the seismic data, a point to be elaborated soon.   

\renewcommand{\nScale}{0.45} 
\begin{figure*}
	\centering
	\subfigure[RMSEs of log PERMX ($c=1$)]{ \label{subfig:rmse_PERMX_boxplot_ensemble_c1_S2}
		\includegraphics[scale=\nScale]{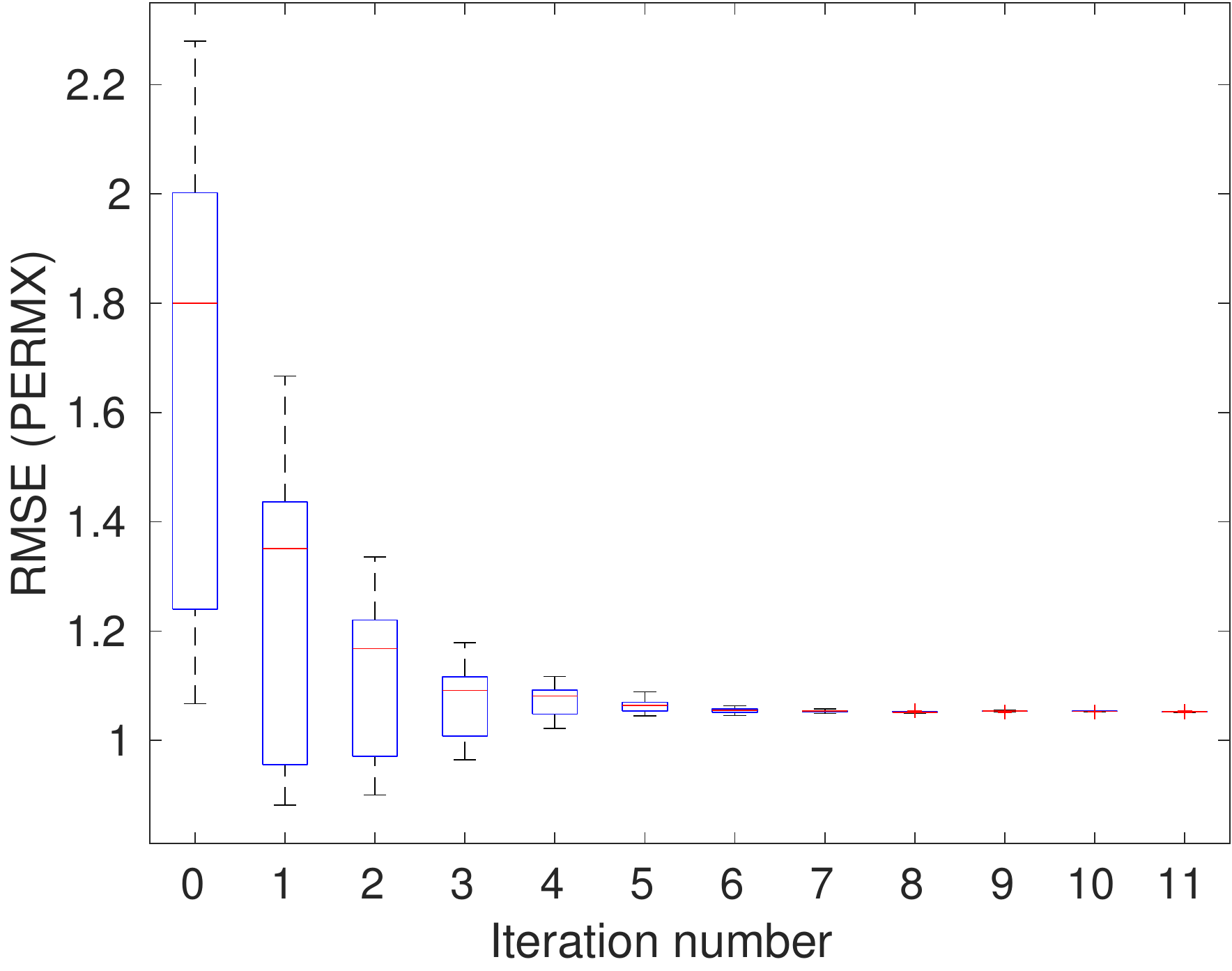}
	}%
	\subfigure[RMSEs of PORO ($c=1$)]{ \label{subfig:rmse_PORO_boxplot_ensemble_c1_S2}
		\includegraphics[scale=\nScale]{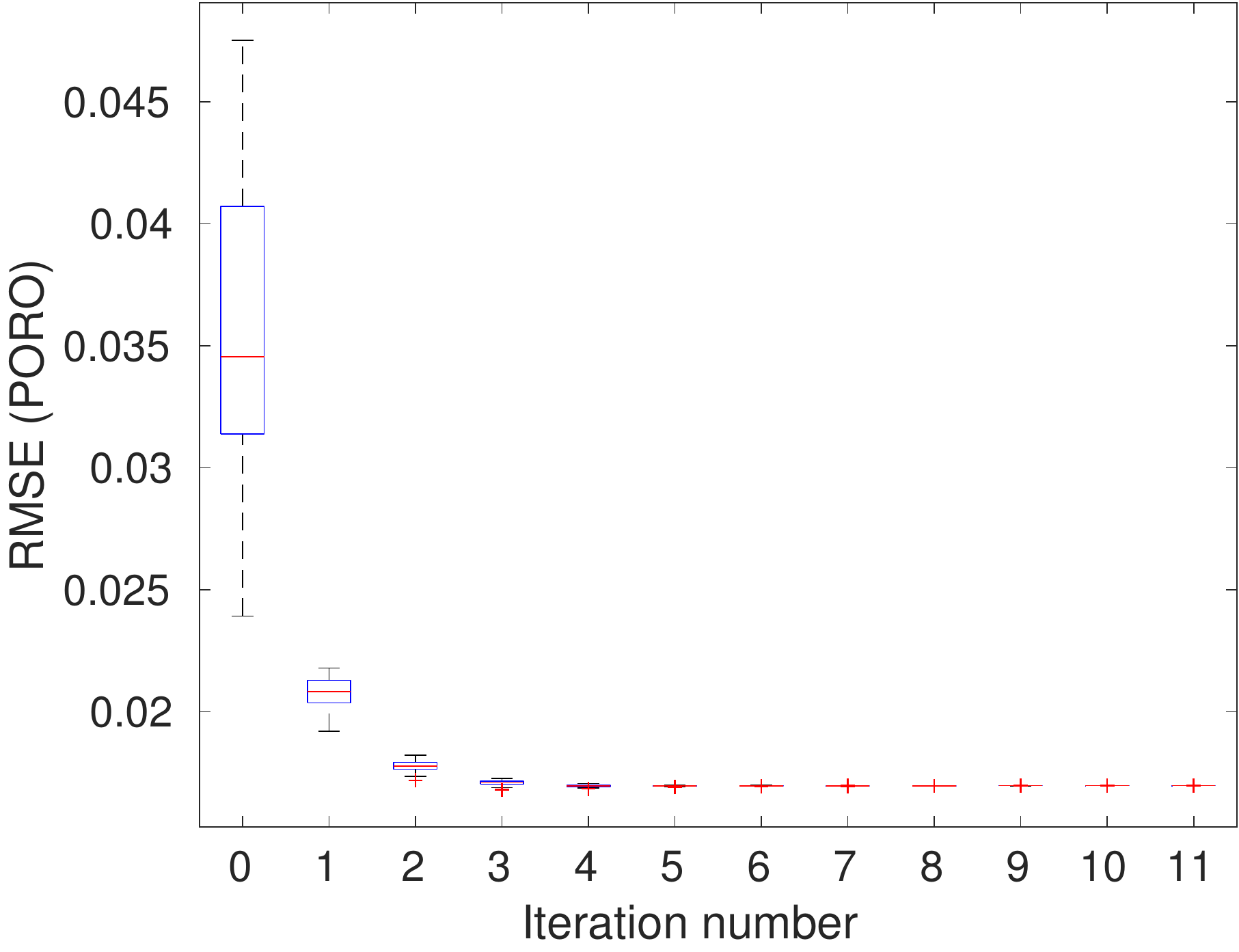}
	}%
	
	\subfigure[RMSEs of log PERMX ($c=5$)]{ \label{subfig:rmse_PERMX_boxplot_ensemble_c5_S2}
		\includegraphics[scale=\nScale]{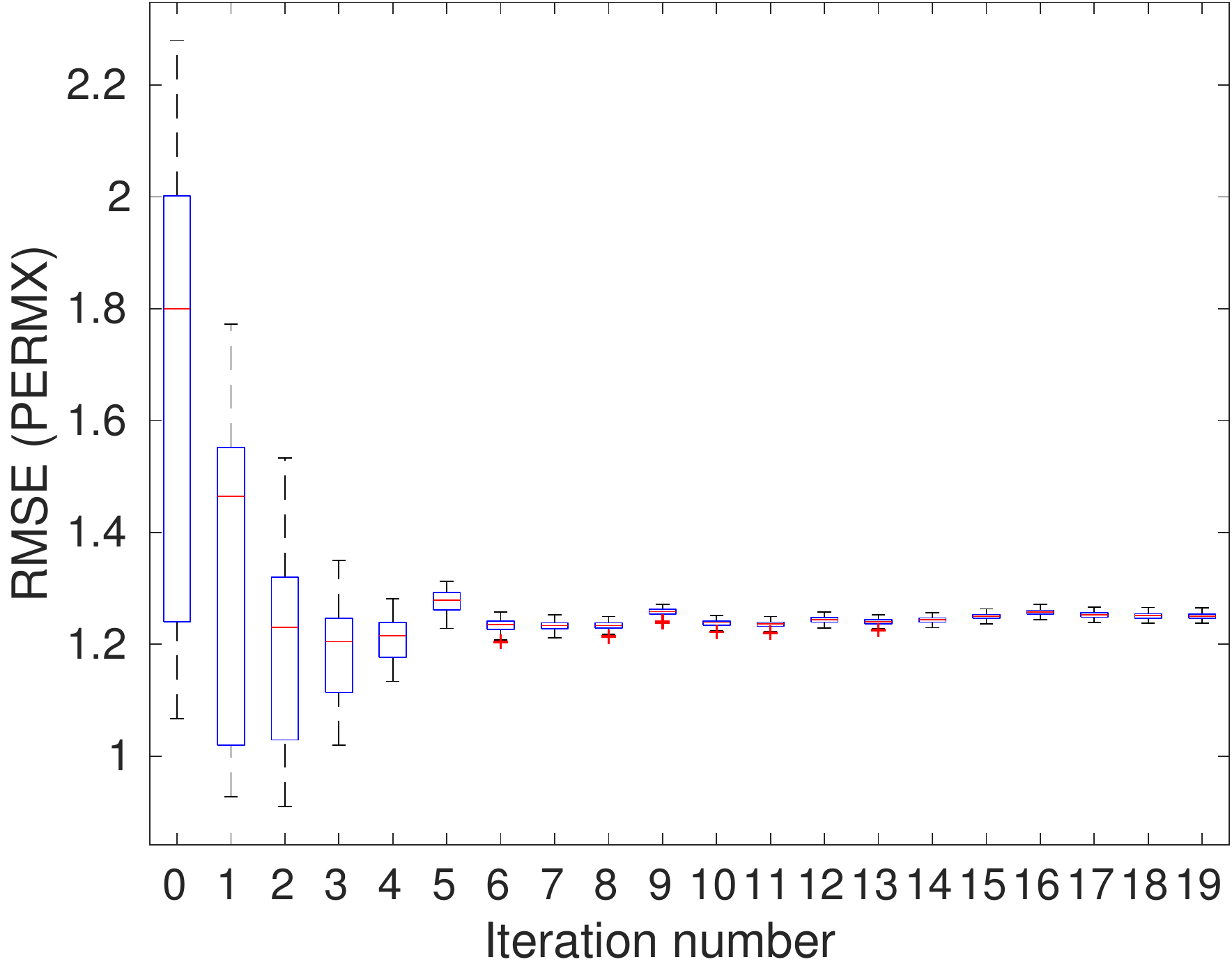}
	}%
	\subfigure[RMSEs of PORO ($c=5$)]{ \label{subfig:rmse_PORO_boxplot_ensemble_c5_S2}
		\includegraphics[scale=\nScale]{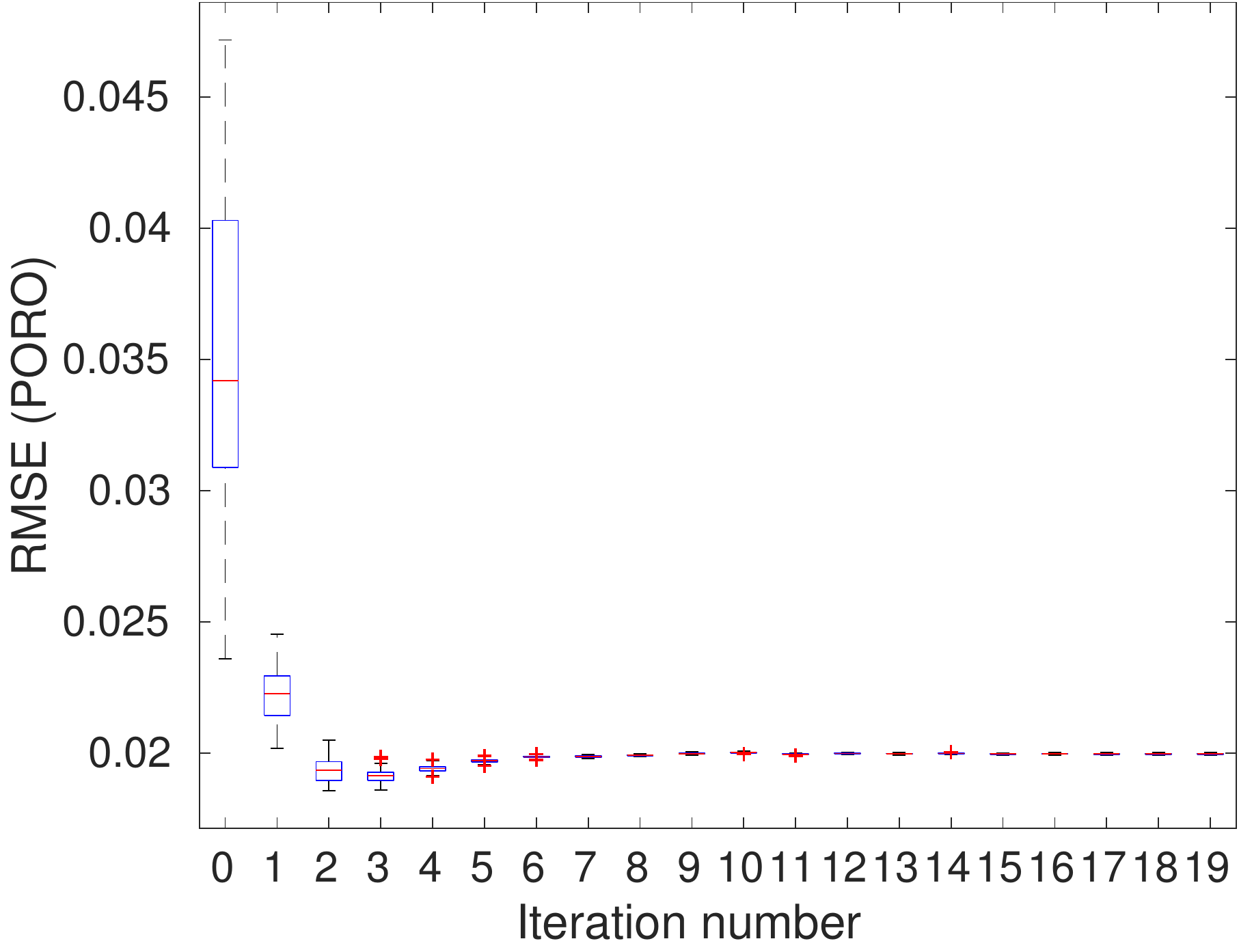}
	}%
	\caption{\label{fig:Brugge_RLM-MAC_RMSE_S2} Boxplots of RMSEs of log PERMX (1st column) and PORO (2nd column) as functions of iteration step, with $c$ being $1$ (top) and $5$ (bottom), respectively (scenario S2).}
\end{figure*}

Figure \ref{fig:Brugge_RLM-MAC_RMSE_S2} shows boxplots of RMSEs of log PERMX (1st column) and PORO (2nd column) as functions of iteration step. It is clear that the RMSEs of the final ensembles are lower than those of the initial ones, even at $c=5$, the case in which data size is reduced more than $4000$ times. On the other hand, when $c=1$ (top row), the RMSEs of both log PERMX and PORO in the final ensemble are lower than those at $c=5$ (bottom row). This indicates that, better history matching performance is achieved at $c=1$, with more information content captured in the leading wavelet coefficients.  

\renewcommand{\nScale}{0.33} 
\begin{figure*}
	\centering
	\subfigure[Observed slice (base)]{ \label{subfig:X80_tstep1_mid_trace_noisy_S2}
		\includegraphics[scale=\nScale]{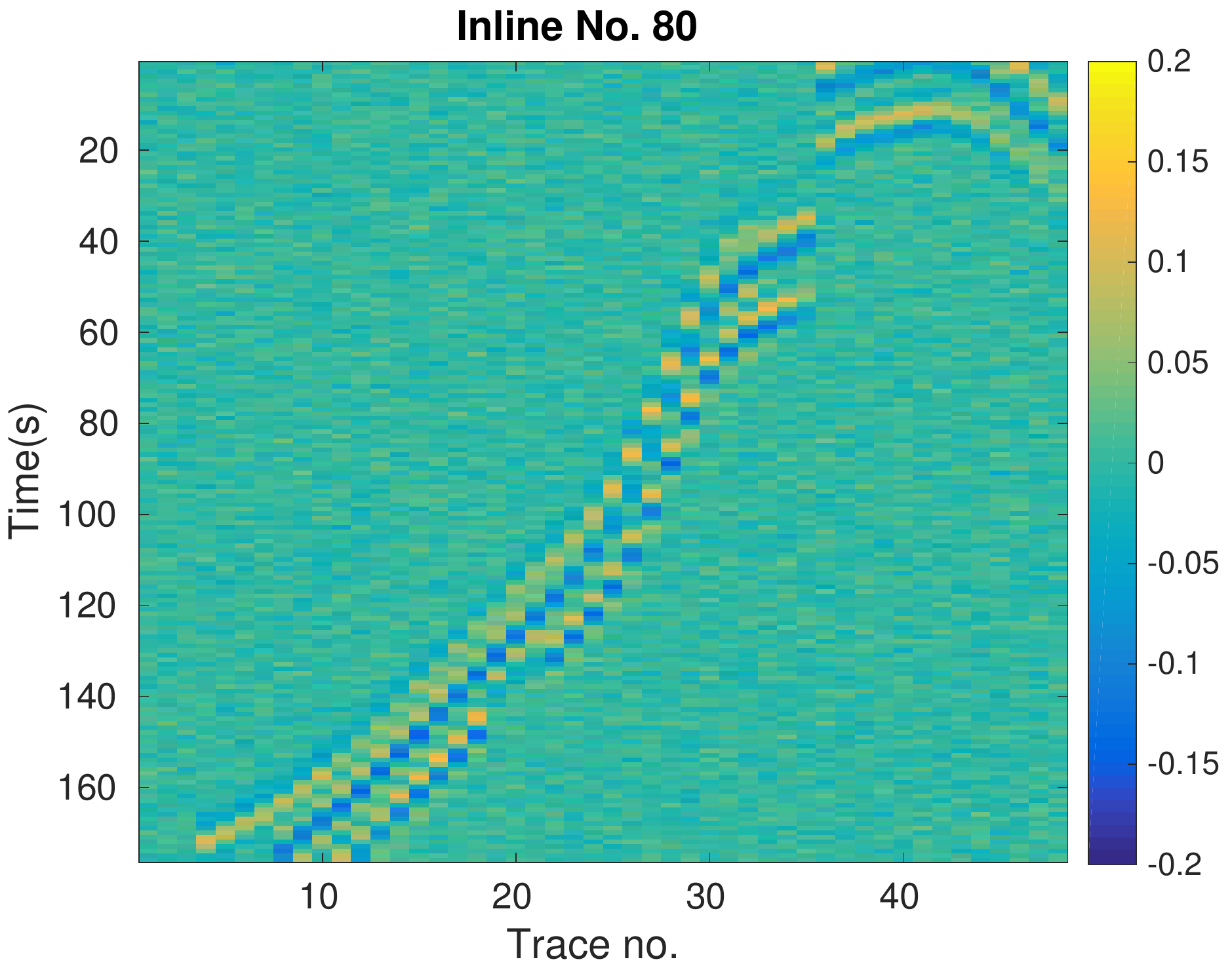}
	}%
	\subfigure[Observed slice (1st mornitor)]{ \label{subfig:X80_tstep2_mid_trace_noisy_S2}
		\includegraphics[scale=\nScale]{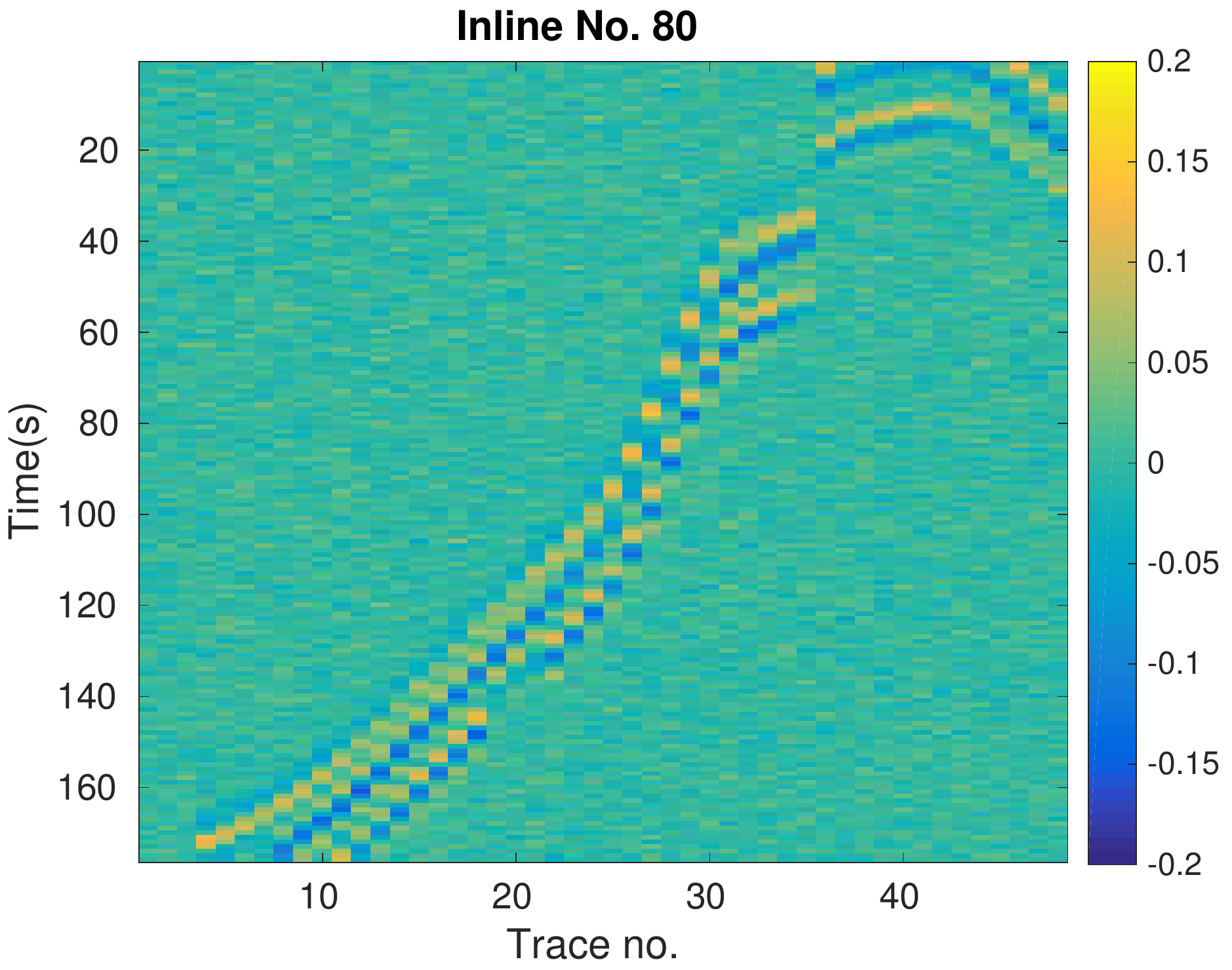}
	}%
	\subfigure[Observed slice (2st mornitor)]{ \label{subfig:X80_tstep3_mid_trace_noisy_S2}
		\includegraphics[scale=\nScale]{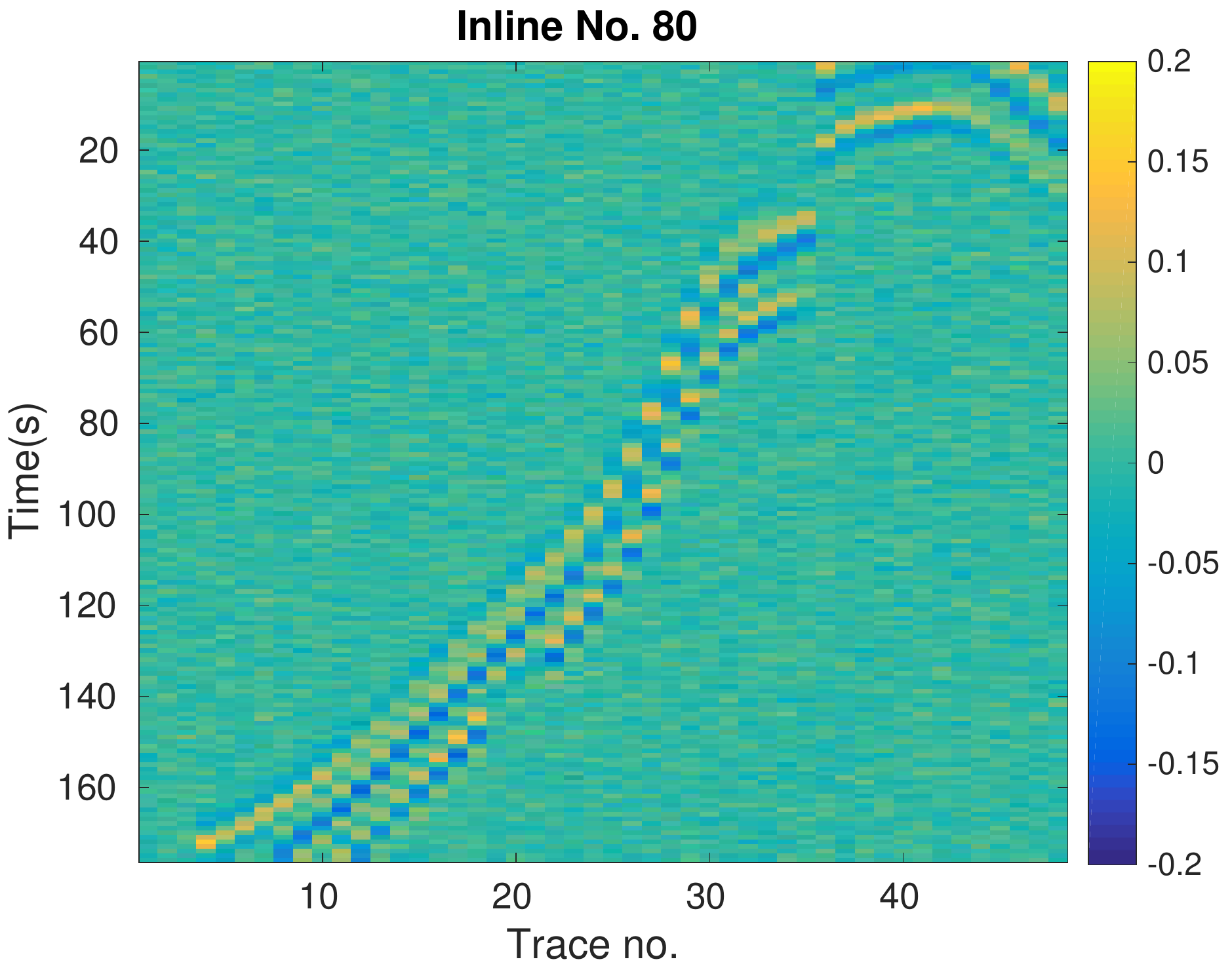}
	}%
	
	\subfigure[Reconstructed (base) with $c=1$]{ \label{subfig:X80_tstep1_mid_trace_rec_C1_S2}
		\includegraphics[scale=\nScale]{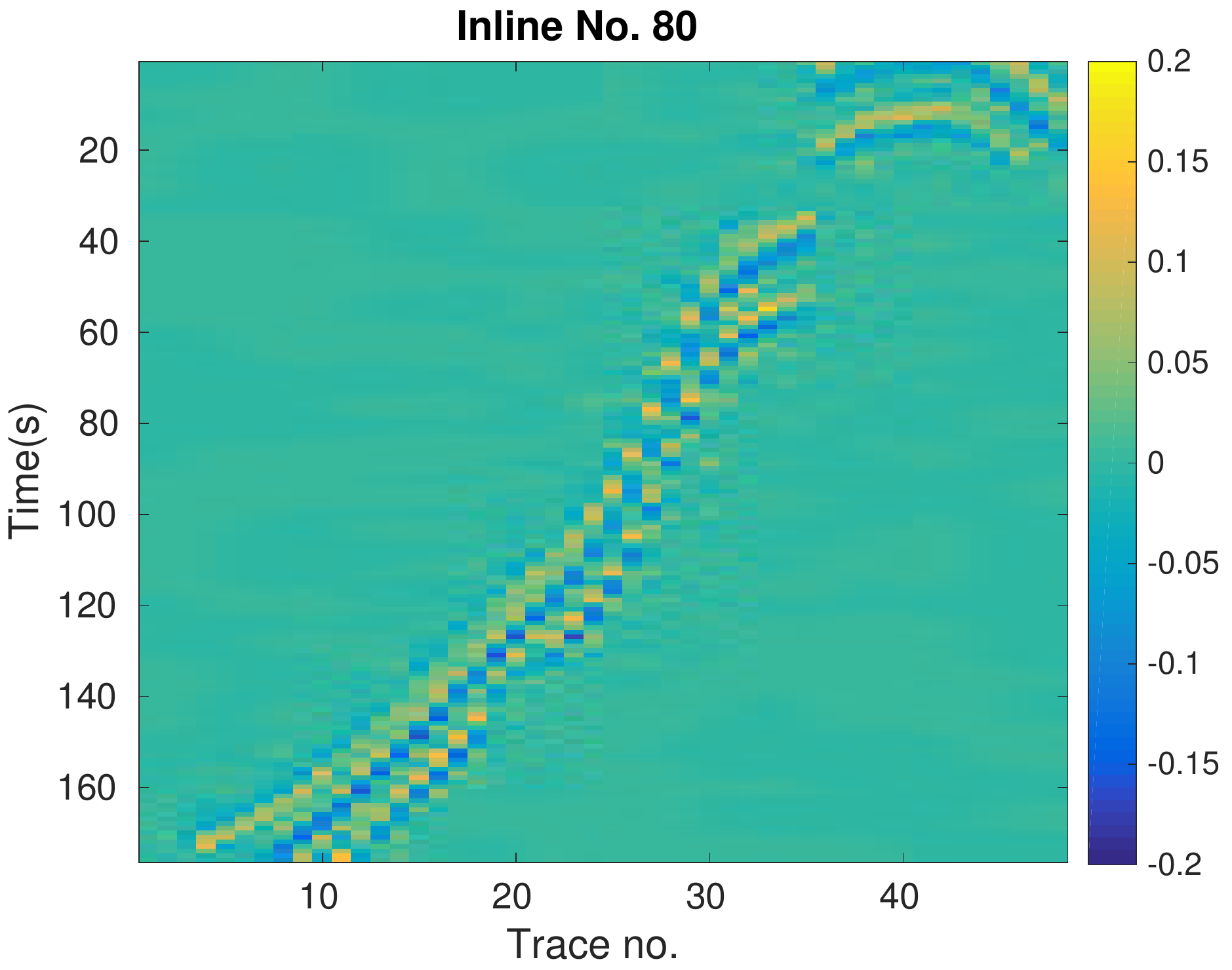}
	}%
	\subfigure[Reconstructed (1st mornitor) with $c=1$]{ \label{subfig:X80_tstep2_mid_trace_rec_C1_S2}
		\includegraphics[scale=\nScale]{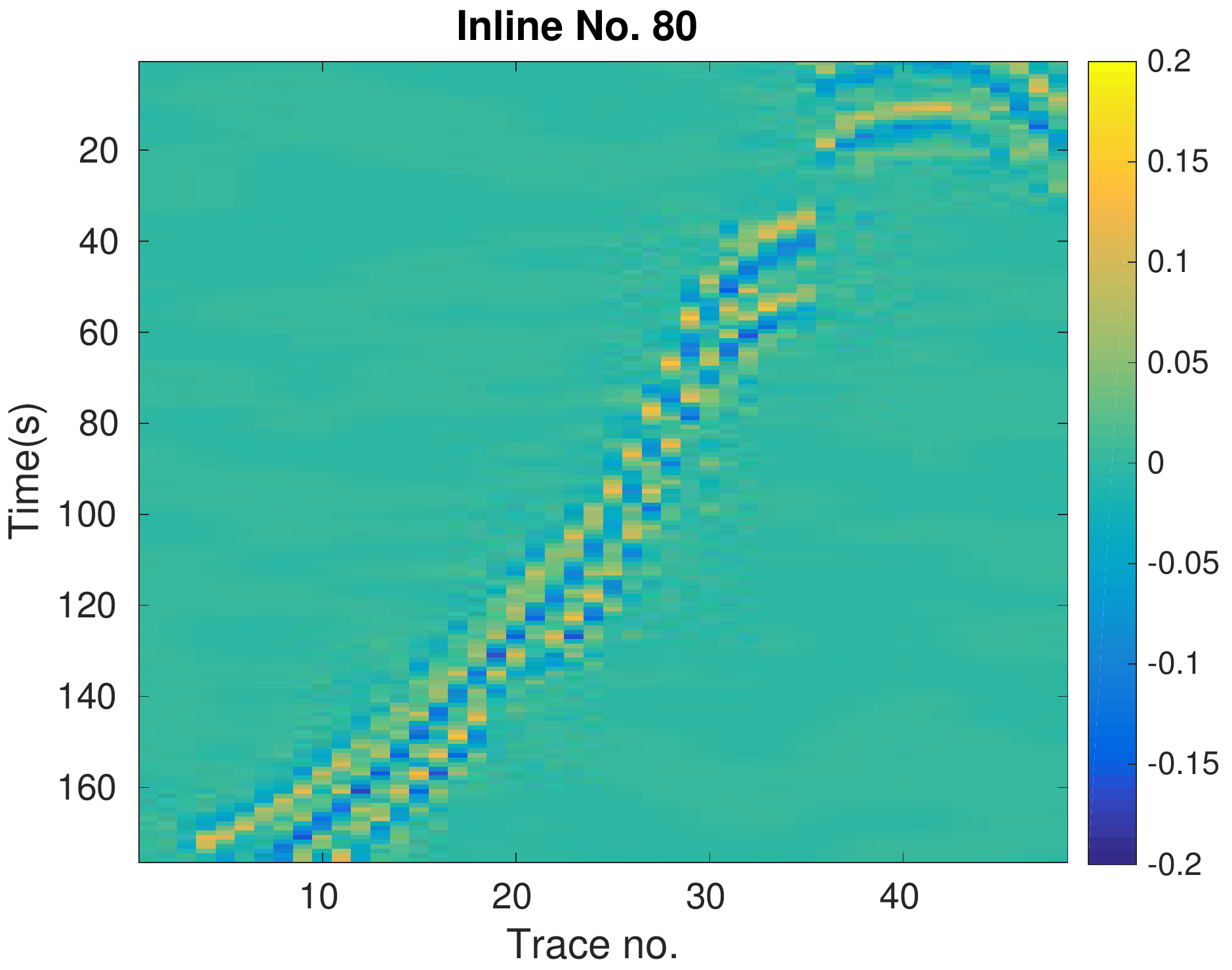}
	}%
	\subfigure[Reconstructed (2st mornitor) with $c=1$]{ \label{subfig:X80_tstep3_mid_trace_rec_C1_S2}
		\includegraphics[scale=\nScale]{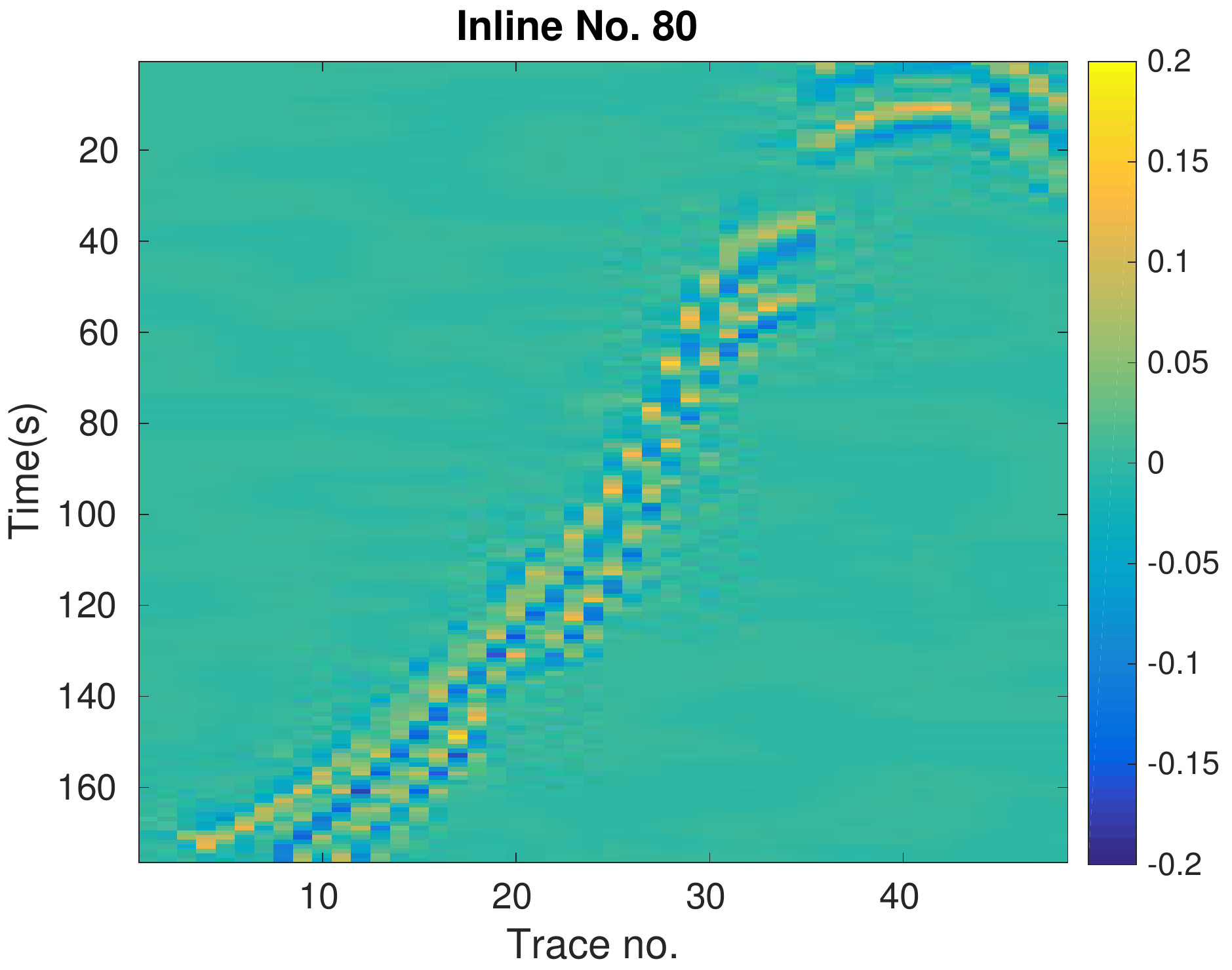}
	}%
	
	\subfigure[Reconstructed (base) with $c=5$]{ \label{subfig:X80_tstep1_mid_trace_rec_C5_S2}
		\includegraphics[scale=\nScale]{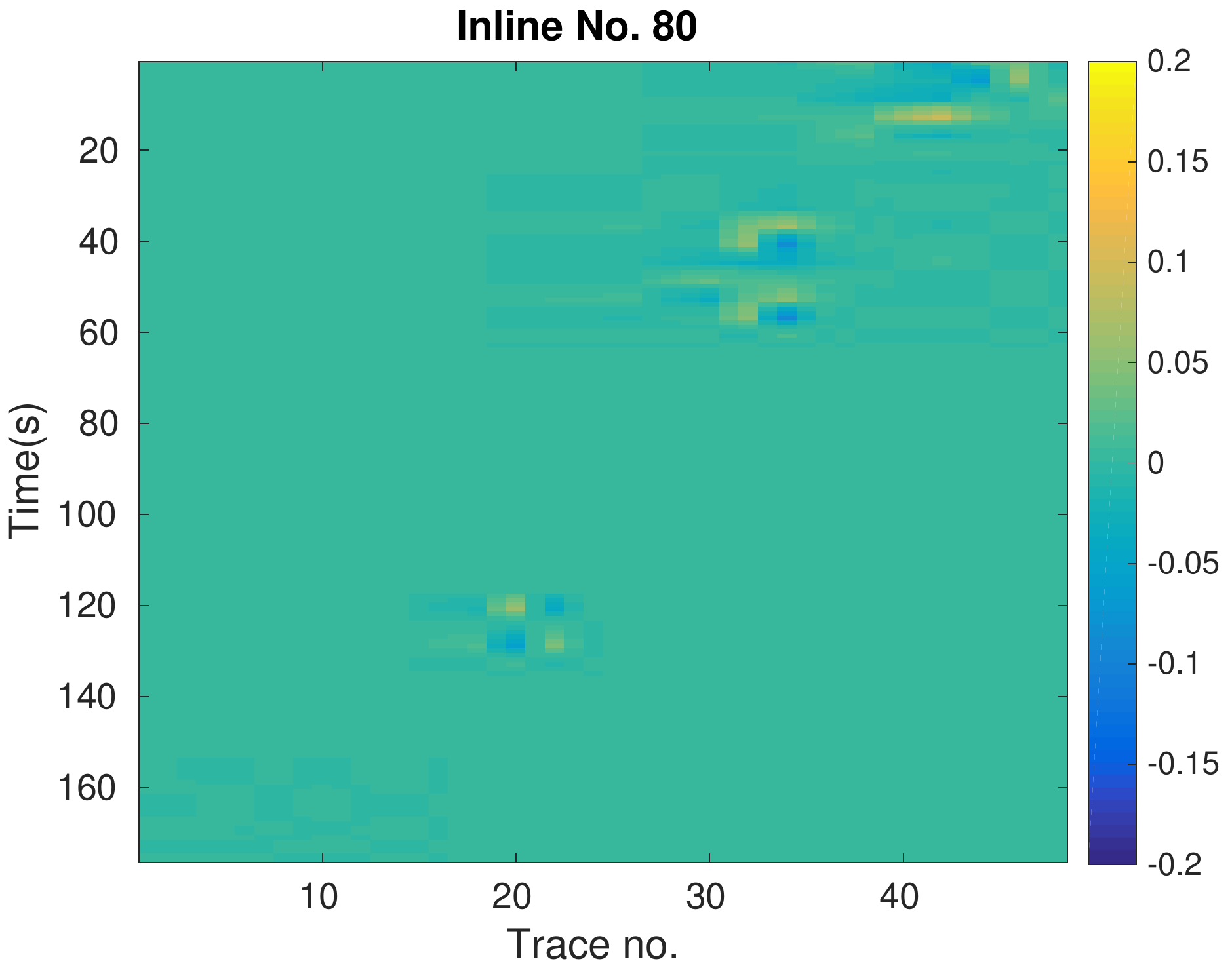}
	}%
	\subfigure[Reconstructed (1st mornitor) with $c=5$]{ \label{subfig:X80_tstep2_mid_trace_rec_C5_S2}
		\includegraphics[scale=\nScale]{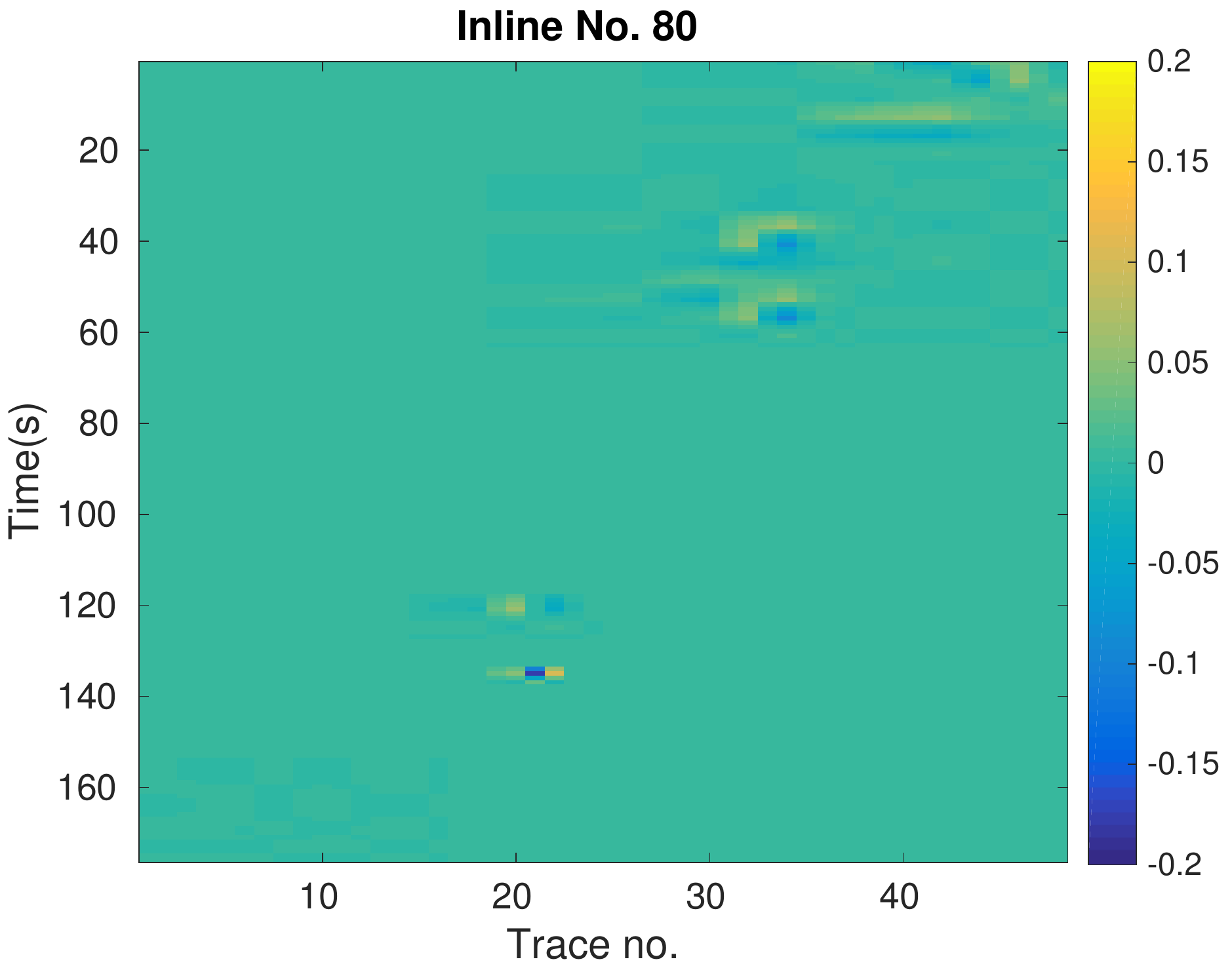}
	}%
	\subfigure[Reconstructed (2st mornitor) with $c=5$]{ \label{subfig:X80_tstep3_mid_trace_rec_C5_S2}
		\includegraphics[scale=\nScale]{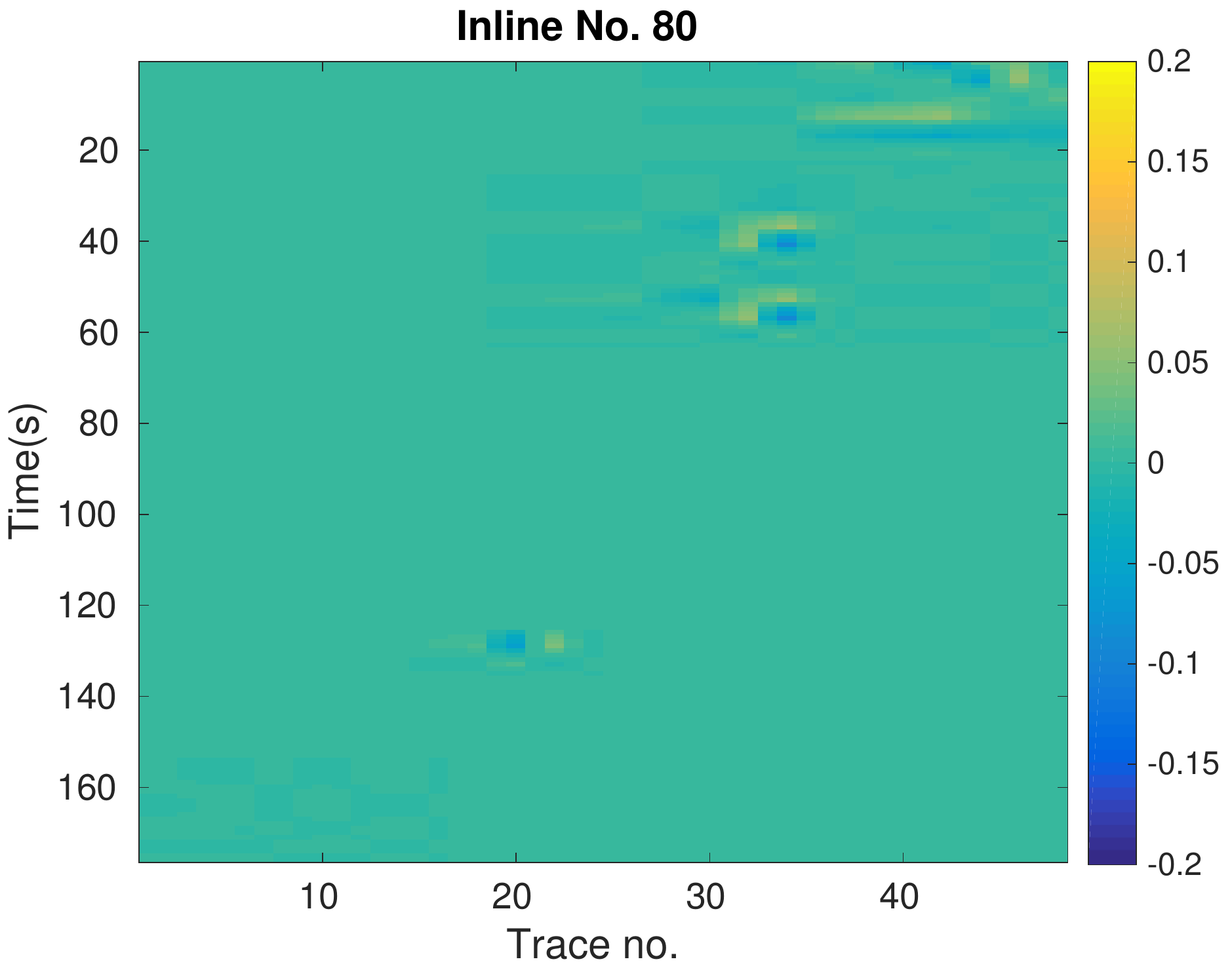}
	}%
	
	\caption{\label{fig:observed_and_rec_slices} Top row: slices of the observed far-offset AVA cubes at $X=80$, with respect to the base survey (1st column), the 1st monitor survey (2nd column) and the 2nd monitor survey (3rd column), respectively. Middle row: corresponding reconstructed slices at $X=80$ using the leading wavelet coefficients at $c=1$ (while all other wavelet coefficients are set to zero). Bottom row:  corresponding reconstructed slices at $X=80$ using the leading wavelet coefficients at $c=5$ (while all other wavelet coefficients are set to zero). }
\end{figure*}

As aforementioned, each 3D AVA cube is in the dimension of $139 \times 48 \times 176$. For illustration, the top row of Figure \ref{fig:observed_and_rec_slices} indicates the slices of far-offset AVA cubes at $X=80$, with respect to the base survey (1st column), the 1st monitor survey (2nd column) and the 2nd monitor survey (3rd column), respectively, whereas the middle and bottom rows show the corresponding slices reconstructed using the leading wavelet coefficients (while setting other coefficients to zero) at $c=1$ and $c=5$, respectively. Compared to figures in the top row, it is clear that the reconstructed ones at $c=1$ capture the main features of the observed slices, while removing the noise component. Therefore in this case, although the universal rule (corresponding to $c=1$) still leads to a relatively large data size, it achieves a good trade-off between data size reduction and feature preservation. In contrast, at $c=5$, the seismic data size is significantly reduced. However, the reconstructed slices in the bottom row only retain a small portion of the strips in the observed slices of the top row, meaning that the data size is reduced at the cost of losing substantial information content of the seismic data. Nevertheless, even with such an information loss, using the leading wavelet coefficients at $c=5$ still leads to significantly improved model estimation in comparison to the initial ensemble, and this will become more evident when both production and seismic data are used in scenario S3. 

\renewcommand{\nScale}{0.33} 
\begin{figure*}
	\centering
	\subfigure[Slice of initial differerence (base)]{ \label{subfig:X80_tstep1_mid_trace_diff_initEns_S2}
		\includegraphics[scale=\nScale]{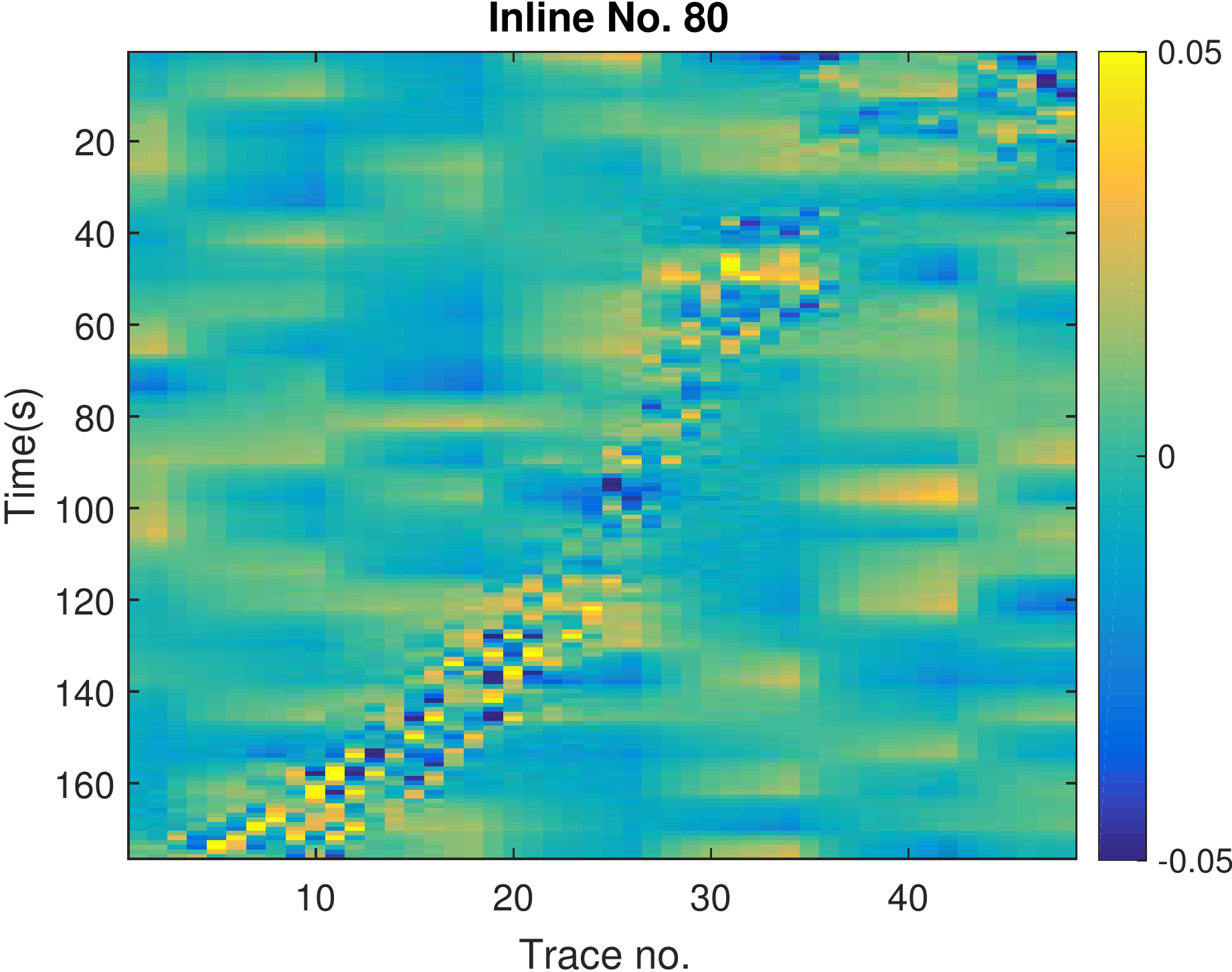}
	}%
	\subfigure[Slice of initial differerence (1st mornitor)]{ \label{subfig:X80_tstep2_mid_trace_diff_initEns_S2}
		\includegraphics[scale=\nScale]{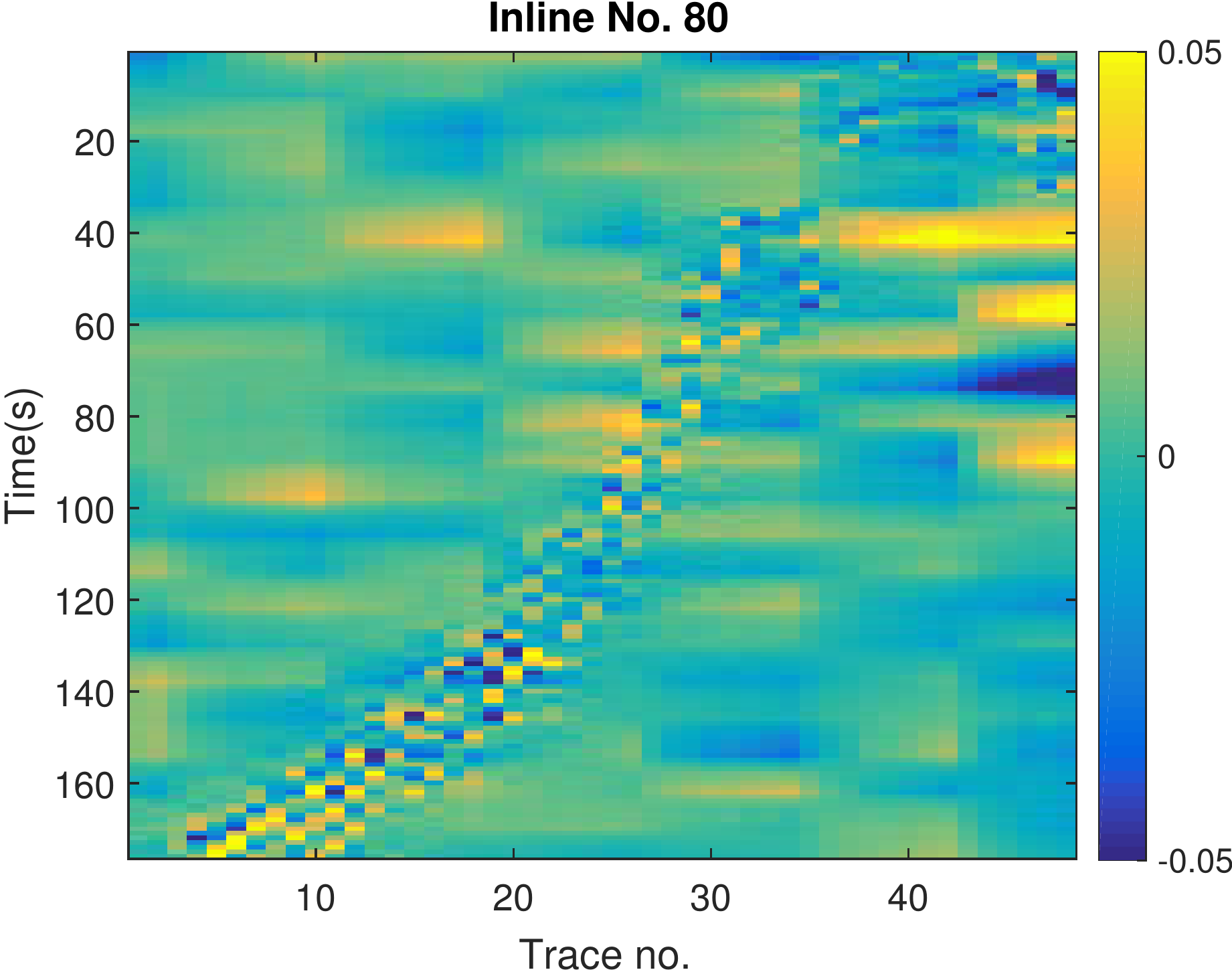}
	}%
	\subfigure[Slice of initial differerence (2st mornitor)]{ \label{subfig:X80_tstep3_mid_trace_diff_initEns_S2}
		\includegraphics[scale=\nScale]{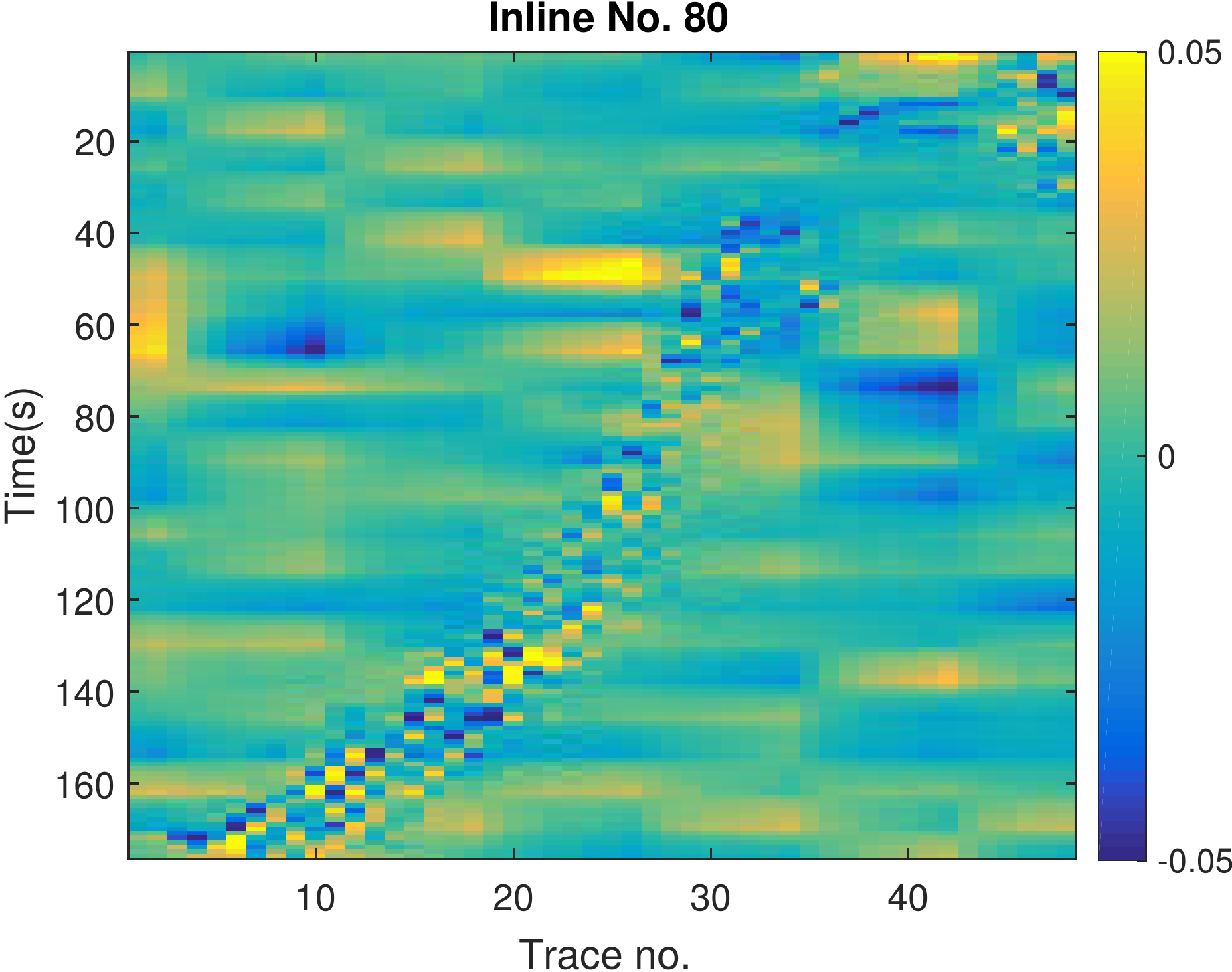}
	}%
	
	\subfigure[Slice of final differerence (base)]{ \label{subfig:X80_tstep1_mid_trace_diff_finalEns_S2}
		\includegraphics[scale=\nScale]{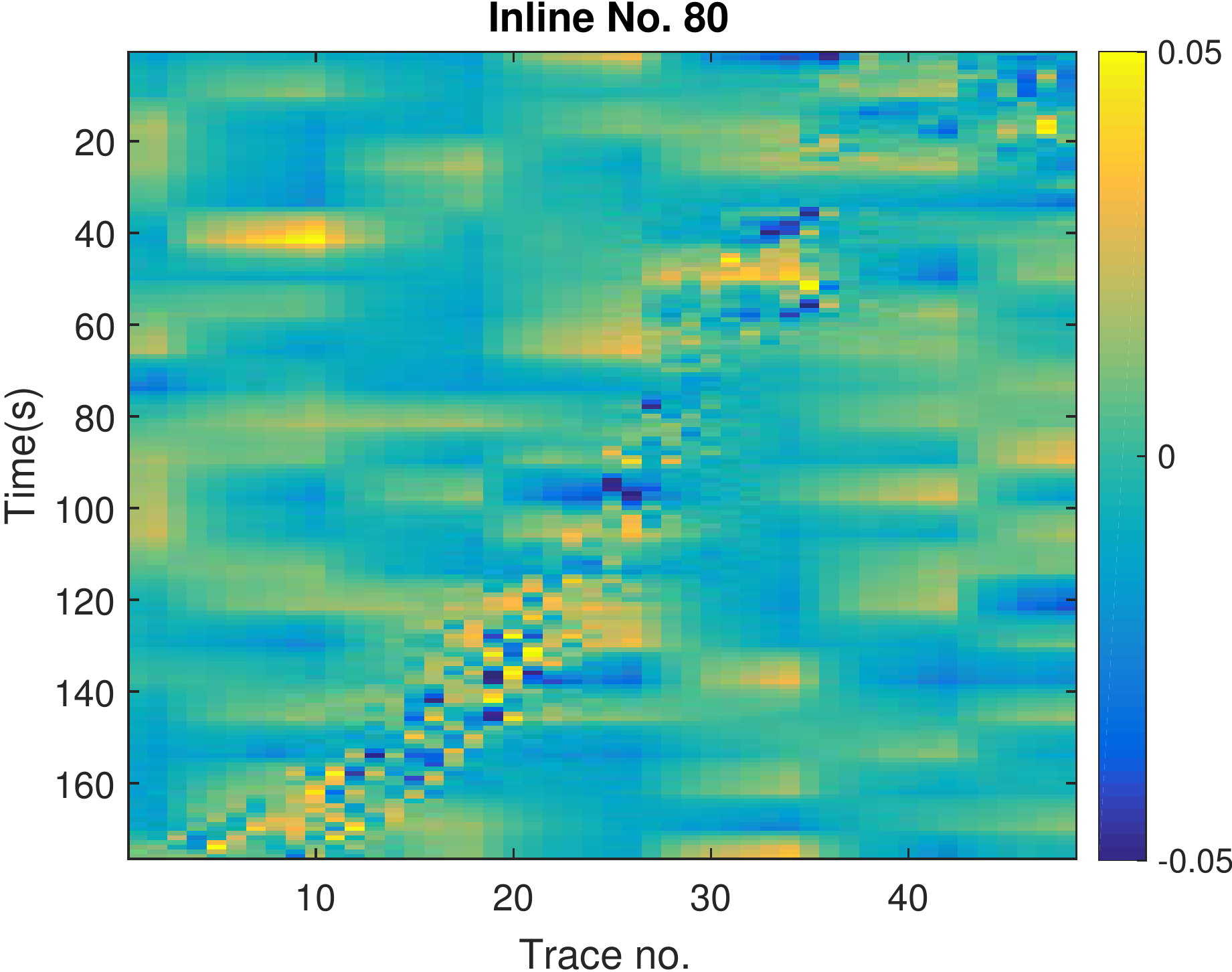}
	}%
	\subfigure[Slice of final differerence (1st mornitor)]{ \label{subfig:X80_tstep2_mid_trace_diff_finalEns_S2}
		\includegraphics[scale=\nScale]{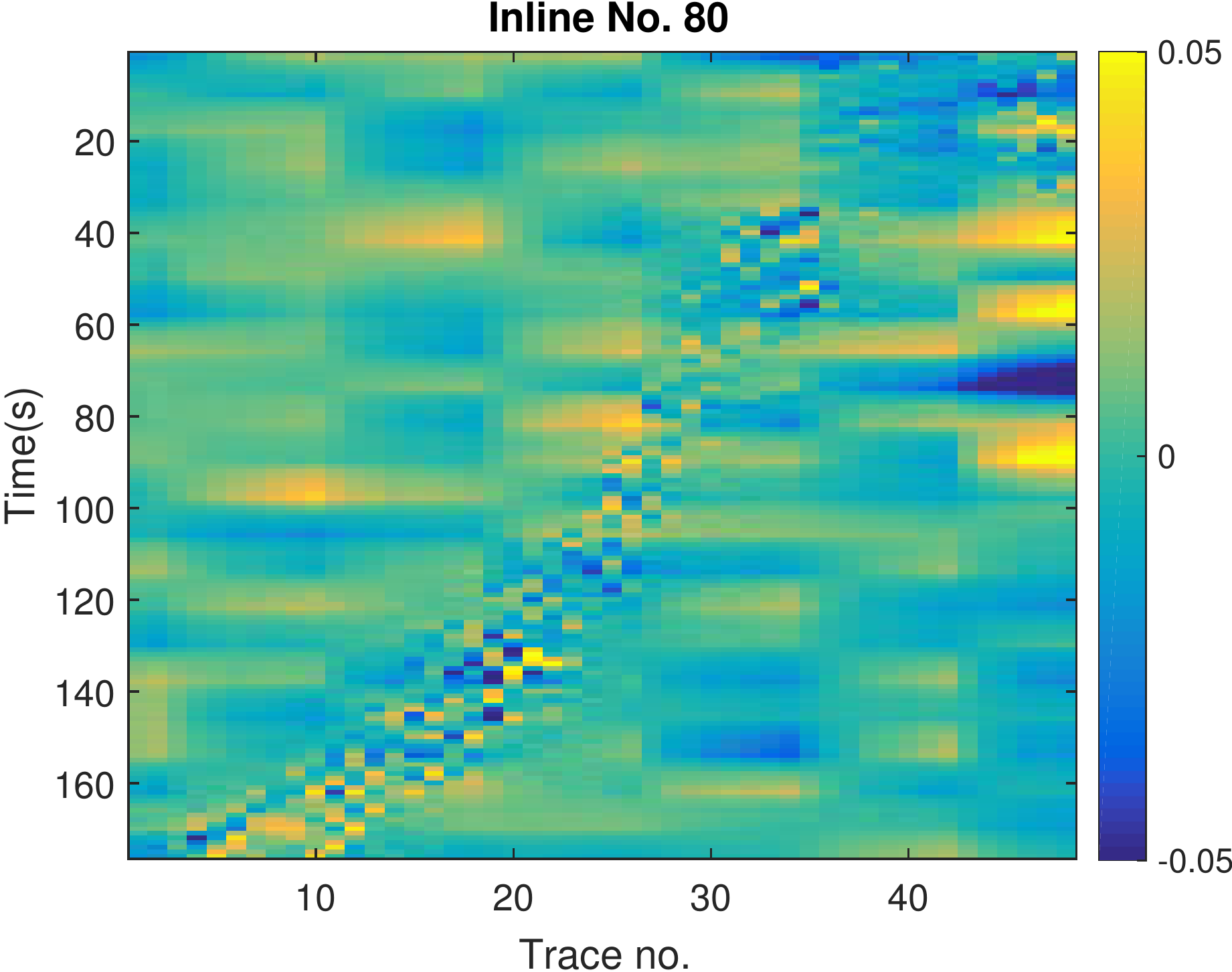}
	}%
	\subfigure[Slice of final differerence (2st mornitor)]{ \label{subfig:X80_tstep3_mid_trace_diff_finalEns_S2}
		\includegraphics[scale=\nScale]{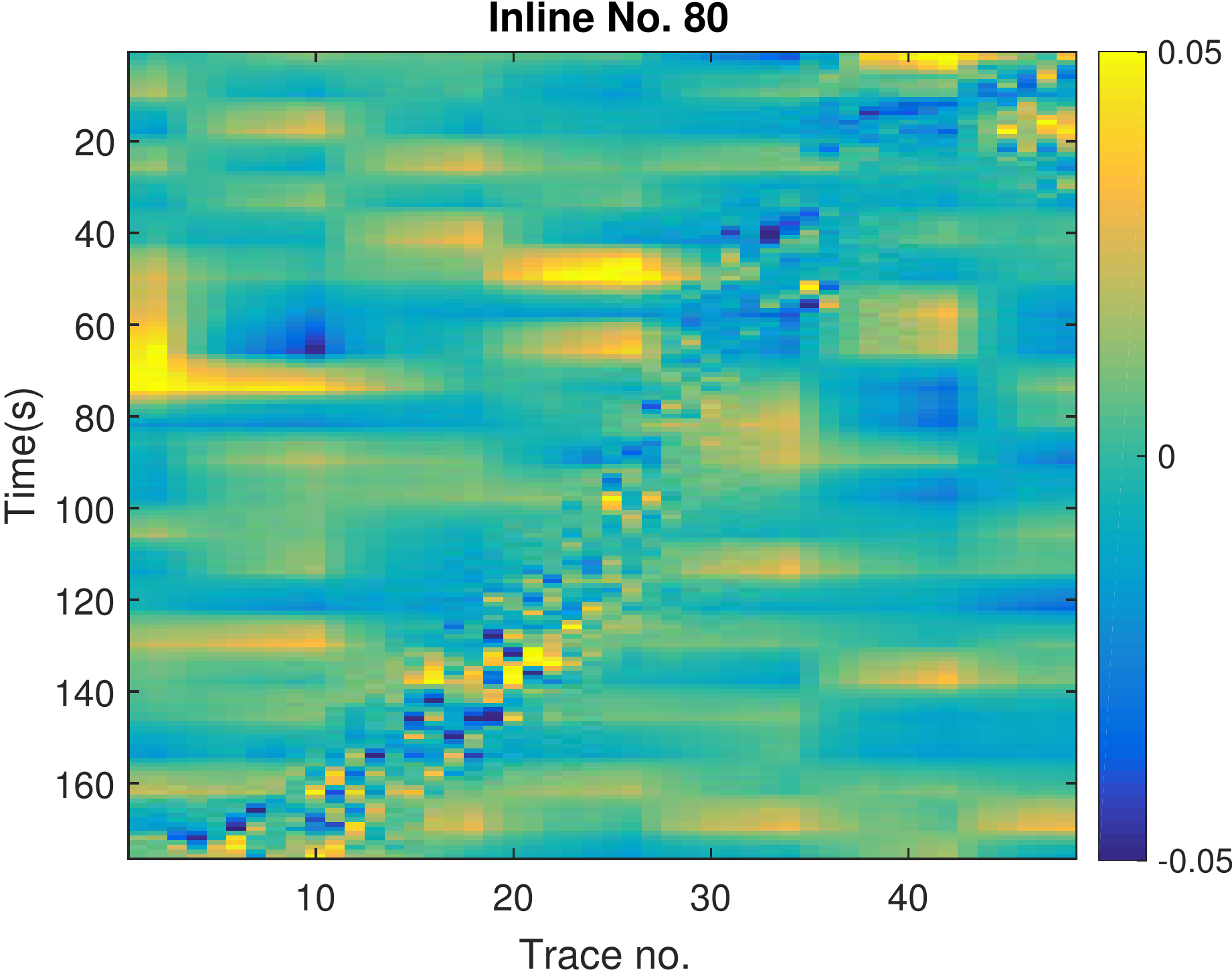}
	}%
	
	\caption{\label{fig:diff_slices} Top row: slices (at $X=80$) of the differences between the reconstructed far-offset AVA cubes using the leading wavelet coefficients ($c=1$) of the observed seismic data, and the reconstructed far-offset AVA cubes using the corresponding leading wavelet coefficients ($c=1$)  of the means of the simulated seismic data of the \textbf{initial} ensemble. From left to right, the three columns correspond to the differences at the base, the 1st monitor, and the 2nd monitor surveys, respectively. Bottom row: as in the top row, except that it is for the differences between the reconstructed far-offset AVA cubes of the observed seismic data, and the reconstructed far-offset AVA cubes of the mean simulated seismic data of the \textbf{final} ensemble. }
\end{figure*}

For brevity, in what follows we only present the results with respect to the case $c=1$. In the top row of Figure \ref{fig:diff_slices}, we show the slices (at X = 80) of differences between two groups of reconstructed far-offset AVA cubes. One group corresponds to the reconstructed far-offset AVA cubes at three survey times, using the leading wavelet coefficients ($c = 1$) of the observed far-offset AVA cubes. The other group contains the reconstructed far-offset AVA cubes at three survey times, using the corresponding leading wavelet coefficients ($c = 1$) of the mean simulated seismic data of the initial ensemble. Therefore the slices of differences in the top row can be considered as a reflection of the initial seismic data mismatch in Figure \ref{subfig:Brugge_boxplot_objRealIter_c1_S2}. Here, we use the slices of differences for ease of visualization, as the reconstructed slices of the observed and the mean simulated AVA cubes look very similar. Similarly, in the bottom row, we show the slices of differences between the reconstructed far-offset AVA cubes of the observed seismic data, and the reconstructed far-offset AVA cubes of the mean simulated seismic data of the final ensemble. In this case, the slices of differences can be considered as a reflection of the final seismic data mismatch in Figure \ref{subfig:Brugge_boxplot_objRealIter_c1_S2}. Comparing the top and bottom rows at a given survey time, one can observe certain distinctions, which, however, are not very significant in general. This is in line with the results in Figure \ref{subfig:Brugge_boxplot_objRealIter_c1_S2}, where the initial and final seismic data mismatch remain in the same order, in contrast to the substantial reduction of production data mismatch in scenario S1 (Figure \ref{fig:Brugge_boxplot_objRealIter_S1}).  

\renewcommand{\nScale}{0.21} 
\begin{figure*}
	\centering
	\subfigure[Rerefrence log PERMX]{ \label{subfig:PERMX_L2_true_S2} %
		\includegraphics[scale=\nScale]{PERMX_L2_true.eps}
	}
	\subfigure[Mean of initial log PERMX]{ \label{subfig:PERMX_L2_Mean_initEns_S2}
		\includegraphics[scale=\nScale]{PERMX_L2_Mean_initEns.eps}
	}
	\subfigure[Mean of final log PERMX]{ \label{subfig:PERMX_L2_Mean_ensemble11_S2}
		\includegraphics[scale=\nScale]{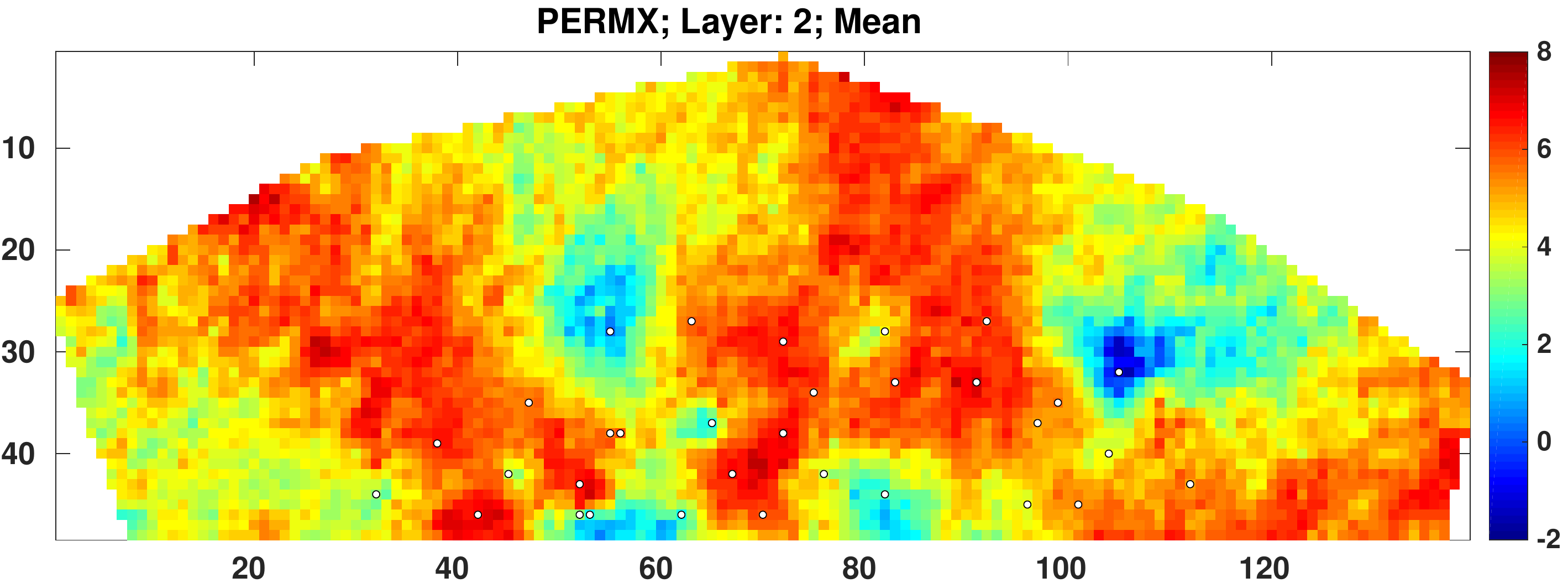}
	}%
	
	\subfigure[Rerefrence PORO]{ \label{subfig:PORO_L2_true_S2} %
		\includegraphics[scale=\nScale]{PORO_L2_true.eps}
	}
	\subfigure[Mean of initial PORO]{ \label{subfig:PORO_L2_Mean_initEns_S2}
		\includegraphics[scale=\nScale]{PORO_L2_Mean_initEns.eps}
	}
	\subfigure[Mean of final PORO]{ \label{subfig:PORO_L2_Mean_ensemble11_S2}
		\includegraphics[scale=\nScale]{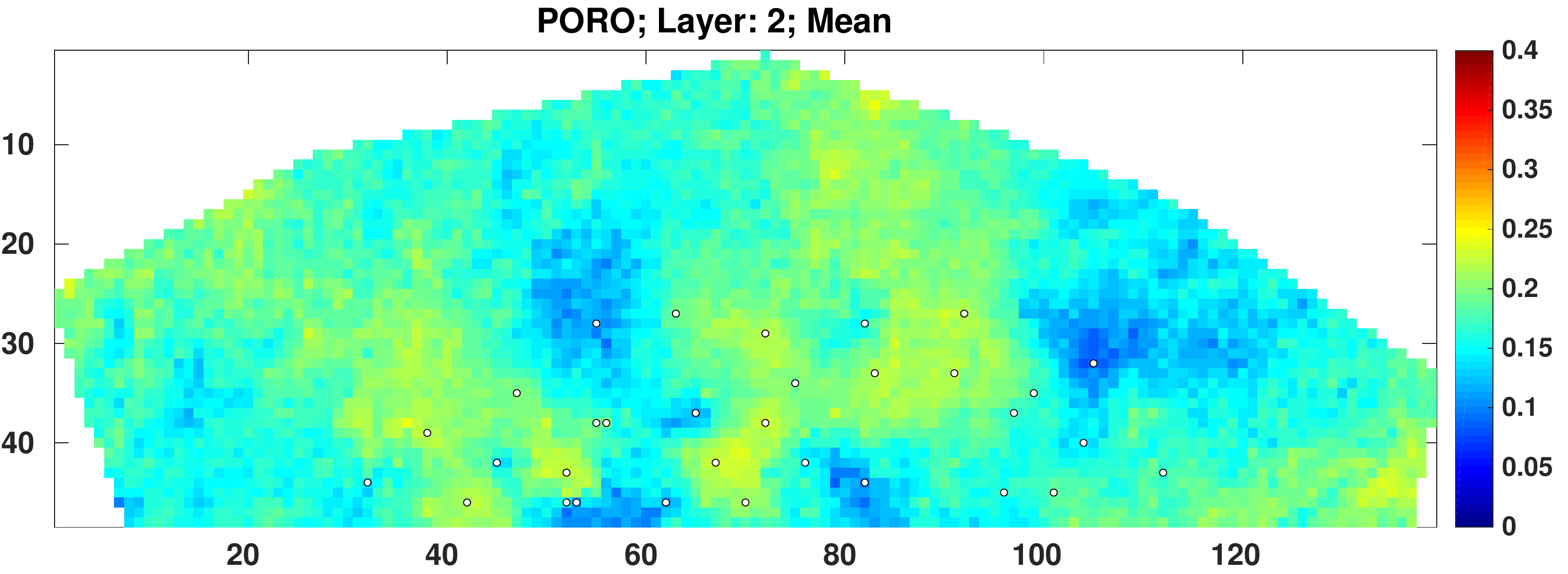}
	}%
	\caption{\label{fig:estimation_S2} As in Figure \ref{fig:estimation_S1}, but for scenario S2 with $c=1$.}
\end{figure*}     

Similar to Figure \ref{fig:estimation_S1}, Figure \ref{fig:estimation_S2} depicts the reference log PERMX and PORO at layer 2 (1st column), the mean of initial log PERMX and PORO at layer 2 (2nd column), and the mean of final log PERMX and PORO at layer 2 (3rd column). Compared to the initial mean estimates, the final mean log PERMX and PORO show clear improvements, in terms of the similarities to the reference fields. In addition, an inspection on the 3rd columns of Figures \ref{fig:estimation_S1} and \ref{fig:estimation_S2} reveals that the final mean estimates in S2 capture the geological structures of the reference fields better, especially in areas where there is neither injection nor production well (well locations are represented by black dots in Figures \ref{fig:estimation_S1} and \ref{fig:estimation_S2}).

\subsection{Results of scenario S3 (using both production and seismic data)}

\renewcommand{\nScale}{0.45} 
\begin{figure*}
	\centering
	\subfigure[Production data mismatch ($c=1$)]{ \label{subfig:Brugge_boxplot_objRealIter_prod_C1_S3} %
		\includegraphics[scale=\nScale]{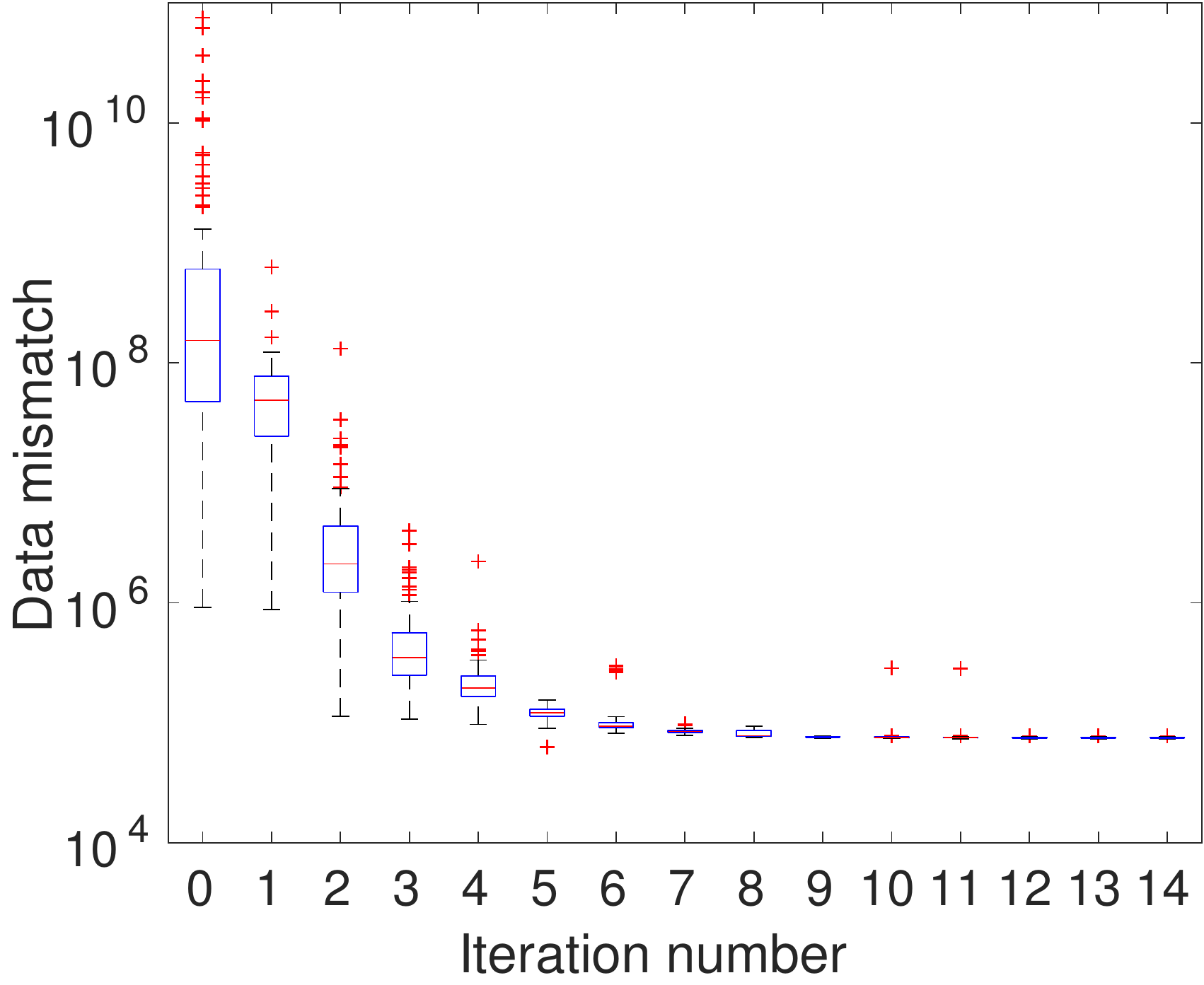}
	}%
	\subfigure[Production data mismatch ($c=5$)]{ \label{subfig:Brugge_boxplot_objRealIter_prod_C5_S3} %
		\includegraphics[scale=\nScale]{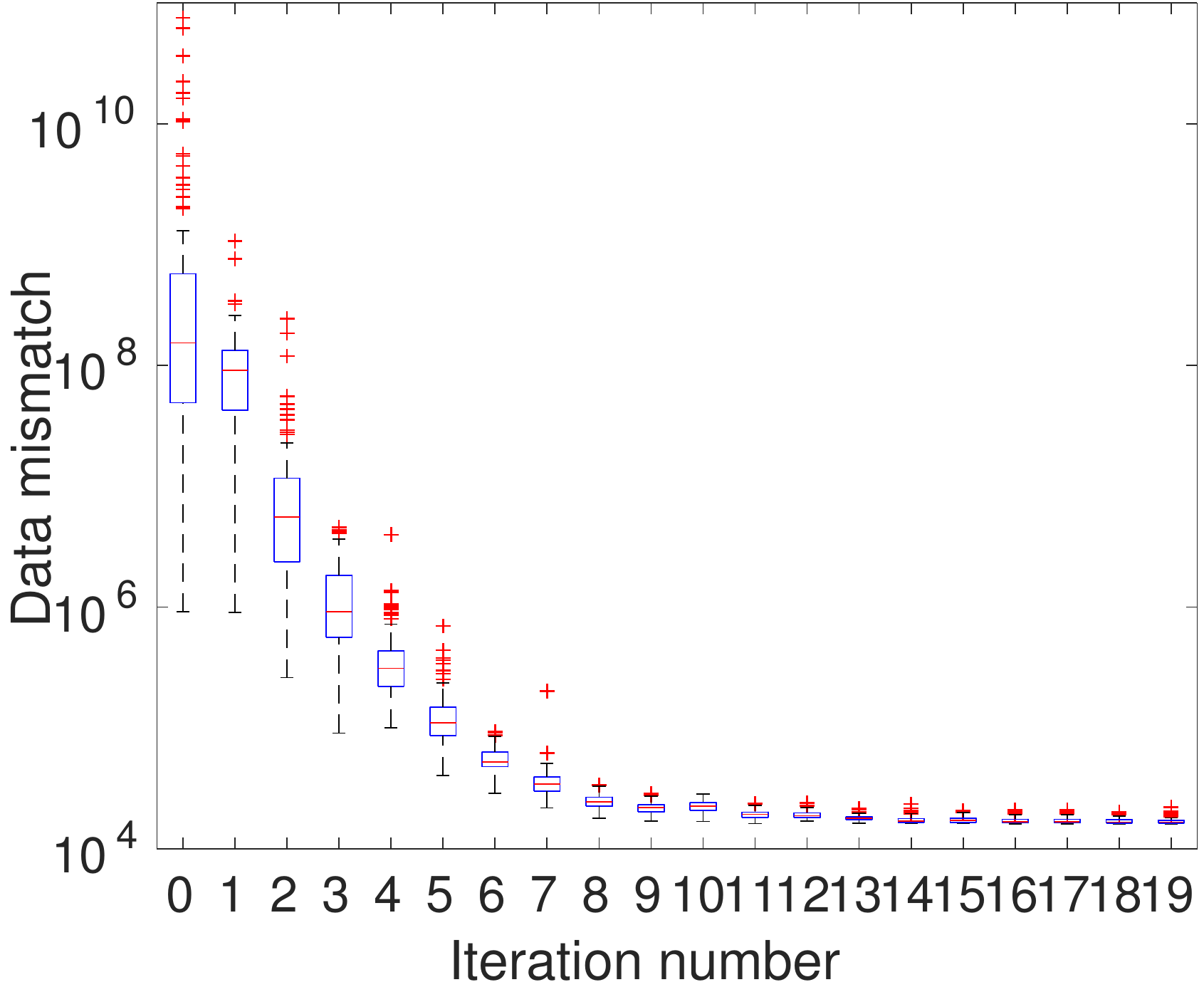}
	}%
	
	\subfigure[Seimic data mismatch ($c=1$)]{ \label{subfig:Brugge_boxplot_objRealIter_seis_C1_S3} %
		\includegraphics[scale=0.425]{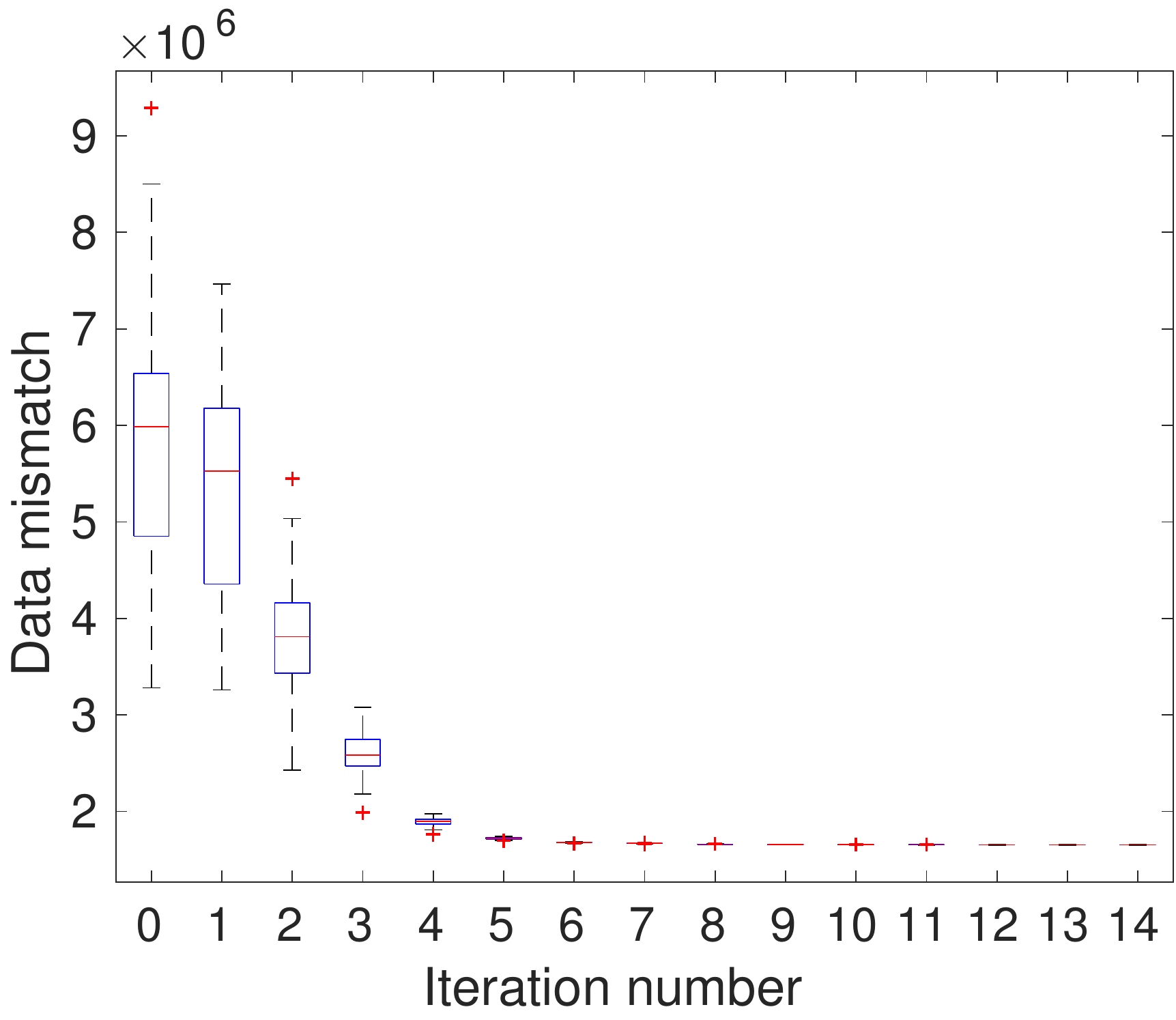}
	}%
	\subfigure[Seimic data mismatch ($c=5$)]{ \label{subfig:Brugge_boxplot_objRealIter_seis_C5_S3} %
		\includegraphics[scale=0.425]{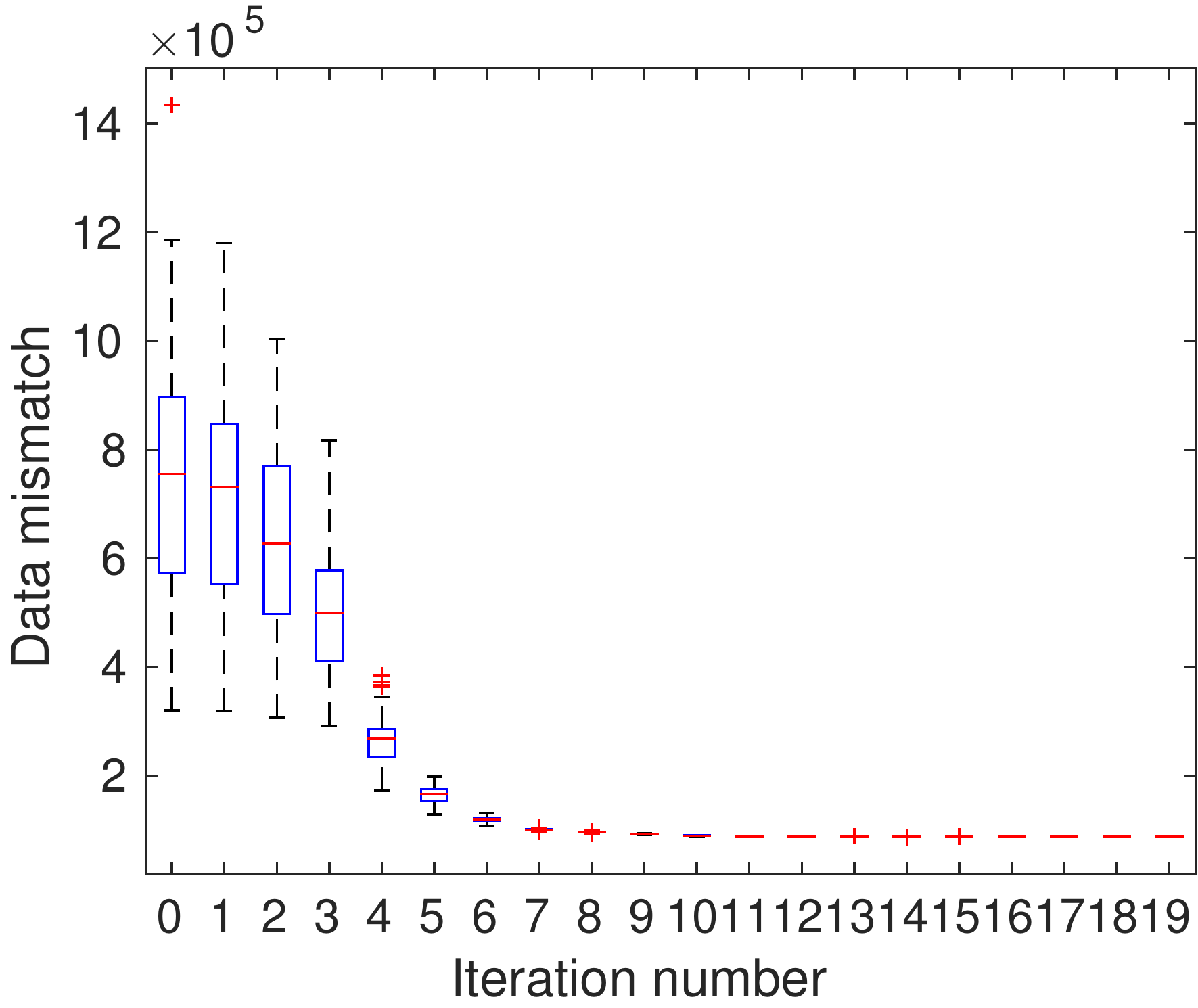}
	}%
	\caption{\label{fig:Brugge_boxplot_objRealIter_S3} Boxplots of production (top) and seismic (bottom) data mismatch as functions of iteration step (scenario S3).}
\end{figure*}  

In scenario S3, production and seismic (in terms of leading wavelet coefficients) data are assimilated simultaneously. Figure \ref{fig:Brugge_boxplot_objRealIter_S3} reports the boxplots of production (top) and seismic (bottom) data mismatch as functions of iteration step. Because of the simultaneous assimilation of production and seismic data, the way to use the seismic data (in terms of the value of $c$ in Eq. (\ref{eq:multiple_universal_rule})) will affect the history matching results. This becomes evident if one compares the first and second columns of Figure \ref{fig:Brugge_boxplot_objRealIter_S3}. Indeed, when $c=1$, because the relatively large data size, it is clear that ensemble collapse takes place in Figures \ref{subfig:Brugge_boxplot_objRealIter_prod_C1_S3} and \ref{subfig:Brugge_boxplot_objRealIter_seis_C1_S3}. Also, the iteration stops at step 14, due to the stopping criterion (C2). By increasing $c$ to 5, the size of seismic data is reduced from 178332 to 1665, and ensemble collapse seems mitigated to some extent, especially for production data, while the final iteration step is 19, due to the stopping criterion (C2). On the other hand, by comparing Figures \ref{fig:Brugge_boxplot_objRealIter_S1}, \ref{fig:Brugge_boxplot_objRealIter_S2} and \ref{fig:Brugge_boxplot_objRealIter_S3}, it is clear that, in S3, the presence of both production and seismic data makes the reduction of data mismatch different from the case of using either production or seismic data only. For instance, in the presence of seismic data, the production data mismatch (see Figures \ref{subfig:Brugge_boxplot_objRealIter_prod_C1_S3} and \ref{subfig:Brugge_boxplot_objRealIter_prod_C5_S3}) tend to be higher than that in Figure \ref{fig:Brugge_boxplot_objRealIter_S1}. On the other hand, with the influence of production data, the occurrence of ensemble collapse seems to be postponed in Figures \ref{subfig:Brugge_boxplot_objRealIter_seis_C1_S3} and \ref{subfig:Brugge_boxplot_objRealIter_seis_C5_S3}, in comparison to those in Figure \ref{fig:Brugge_boxplot_objRealIter_S2}.       

\renewcommand{\nScale}{0.45} 
\begin{figure*}
	\centering
	\subfigure[RMSEs of log PERMX ($c=1$)]{ \label{subfig:rmse_PERMX_boxplot_ensemble_C1_S3}
		\includegraphics[scale=\nScale]{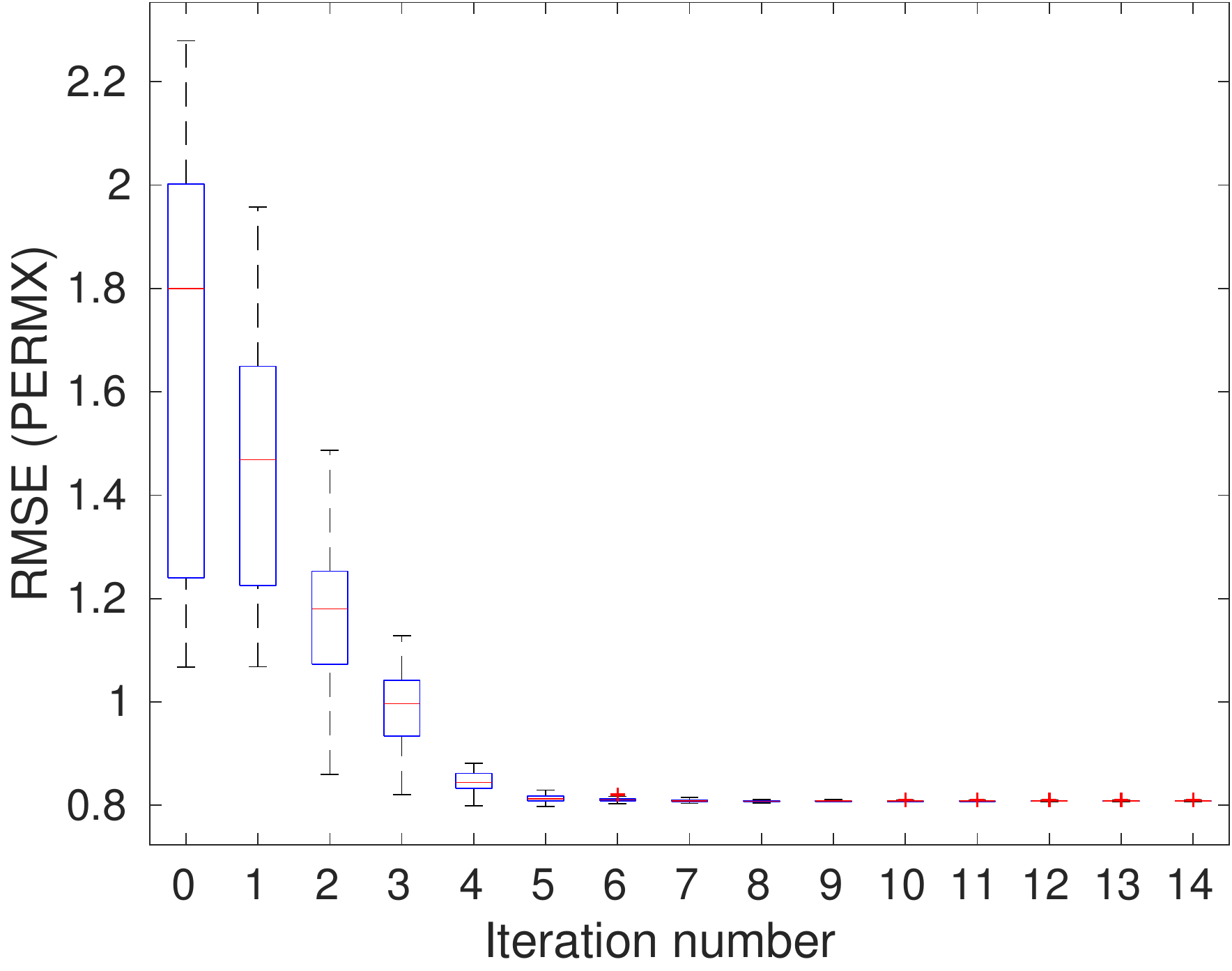}
	}%
	\subfigure[RMSEs of PORO ($c=1$)]{ \label{subfig:rmse_PORO_boxplot_ensemble_C1_S3}
		\includegraphics[scale=\nScale]{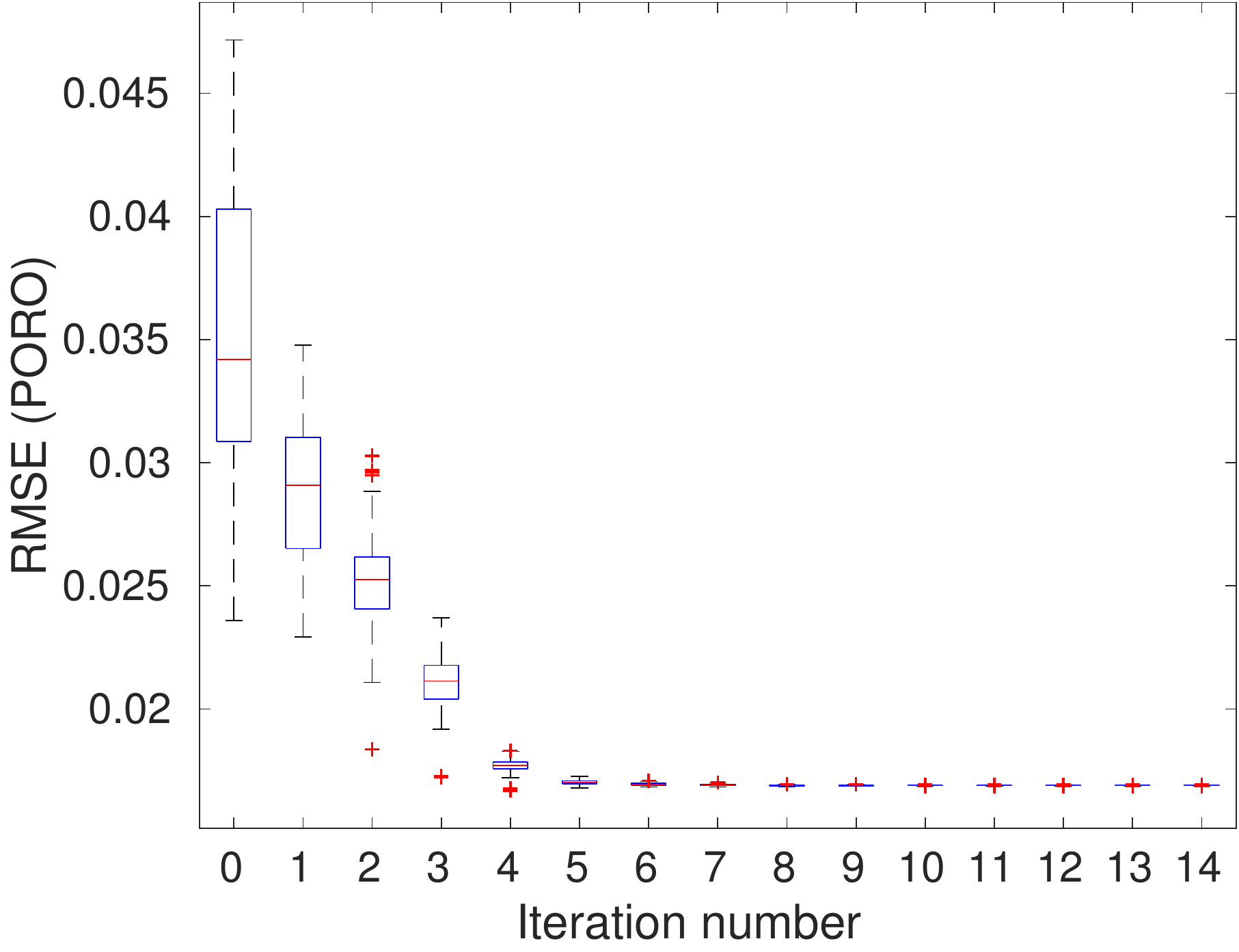}
	}%
	
	\subfigure[RMSEs of log PERMX ($c=5$)]{ \label{subfig:rmse_PERMX_boxplot_ensemble_C5_S3}
		\includegraphics[scale=\nScale]{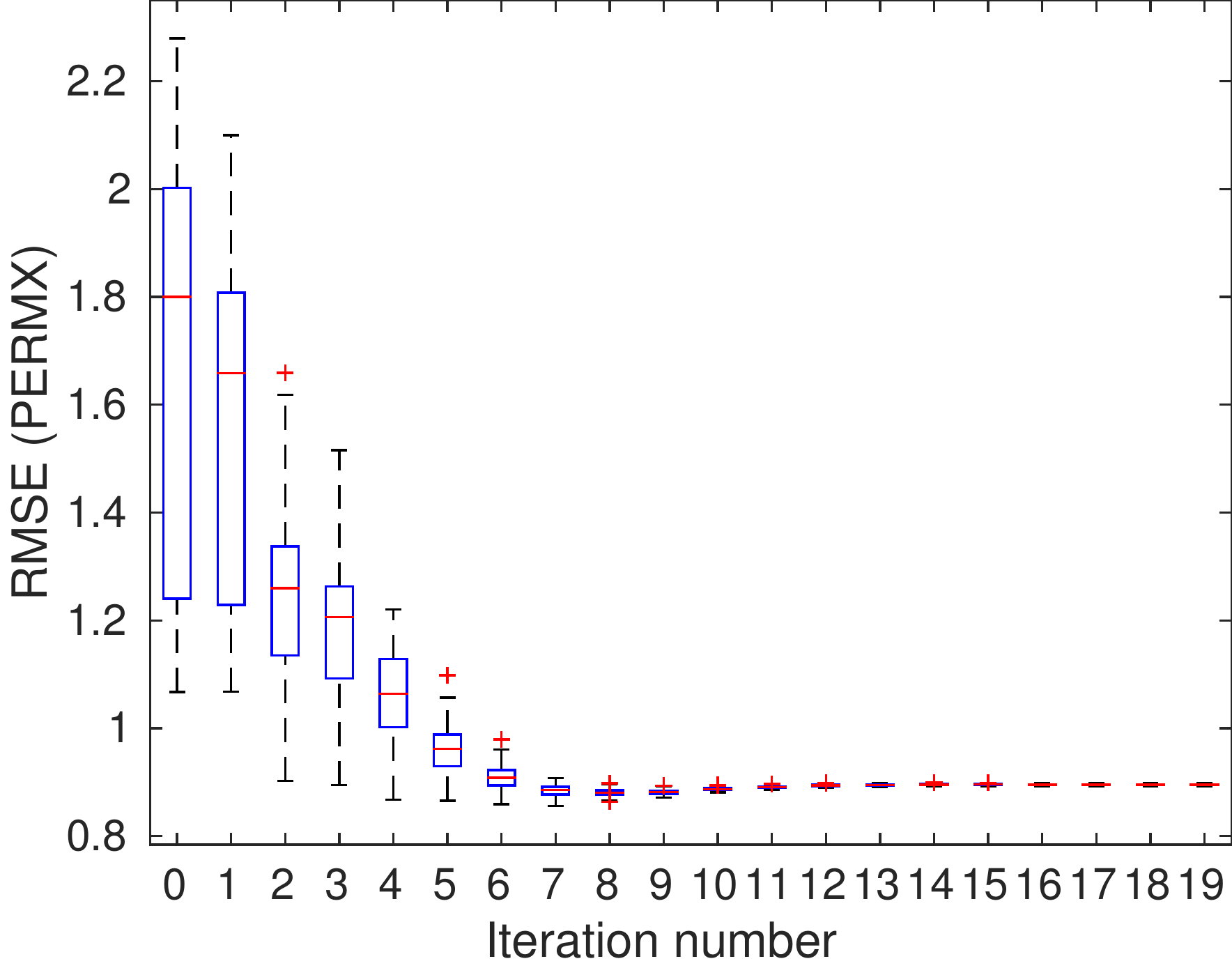}
	}%
	\subfigure[RMSEs of PORO ($c=5$)]{ \label{subfig:rmse_PORO_boxplot_ensemble_C5_S3}
		\includegraphics[scale=\nScale]{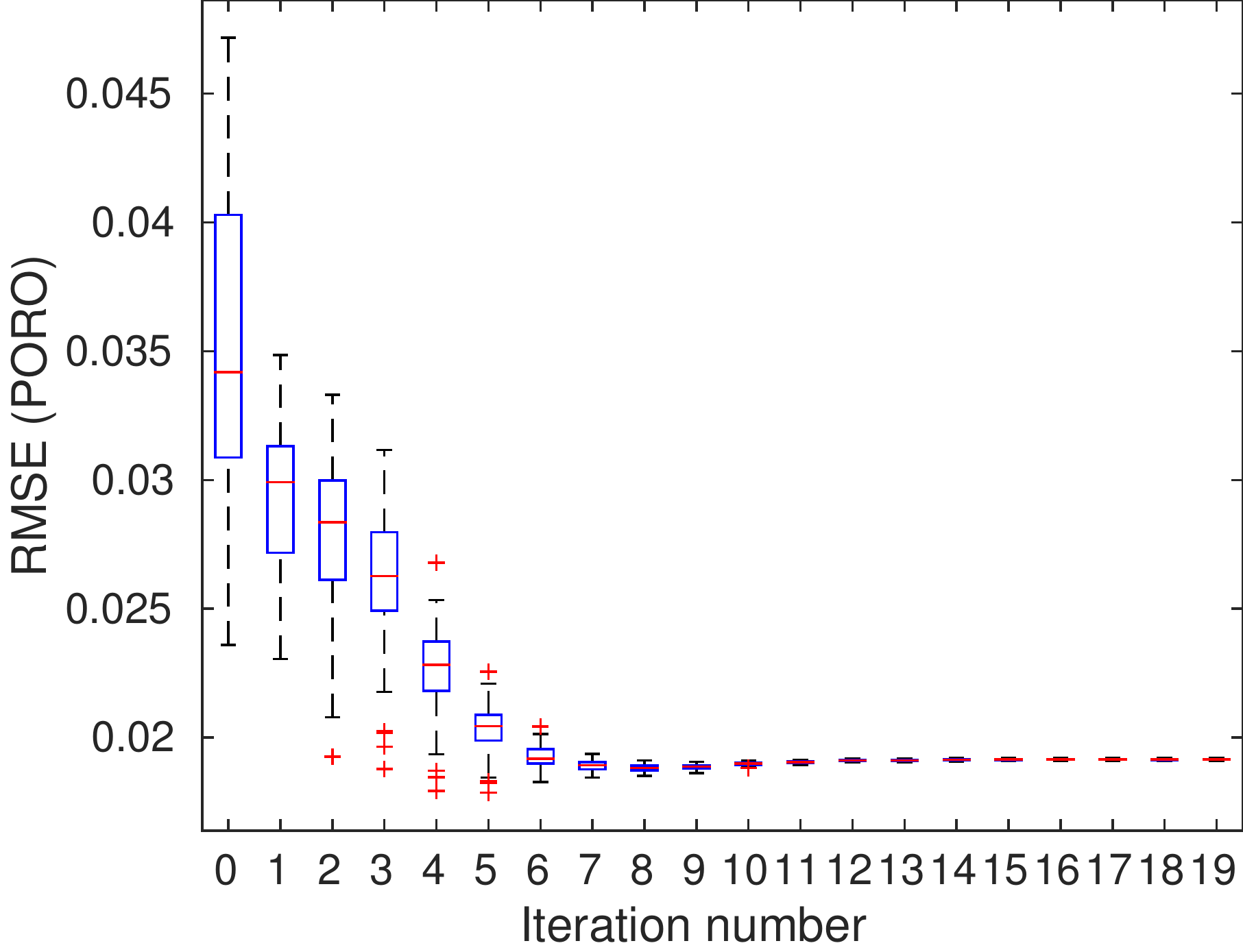}
	}%
	\caption{\label{fig:Brugge_RLM-MAC_RMSE_S3} Boxplots of RMSEs of log PERMX (1st column) and PORO (2nd column) as functions of iteration step, with $c$ being $1$ (top) and $5$ (bottom), respectively (scenario S3).}
\end{figure*}

Figure \ref{fig:Brugge_RLM-MAC_RMSE_S3} shows boxplots of RMSEs of log PERMX (1st column) and PORO (2nd column) as functions of iteration step. Again, the RMSEs of the final ensembles are lower than those of the initial ones, for either $c=1$ or $c=5$. On the other hand, a comparison of Figures \ref{fig:Brugge_RLM-MAC_RMSE_S1}, \ref{fig:Brugge_RLM-MAC_RMSE_S2} and \ref{fig:Brugge_RLM-MAC_RMSE_S3} indicates that the RMSEs of log PERMX and PORO (and similarly, the RMSEs of log PERMY and log PERMZ) are the lowest when using both production and seismic data in history matching. Using $c=5$ in scenario S3 (Figures \ref{subfig:rmse_PERMX_boxplot_ensemble_C5_S3} and \ref{subfig:rmse_PORO_boxplot_ensemble_C5_S3}) leads to higher RMSEs than using $c=1$. Nevertheless, they are still better than the RMSEs in scenario S1, and close to (for PORO) or better than (for log PERMX) those in scenario S2 with $c=1$ (see Figures \ref{subfig:rmse_PERMX_boxplot_ensemble_c1_S2} and \ref{subfig:rmse_PORO_boxplot_ensemble_c1_S2}). This suggests that, in this particular case, reasonably good history matching performance can still be achieved, even though the data size is reduced more than $4000$ times (at $c=5$) through the wavelet-based sparse representation procedure. 

\renewcommand{\nScale}{0.33} 
\begin{figure*}
	\centering
	\subfigure[WBHP of the initial ensemble]{ \label{subfig:WBHP_BR-P-5_initial_S3} 
		\includegraphics[scale=\nScale]{WBHP_BR-P-5_initial.eps}
	}%
	\subfigure[WOPR of the initial ensemble]{ \label{subfig:WOPR_BR-P-5__initial_S3}
		\includegraphics[scale=\nScale]{WOPR_BR-P-5_initial.eps}
	}%
	\subfigure[WWCT of the initial ensemble]{ \label{subfig:WWCT_BR-P-5__initial_S3}
		\includegraphics[scale=\nScale]{WWCT_BR-P-5_initial.eps}
	}%
	
	\subfigure[WBHP of the final ensemble]{ \label{subfig:WBHP_BR-P-5_S3} 
		\includegraphics[scale=\nScale]{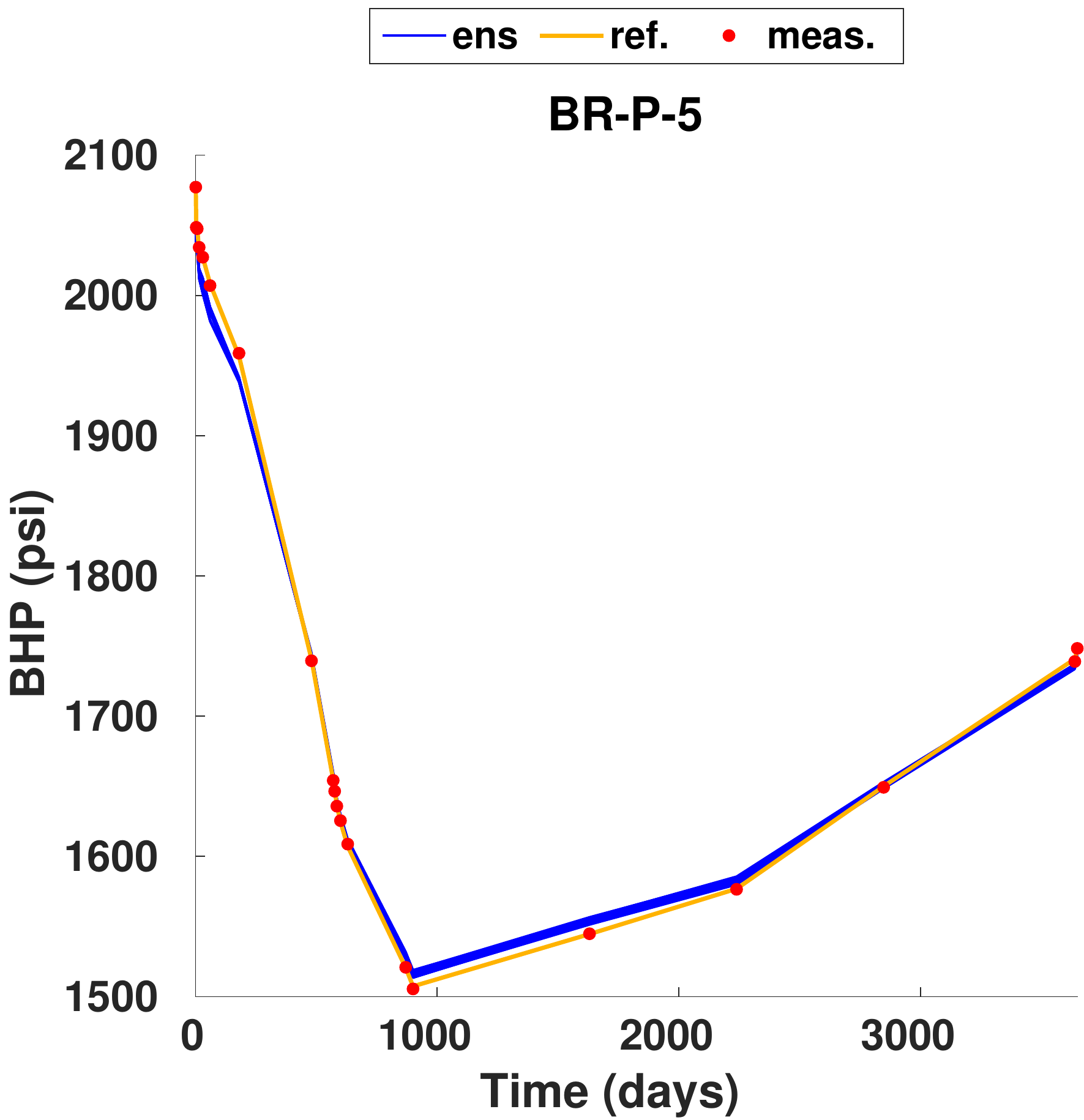}
	}%
	\subfigure[WOPR of the final ensemble]{ \label{subfig:WOPR_BR-P-5_S3}
		\includegraphics[scale=\nScale]{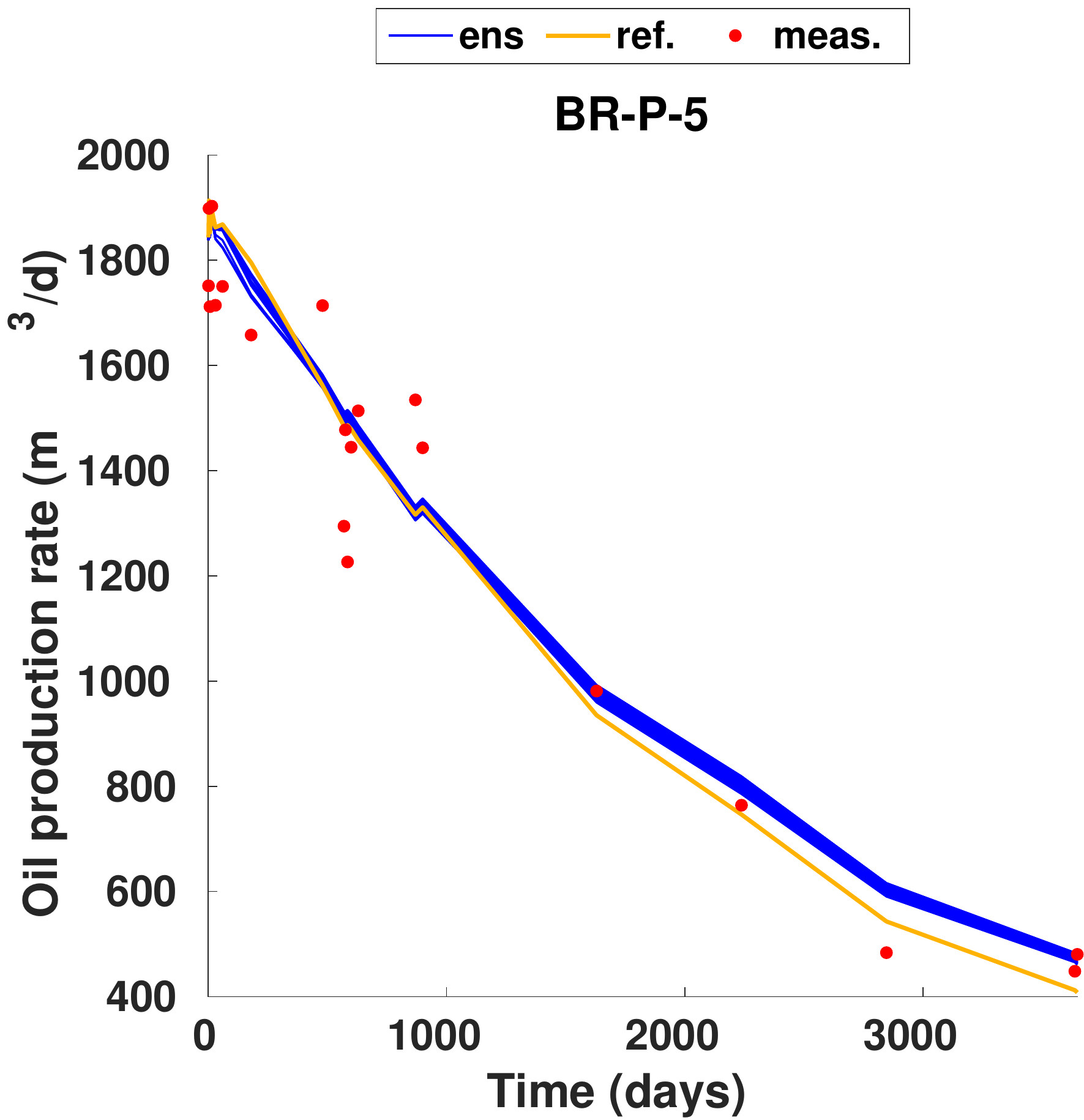}
	}%
	\subfigure[WWCT of the final ensemble]{ \label{subfig:WWCT_BR-P-5_S3}
		\includegraphics[scale=\nScale]{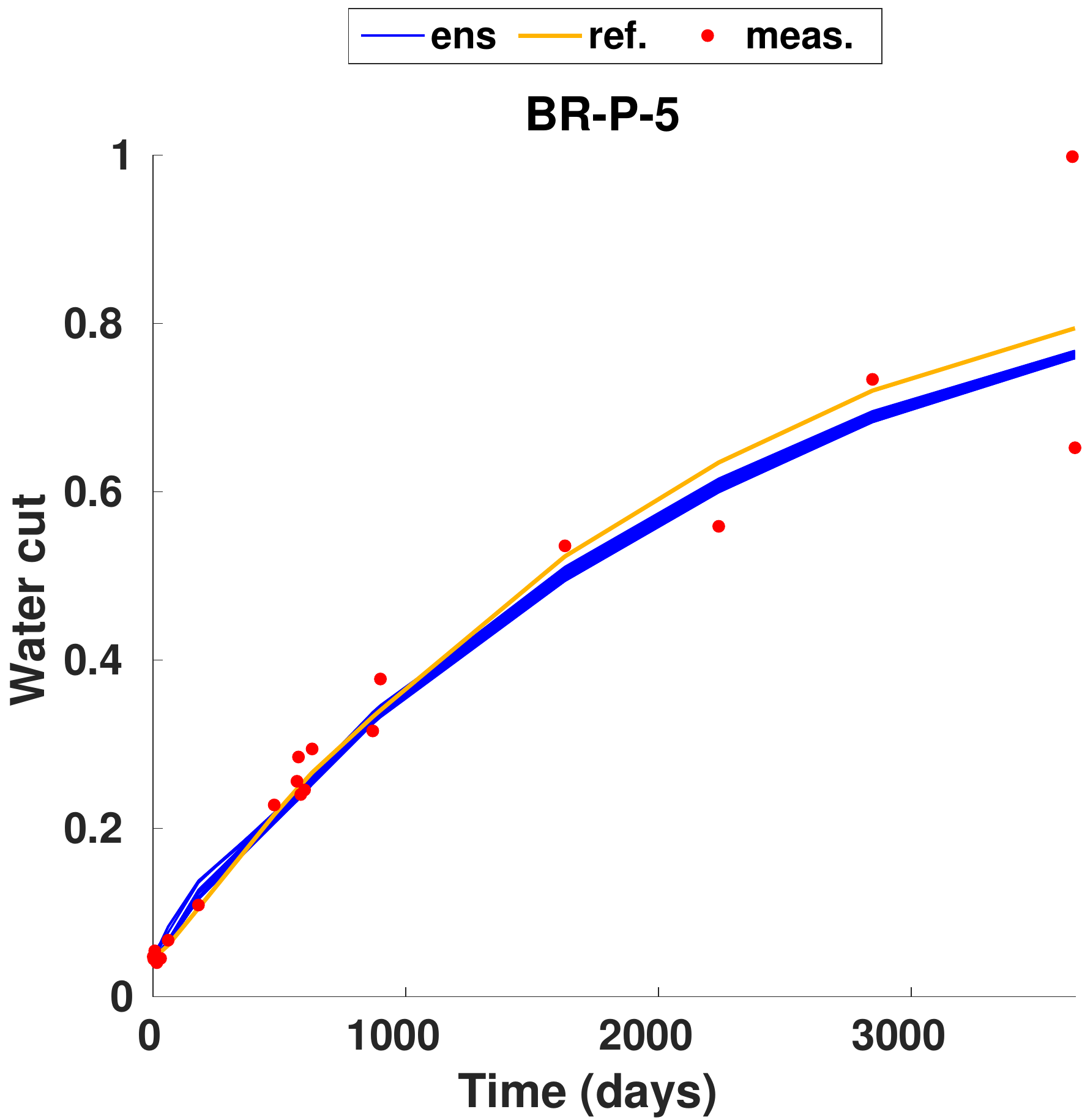}
	}%
	
	\caption{\label{fig:production_P5_S3} As in Figure \ref{fig:production_P5_S1}, but for the production data profiles in scenario S3 with $c=1$.}
\end{figure*} 

Again, for brevity, in what follows we only report the results with respect to the case $c=1$. Figure \ref{fig:production_P5_S3} shows the production data profiles at the producer BR-P-5 in scenario S3 with $c=1$. Clearly, compared to the initial ensemble, the final one matches the observed production data (red dots) better. Nevertheless, a comparison of the bottom rows of Figures \ref{fig:production_P5_S1} and \ref{fig:production_P5_S3} indicates that, the ensemble spreads of simulated production data (blue curves) tend to be under-estimated, such that the reference production data (yellow curves) are outside the profiles of simulated production data at certain time instances. 

\renewcommand{\nScale}{0.33} 
\begin{figure*}
	\centering
	\subfigure[Slice of initial differerence (base)]{ \label{subfig:X80_tstep1_mid_trace_diff_initEns_S3}
		\includegraphics[scale=\nScale]{X80_tstep1_mid_trace_diff_initEns_S2.eps}
	}%
	\subfigure[Slice of initial differerence (1st mornitor)]{ \label{subfig:X80_tstep2_mid_trace_diff_initEns_S3}
		\includegraphics[scale=\nScale]{X80_tstep2_mid_trace_diff_initEns_S2.eps}
	}%
	\subfigure[Slice of initial differerence (2st mornitor)]{ \label{subfig:X80_tstep3_mid_trace_diff_initEns_S3}
		\includegraphics[scale=\nScale]{X80_tstep3_mid_trace_diff_initEns_S2.eps}
	}%
	
	\subfigure[Slice of final differerence (base)]{ \label{subfig:X80_tstep1_mid_trace_diff_finalEns_S3}
		\includegraphics[scale=\nScale]{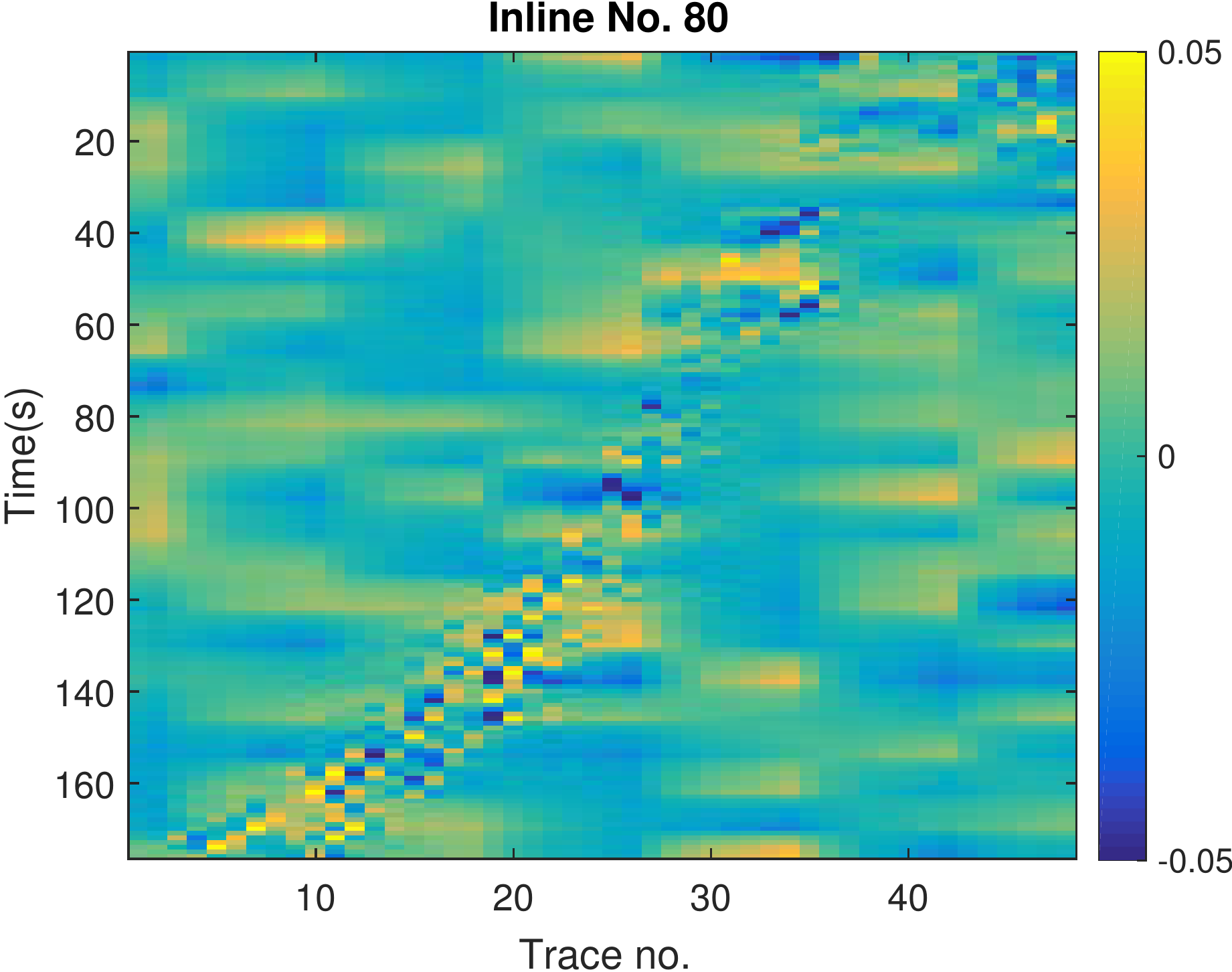}
	}%
	\subfigure[Slice of final differerence (1st mornitor)]{ \label{subfig:X80_tstep2_mid_trace_diff_finalEns_S3}
		\includegraphics[scale=\nScale]{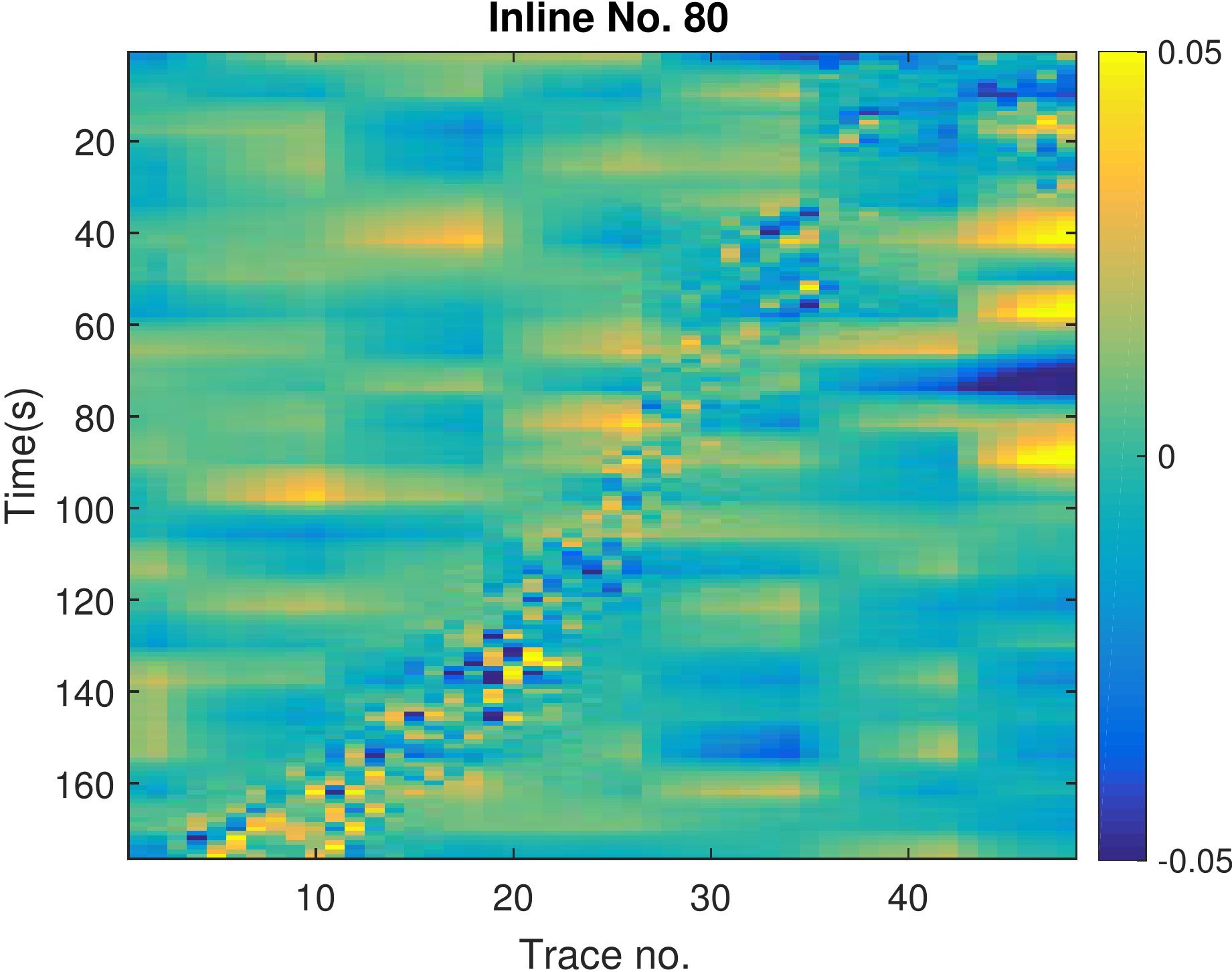}
	}%
	\subfigure[Slice of final differerence (2st mornitor)]{ \label{subfig:X80_tstep3_mid_trace_diff_finalEns_S3}
		\includegraphics[scale=\nScale]{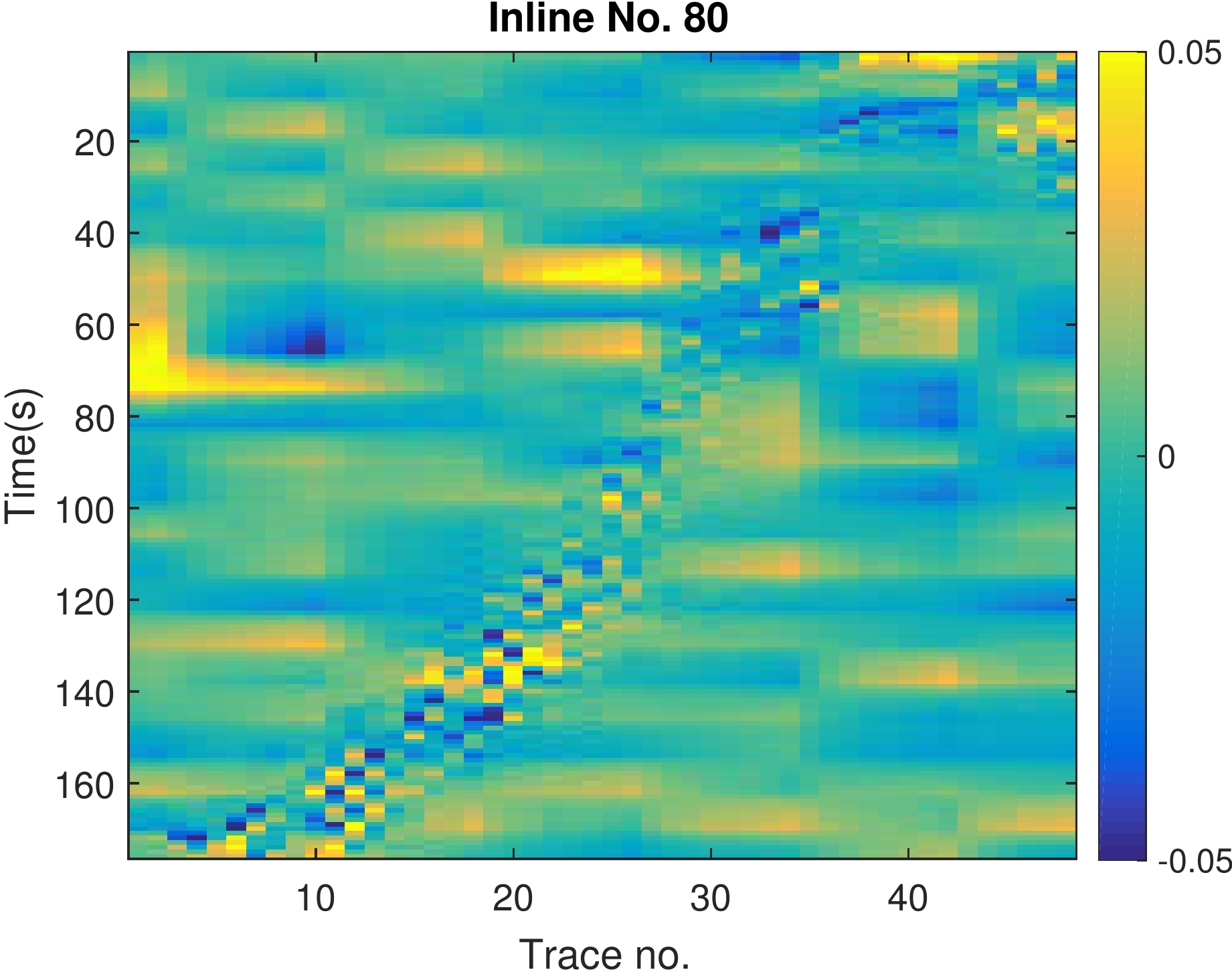}
	}%
	
	\caption{\label{fig:diff_slices_S3} As in Figure \ref{fig:diff_slices_S3}, but for the slices (at $X=80$) of differences in scenario S3 with $c=1$. }
\end{figure*}

Similar to Figure \ref{fig:diff_slices}, in Figure \ref{fig:diff_slices_S3} we show the slices (at X = 80) of differences between the reconstructed far-offset AVA cubes of observed and mean simulated seismic data, at three survey times. Again, compared to the slices with respect to the initial ensemble (top), there are some visible distinctions in the slices with respect to the final ensemble (bottom). However, if one compares the bottom rows of Figures \ref{fig:diff_slices} and \ref{fig:diff_slices_S3}, it seems that these slices look very similar to each other. This is consistent with the results in Figures \ref{subfig:Brugge_boxplot_objRealIter_c1_S2} and \ref{subfig:Brugge_boxplot_objRealIter_seis_C1_S3}, where the final seismic data mismatch of S2 and S3 remains close.    

\renewcommand{\nScale}{0.21} 
\begin{figure*}
	\centering
	\subfigure[Rerefrence log PERMX]{ \label{subfig:PERMX_L2_true_S3} %
		\includegraphics[scale=\nScale]{PERMX_L2_true.eps}
	}
	\subfigure[Mean of initial log PERMX]{ \label{subfig:PERMX_L2_Mean_initEns_S3}
		\includegraphics[scale=\nScale]{PERMX_L2_Mean_initEns.eps}
	}
	\subfigure[Mean of final log PERMX]{ \label{subfig:PERMX_L2_Mean_ensemble11_S3}
		\includegraphics[scale=\nScale]{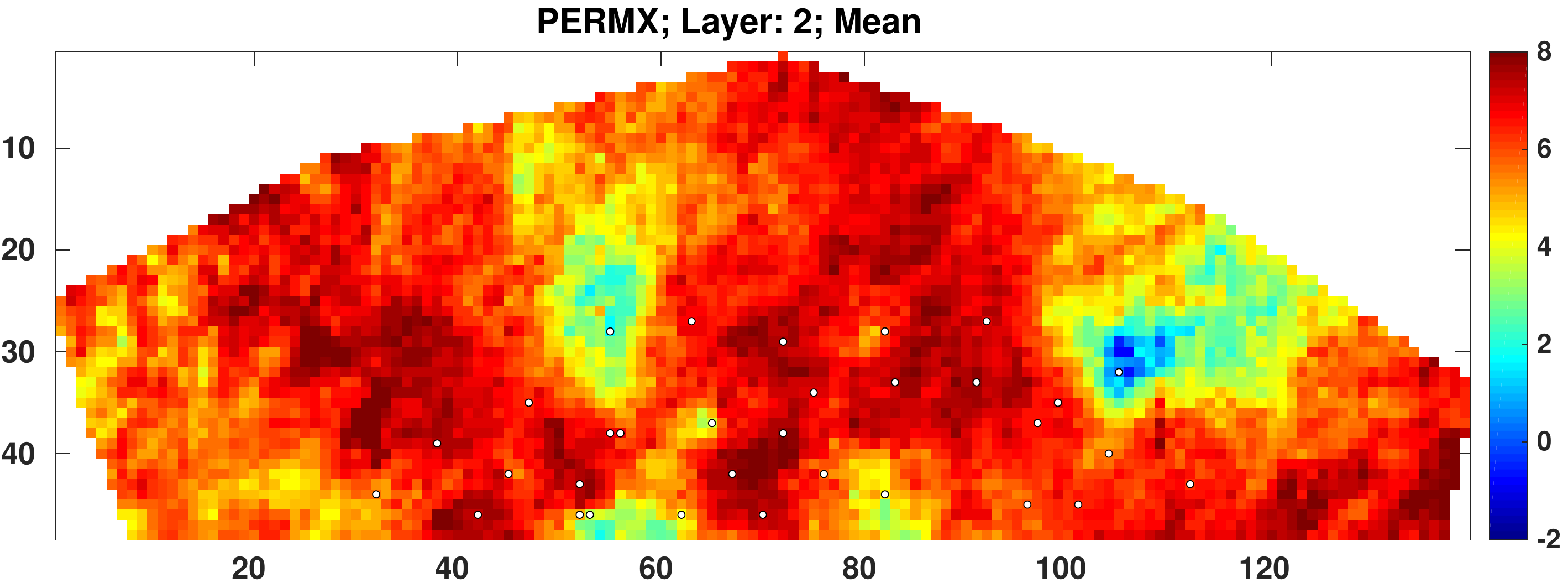}
	}%
	
	\subfigure[Rerefrence PORO]{ \label{subfig:PORO_L2_true_S3} %
		\includegraphics[scale=\nScale]{PORO_L2_true.eps}
	}
	\subfigure[Mean of initial PORO]{ \label{subfig:PORO_L2_Mean_initEns_S3}
		\includegraphics[scale=\nScale]{PORO_L2_Mean_initEns.eps}
	}
	\subfigure[Mean of final PORO]{ \label{subfig:PORO_L2_Mean_ensemble11_S3}
		\includegraphics[scale=\nScale]{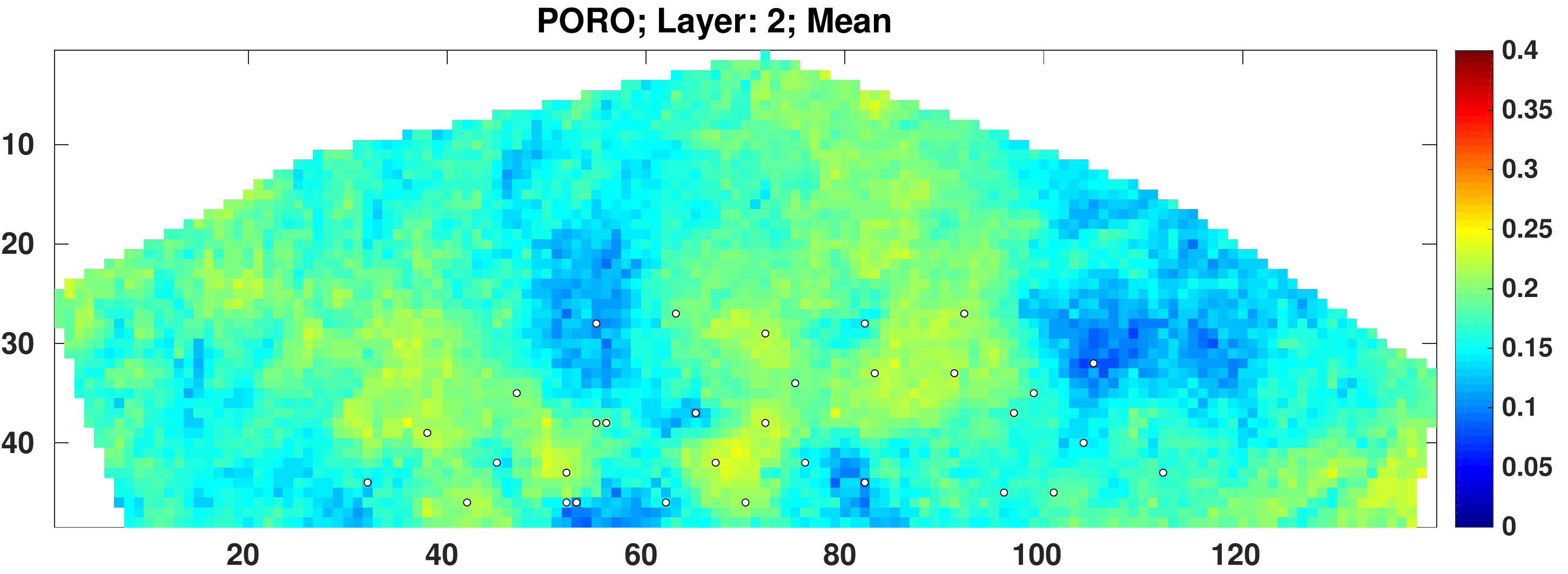}
	}%
	\caption{\label{fig:estimation_S3} As in Figure \ref{fig:estimation_S1}, but for scenario S3 with $c=1$.}
\end{figure*}  

Finally, Figure \ref{fig:estimation_S3} compares the reference, initial and final mean log PERMX and PORO fields at layer 2. Again, the final mean estimates improve over the initial mean fields, in terms of the closeness to the references. In addition, a comparison of the final estimated fields (the 3rd columns) of Figures \ref{fig:estimation_S1}, \ref{fig:estimation_S2} and \ref{fig:estimation_S3} shows that the final mean estimates in S3 best capture the geological structures of the reference fields (the same observation is also obtained at $c = 5$). This indicates the benefits of using both production and seismic data in history matching.

\section{Discussions and conclusions}\label{sec:conclusion}

In this work, we apply an efficient, ensemble-based seismic history matching framework to the 3D Brugge field case. The seismic data used in this study are near- and far-offset amplitude versus angle (AVA) attributes, with the data size more than 7 million. To handle the big data, we introduce a wavelet-based sparse representation procedure to substantially reduce the data size, while capturing the main features of the seismic data as many as possible. Through numerical experiments, we demonstrate the efficacy of the proposed history matching framework with the sparse representation procedure, even in the case that the seismic data size is reduced more than 4000 times. The size of seismic data (in the form of leading wavelet coefficients) can be conveniently controlled through a threshold value. A relatively large threshold value means more reduction in data size, which is desirable for the history matching algorithm, but at the cost of extra information loss. In contrast, a relatively small threshold value results in a larger number of leading wavelet coefficients, hence better preserves the information content of observed data. In this case, however, the history matching algorithm may become more vulnerable to certain practical issues like ensemble collapse. As a result, the best practice would need to achieve a trade-off between reduction of data size and preservation of data information. Another observation from the experiment results is that, in this particular case, a combined use of production and seismic data in history matching leads to better estimation results than the cases of using either production or seismic data only.

Ensemble collapse is clearly visible when seismic data is used in history matching. This phenomenon can be mitigated to some extent by increasing the threshold value (hence reducing the seismic data size), but it cannot be completely avoided. A possible remedy to this problem is to also introduce localization (see, for example, \citealp{Emerick2011combining,chen2010cross}) to the iterative ensemble smoother. In the presence of the sparse presentation procedure, however, seismic data are transformed into wavelet domain, and the concept of ``physical distance'' may not be valid any more. As a result, localization will need to be adapted to this change. We will investigate this issue in our future study.        

\section*{Acknowledgments}
\label{sec:acknowledgments}
\noindent
We would like to thank Schlumberger for providing us academic software licenses to ECLIPSE$^\copyright$. XL acknowledges partial financial supports from the CIPR/IRIS cooperative research project ``4D Seismic History Matching'' which is funded by industry partners Eni,  Petrobras, and Total, as well as the Research Council of Norway (PETROMAKS). All authors acknowledge the Research Council of Norway and the industry partners -- ConocoPhillips Skandinavia AS, BP Norge AS, Det Norske Oljeselskap AS, Eni Norge AS, Maersk Oil Norway AS, DONG Energy A/S, Denmark, Statoil Petroleum AS, ENGIE E\&P NORGE AS, Lundin Norway AS, Halliburton AS, Schlumberger Norge AS, Wintershall Norge AS -- of The National IOR Centre of Norway for financial supports.

\bibliography{references,tuhin}
\bibliographystyle{firstbreak}

\end{document}